\documentclass[preprint,prc,
tightenlines,
nofootinbib
,superscriptaddress]{revtex4}
\usepackage{amssymb}
\usepackage{amsmath}
\usepackage{graphicx}
\usepackage{bm} 
\usepackage[loose]{subfigure}
\usepackage{xcolor}
\usepackage[squaren]{SIunits}
\usepackage{multirow}
\usepackage{braket}
\usepackage{tensor}
\usepackage{slashed}
\usepackage{array}

\newcommand\numberthis{\addtocounter{equation}{1}\tag{\theequation}}

\newcommand{\di}{\mathrm{d}}

\newcommand{\punkt}{\;\text{.}}
\renewcommand{\vec}[1]{\boldsymbol{#1}}
\newcommand{\komma}{\;\text{,}}
\renewcommand{\i}{\mathrm{i}}

\allowdisplaybreaks

\begin{document}
\title{Elastic pion-nucleon scattering in chiral perturbation theory:\\
  Explicit $\Delta$(1232) degrees of freedom }

\author{D.~Siemens}
\email[]{dmitrij.siemens@rub.de}
\affiliation{Independent}

\begin{abstract}
For the first time, elastic pion-nucleon scattering is analyzed in the framework of
chiral perturbation theory up to fourth order with explicit $\Delta$ degrees of freedom.
The analysis is performed within the heavy-baryon expansion as well as
in a covariant approach based on an extended on-mass-shell renormalization
scheme. The renormalization of low-energy constants in both chiral
approaches is discussed in detail and the explicit expressions to
cancel both power-counting breaking as well as decoupling breaking terms are given.
The low-energy constants from the $2\pi \bar{N}N$ interaction as well as
additional constants from the $\Delta$ sector are reliably constrained by fits to
experimental data. The traditional $K$-matrix unitarization is
employed in the near threshold region, whereas a complex mass approach
is used to extend the applicability of the theory up to the $\Delta$ pole region.
Additionally, we estimate a theoretical error based
on the truncation of the chiral series as employed in recent
analyses of nuclear forces as well as pion-nucleon scattering. 
The obtained results provide a clear evidence that the explicit
inclusion of $\Delta$ degrees of freedom is fundamental to describe
pion-nucleon physics at threshold. The resulting predictions for the
subthreshold and threshold parameters as well as phase shifts are in
excellent agreement with the ones determined by the recent Roy-Steiner
analysis of pion-nucleon scattering.
\end{abstract}

\maketitle

\section{Introduction}
Relying on the approximate chiral symmetry of QCD and its strong constraints on low-energy hadronic dynamics,
Chiral perturbation theory ($\chi$PT), an effective field theory of the strong interactions, provides the toolkit to
perform a systematically improvable expansion of low-energy hadronic observables around the chiral and zero-energy limit.
Starting with the pioneering work in the meson sector \cite{Weinberg:1978kz,Gasser:1983yg,Gasser:1984gg}, 
$\chi$PT has been extended to the single-baryon and few baryon sectors 
\cite{Gasser:1987rb,Bernard:1995dp,Bernard:2006gx,Bernard:2007zu,Weinberg:1990rz,Ordonez:1995rz,Epelbaum:2008ga,Machleidt:2011zz},
including numerous applications and extensions like the heavy baryon (HB) approach, 
the infrared renormalization scheme or the extended on-mass shell scheme \cite{Jenkins:1990jv,Bernard:1992qa,Becher:1999he,Ellis:1997kc,Gegelia:1999gf,Fuchs:2003qc}.  
The interested reader is referred to Ref.~\cite{Bernard:2007zu} for a detailed discussion and comparison of the various formulations of $\chi$PT.
\\\\
One of the most studied hadronic processes in $\chi$PT is low-energy pion-nucleon scattering. 
The recent interest in this reaction stems from the observation that the pion-nucleon LECs enter
the two-pion exchange contributions to the two- and three-nucleon forces 
\cite{Krebs:2012yv,Wendt:2014lja,Entem:2014msa,Entem:2015xwa,Epelbaum:2014sza}. 
Thus, these pion-nucleon LECs are vital inputs for nuclear chiral EFT \cite{Epelbaum:2015pfa} and
a reliable extraction of these LECs becomes a crucial task in understanding the long-range
behavior of nuclear forces 
\\\\
In this paper, elastic low-energy pion-nucleon scattering is analyzed in detail
within the HB and covariant baryon $\chi$PT formulation at the full one-loop order. 
In particular, the effects of including the $\Delta$ resonance as an explicit degree of freedom
in a consistent power counting is the focus of this analysis. In this work, we employ the so called small scale
expansion $\varepsilon$, where the difference between the $\Delta$(1232) mass and the nucleon mass is
counted as of the same order as the pion mass \cite{Hemmert:1997ye}. Mainly due to the analytic complexity of 
calculating loop graphs including the $\Delta$ resonance, previous analyses of 
pion-nucleon scattering in the framework of baryon $\chi$PT are shy of 
including explicit $\Delta$ degrees of freedom or only study the effects of leading-order $\Delta$ contributions
~\cite{Bernard:1992qa,Mojzis:1997tu,Fettes:1998ud,
Buettiker:1999ap,Fettes:2000gb,Fettes:2000xg,Becher:2001hv,Hoferichter:2009gn,
Gasparyan:2010xz,Alarcon:2012kn,Chen:2012nx,Siemens:2016hdi}.
Within the small scale expansion, the first calculations of loop contributions to the $\pi$N amplitude 
were performed in the HB approach two decades ago \cite{Fettes:2000bb} and only a couple years ago
in the covariant approach as well \cite{Yao:2016vbz}. In both analyses, the amplitudes are determined up to
the leading-loop order $\varepsilon^3$ and fits to the $S$- and $P$-wave phase shifts are performed to extract
the $\pi$N-LECs.
\\\\
The analysis carried out in this paper is strongly motivated by the obstacles encountered in Ref.~\cite{Yao:2016vbz}.
Thus, we give a brief summary of this paper in the following.
In Ref.~\cite{Yao:2016vbz}, we calculated the pion-nucleon scattering amplitude in the small scale expansion
up to order $\varepsilon^3$ and renormalized $2\pi\bar N N$-LECs within the EOMS scheme \cite{Gegelia:1999gf,Fuchs:2003qc}, such that 
only the LECs $c_i$ are shifted to absorb power counting breaking terms (PCBTs). In addition, we employed a complex $\Delta$
mass in the amplitudes. Finally, the unknown LECs are extracted from fits to the Roy-Steiner (RS) phase shifts including their uncertainties.
Given that the RS analysis provides parametrization for the $S$- and $P$-waves, equidistant points are generated and then the corresponding
mean values are normally distributed, such that the reduced $\chi^2$ gets the proper definition of $\chi^2/$dof $\sim1$.\linebreak
The outcome of this analysis is a good agreement with the RS values \cite{Hoferichter:2015hva} regarding the fitted phase shifts and predicted
threshold parameters. Unfortunately, the LECs extracted from the fits to the phase shifts turn out to be strongly correlated.
Note that the above mentioned $\Delta$-less analyses exhibited strong correlations only at order $Q^4$. 
Compared to a $\Delta$-less analysis at order $Q^3$, only
two more LECs are introduced by the explicit $\Delta$ degrees of freedom at order $\varepsilon^3$. 
All in all, a reliable determination of LECs from fitting to phase shifts
seems problematic, especially when considering even higher-order calculations like the one carried out in this work.
\\\\
Encouraged by our analysis in Ref.~\cite{Siemens:2016hdi}, this paper follows the more tedious path
of studying the large amount of available data on pion-nucleon scattering observables at low energies.
Furthermore, we employ a systematic approach of including the theoretical uncertainty stemming from
the truncation of the chiral expansion into the fitting procedure, which was motivated in Ref.~\cite{Epelbaum:2014efa}.
Note that recently a more sophisticated approach for estimating the truncation errors was introduced in Refs.~\cite{Epelbaum:2019zqc,Melendez:2017phj}.
In contrast to the analyses where phase shifts are used as inputs in the fitting routine, we are able to 
give predictions for the phase shifts based on experimental scattering data employing $\chi$PT amplitudes.
In addition, two different unitarization prescriptions for the $\pi$N amplitudes are employed in this paper.
First, we study the near threshold region and employ the standard $K$-matrix unitarization. Second, when increasing
the energy up to the $\Delta$ pole region, we employ a complex mass for the $\Delta$ resonance instead.
To the best of our knowledge, this is the first full one-loop order calculation including
$\Delta$ degrees of freedom within the HB and covariant baryon framework of
$\chi$PT. In the small scale expansion this corresponds to order $\varepsilon^4$.
At this time, it is practically not feasible to perform calculations beyond one-loop level, mainly
due to the complexity of multi-loop amplitudes.
Another achievement of this work is the first time discussion of the complete renormalization procedure of the
leading-order coupling constants and the pion-nucleon amplitude up to order $\varepsilon^4$ in both chiral approaches.
\\\\
Given the motivation above, it is natural that we extend the $\Delta$-less analysis in Ref.~\cite{Siemens:2016hdi} in this paper.
Thus, we refer the unfamiliar reader for details on kinematics, observables, renormalization and fitting procedure to this reference.
For the sake of brevity, only the fundamental differences due to the explicit inclusion of the $\Delta$ degrees of freedom
are discussed in the following sections.
\\\\
The organization of this paper is as follows.
In section~\ref{sec:basicdef-Delta}, we describe the two different unitarization
prescriptions, $K$-matrix unitarization and complex mass approach. Then, in section~\ref{sec:power-count-renorm-Delta}, 
the renormalization schemes including explicit $\Delta$ dynamics are discussed in detail for both chiral
approaches. The specifics of the fitting procedure are explained in section~\ref{sec:fitting-procedure-Delta}. 
In section~\ref{sec:predictions-Delta}, the predicted observables in both unitarization schemes are visualized and discussed. 
Finally, we summarize the main results of this analysis in section \ref{sec:sum-Delta}.
In the appendix we give explicit expressions for the renormalized LECs in the HB framework
and refer to the supplementary material in the form of a Mathematica notebook for the covariant expressions.

\section{Basic Definitions}
\label{sec:basicdef-Delta}
The extraction of real-valued phase shifts demands an unitarization
prescription for the perturbative partial wave amplitudes.
Whereas the standard approach is to use the $K$-matrix unitarization for all
partial waves, we employ a second additional unitarization prescription
in the analysis presented here. In the following, we review the
general idea of an unitarized perturbative amplitude. In particular,
we demonstrate the breakdown of the $K$-matrix
unitarization and emphasize the need of another more general
unitarization prescription.
\\\\
For an elastic scattering process, the transition matrix is given by
\begin{equation}
  \label{eq:7a}
  T_{l\pm}(s)=|\vec{q}|f_{l\pm}(s)=\frac{1}{2\i}\left(e^{2\i\delta_{l\pm}(s)}-1\right)\komma
\end{equation}
where $\vec{q}$ is the center-of-mass three-momentum. 
Phase shifts can be extracted from the partial wave amplitude via
\begin{equation}
  \label{eq:5a}
  \delta^I_{l \pm}(s)=\mathrm{Arg}(f^I_{l\pm}(s))\punkt
\end{equation}
Below the inelastic threshold, partial wave unitarity 
\begin{equation}
  \label{eq:6a}
  \mathrm{Im}\; f^I_{l\pm}(s)= |\vec{q}| |f^I_{l\pm}(s)|^2
\end{equation}
can be used to write Eq.~\eqref{eq:5a} as
\begin{equation}
  \label{eq:2a}
  \delta^I_{l \pm}(s) =\arctan\left(\frac{|\vec{q}| (\mathrm{Re}\, (f^I_{l\pm}(s))^2+\mathrm{Im}\, (f^I_{l\pm}(s))^2)} {\mathrm{Re}\, (f^I_{l\pm}(s))}\right)
=\arctan\left(\frac{|\vec{q}|} {\mathrm{Re}\, (f^I_{l\pm}(s) ^{-1})}\right)\punkt
\end{equation}
In $\chi$PT, however, the partial wave amplitude up to chiral order $n$
\begin{equation}
  \label{eq:9a}
  f=f^{(1)}+f^{(2)}+\dots+f^{(n)}
\end{equation}
fulfils the unitarity condition in
Eq.~\eqref{eq:6a} not exactly but only perturbatively, e.g., for $n=4$ one has
\begin{equation}
  \label{eq:8a}
  \begin{aligned}
    \mathrm{Im}\; f^{(3)}+\mathrm{Im}\;
    f^{(4)}&= |\vec{q}|
    |f^{(1)}+f^{(2)}+f^{(3)}+f^{(4)}|^2\\
    &=|\vec{q}|[(\mathrm{Re}f^{(1)})^2+2\,\mathrm{Re}f^{(1)}\mathrm{Re}f^{(2)}]+\dots\punkt
  \end{aligned}
\end{equation}
Thus, the idea is to enforce the unitarity condition on the
perturbative amplitude by employing Eq.~\eqref{eq:2a} instead of
Eq.~\eqref{eq:5a} such that the extracted phase shifts are real-valued. 
For the special case of non-resonant partial waves, 
Eq.~\eqref{eq:2a} can be expanded in $\mathrm{Im}\, f^I_{l\pm}(s)$
such that to leading-order one gets
\begin{equation}
  \label{eq:1a}
  \delta^I_{l \pm}(s) =\arctan(|\vec{q}|\mathrm{Re}\, f^I_{l\pm}(s))\punkt
\end{equation}
The above equality assumes that imaginary parts do not need to be
resummed and are always suppressed compared to the real parts.
To be more precise, Eq. \eqref{eq:8a} is equivalent to the so-called $K$-matrix
unitarization, where one chooses
\begin{equation}
  \label{eq:10a}
  f^K=\frac{\mathrm{Re}f}{1-\i |\vec{q}|\mathrm{Re}f}
\end{equation}
such that inserting in Eq.~\eqref{eq:5a} yields
\begin{equation}
  \label{eq:11a}
  \delta= \mathrm{Arg}(f^K)=\arctan \left(\frac{\mathrm{Im}f^K}{\mathrm{Re}f^K}\right)=
  \arctan(|\vec{q}|\mathrm{Re} f)\punkt
\end{equation}
Note that for the sake of simplicity, all indices have been
suppressed in the last two equations. We will use the same notation in the following as well.
In the case of resonant partial waves, the $K$-matrix unitarization prescription in
Eq.~\eqref{eq:10a} generates the respective resonance width
by an infinite resummation of self-energy contributions. 
This can be seen by considering only the pole contribution $f^{Pole}$
of a resonance and employing
the geometric series for Eq.~\eqref{eq:10a}, such that one has
\begin{equation}
  \label{eq:12a}
  f^K=\frac{\mathrm{Re}f^{Pole}}{1-\i |\vec{q}|\mathrm{Re}f^{Pole}}\simeq 
  \mathrm{Re}f^{Pole}(1+\i|\vec{q}| \mathrm{Re}f^{Pole}+(\i|\vec{q}| \mathrm{Re}f^{Pole})^2+\dots)\punkt
\end{equation}
One can also take into account further contributions from the
amplitude $f$, which will give an infinite resummation of many
different topologies. This approach is clearly not based
on a power counting scheme but non-perturbatively
resums higher-order contributions. 
However, for smaller values of the phase shifts this modification is
negligible. To be more precise, the condition is
\begin{equation}
  \label{eq:13a}
  |\vec{q}| \mathrm{Re}f= \tan \delta\ll 1
\end{equation}
or translating into a rule of thumb for the phase shifts
\begin{equation}
  \label{eq:14a}
  \tan \delta\simeq \delta\komma
\end{equation}
which is a good approximation for $|\delta|\ll\pi/6$.
\\\\
In the following, to extract real-valued phase shifts the partial wave amplitudes
deduced from the $T$-matrix of $\pi N\to\pi N$
are unitarized in two different ways:
\begin{itemize}
\item { $K$-matrix approach}:
The standard $K$-matrix unitarization given in
Eq.~\eqref{eq:1a} is used for all partial waves. This approach is used
in the threshold region, especially far below the $\Delta$ pole region. 
Here, we employ the
real-valued Breit-Wigner mass of the $\Delta(1232)$ in our
amplitudes.
\item Complex mass approach:
To extend the applicability of the theory to the $\Delta$ pole region the complex
mass renormalization scheme \cite{Stuart:1990vk,Denner:1999gp,Jambul}
is employed. Like in Ref.~\cite{Yao:2016vbz}, only the $P_{33}$ partial wave is unitarized by
the prescription in Eq.~\eqref{eq:2a}, whereas for all the
remaining non-resonant partial waves the $K$-matrix unitarization in Eq.~\eqref{eq:1a} is used. 
Here, we employ the complex-valued
pole mass of the $\Delta(1232)$ in our amplitudes, which is equivalent
to a resummation of graphs corresponding to the $\Delta$ width based on a
consistent power counting.
\end{itemize}
At this point, we emphasize that the $K$-matrix unitarization
should not be used for the $P_{33}$ partial wave in combination with a
complex-valued $\Delta$ mass in the amplitudes. This would lead to a
double counting of particular graphs due to the additional resummation
by the $K$-matrix unitarization, see Eq.~\eqref{eq:12a}. However, its fine
to employ the $K$-matrix unitarization for non-resonant partial waves, given
that for these partial waves the difference between Eq.~\eqref{eq:2a} and Eq.~\eqref{eq:1a} is of higher
order only.

\section{Renormalization Procedure}
\label{sec:power-count-renorm-Delta}

\subsection{Pion-Nucleon Scattering Amplitudes}
In this work, the chiral amplitudes for pion-nucleon scattering including explicit $\Delta$ degrees of freedom 
are calculated in the small scale expansion with the expansion parameter
\begin{equation}
  \label{eq2:3}
  \varepsilon=\left\{\frac{q}{\Lambda_b},\frac{M_\pi}{\Lambda_b},\frac{\Delta}{\Lambda_b}\right\}\komma\quad\text{where}\quad \Lambda_b\in\{\Lambda_\chi,4\pi F,m\}\punkt
\end{equation}
Note that the explicit expansion
in inverse powers of $m$ is only carried out in the HB framework.
To be precise the $T$-matrix for $\pi^a(q) \, N(p) \; \to \; \pi^b(q') \, N^\prime(p^\prime)$ can be decomposed in the following way
\begin{equation}
  \label{eq:1}
  T^{ba}=\chi^\dagger_{N^\prime}\left(\delta^{ab} T^+ +\i
    \epsilon^{bac} \tau_c T^-\right)\chi_N \komma
\end{equation}
where in the covariant approach
\begin{equation}
  \label{eq:3a}
  T^\pm= \bar{u}^{(s^\prime)}\left(D^\pm-\frac{1}{4m_N}[\slashed{q}^\prime,\slashed{q}]B^\pm\right)u^{(s)}
\end{equation}
and in the HB approach
\begin{equation}
  \label{eq:3b}
  T^\pm=\bar{u}_v^{(s^\prime)}\left(g^\pm+\,2\i\,\vec S\cdot \vec
    q\times \vec q^\prime h^\pm\right)u_v^{(s)}\punkt
\end{equation}
The Mandelstam variables are defined in the standard way
\begin{equation}
  \label{eq:17}
     s=(p+q)^2\komma\quad t=(q-q^\prime)^2\komma\quad
  u=(p^\prime-q)^2\komma\quad s+t+u=2m_N^2+2M_\pi^2\punkt
\end{equation}
The effective chiral Lagrangian to describe
pion-nucleon dynamics at the
full one-loop level with explicit $\Delta$ resonances thus reads
\begin{equation}
  \label{eq:L1}
  \begin{aligned}
    \mathcal{L}_{\mathrm{eff}}&=\mathcal{L}_{\pi\pi}^{(2)}+\mathcal{L}_{\pi\pi}^{(4)}
    +\mathcal{L}_{\pi N}^{(1)}+\mathcal{L}_{\pi
      N}^{(2)}+\mathcal{L}_{\pi N}^{(3)} +\mathcal{L}_{\pi
      N}^{(4)}\\&+\mathcal{L}_{\pi \Delta}^{(1)}+\mathcal{L}_{\pi
      \Delta}^{(2)} +\mathcal{L}_{\pi
      \Delta}^{(4)}\\&+\mathcal{L}_{\pi N\Delta}^{(1)}+\mathcal{L}_{\pi
      N\Delta}^{(2)} +\mathcal{L}_{\pi N\Delta}^{(3)}+\mathcal{L}_{\pi
      N\Delta}^{(4)}
  \end{aligned}
\end{equation}
with the individual terms presented in appendix \ref{sec:effective-lagrangian}.
Note that the notation for coupling constants and other quantities used in the following is explained in the above mentioned appendix as well.
Thus, before continuing reading this section, we refer the reader to look through this appendix first.
\\\\
In the following, we sketch the routine employed
to calculate the full $\varepsilon^4$ results in both chiral frameworks.
At tree-level, the covariant amplitudes were determined based on the
corresponding covariant Langrangian. Then, to get the HB expressions, 
we expanded those covariant amplitudes in inverse powers of the
nucleon mass $m$. In principle, one could proceed at
loop-level in analogy and perform a strict expansion of loop-level amplitudes in small parameters.
However, this is a non-trivial task, in particular, for loop functions
including several propagators.
Instead, we calculated both sets of loop-level amplitudes
based on the corresponding effective Lagrangians.
At this point, we emphasize the redundant dependence of the Lagrangian
presented in appendix \ref{sec:effective-lagrangian}
on the off-shell parameters $z_i$ and $y_i$ \cite{Tang:1996sq,Pascalutsa:2000kd,Krebs:2009bf}. 
To be precise, in the loop-level amplitudes, we only kept track of $z_0$, whereas the
dependence on $y_1$ and $z_1$ is a higher-order effect and thus was neglected. 
In the tree-level amplitudes, the off-shell parameters $y_i$ and $z_i$
would show up in $1/m$ corrections to the LECs $b_i$ and $h_i$. 
However, these LECs can be absorbed into the $2\pi\bar N N$-LECs
and the renormalization of $h$ and thus are redundant in the $\pi N$ amplitude.
In Appendix \ref{cha:RedShiftsAmpl}, we present all the necessary shifts
to cancel the redundant LECs $b_3$, $b_6$, $c_i^\Delta$ and $h_i$
and the off-shell parameter $z_0$ in the amplitudes.
Note that the shifts are given in the strict HB expansion only.
In the covariant approach, we just set $z_0=z_1=z_2=y_1=y_3=0$ and
$b_3=b_6=h_i=c_i^\Delta=0$. 

\subsection{Coupling Constants}
As a next step, we consider the renormalization of coupling constants.
We start with the bare quantities in the leading-order Lagrangian 
which are renormalized on mass-shell. The renormalization of the
quantities from the nucleon sector, mass $m$ and coupling $g$, is extended by explicit 
$\Delta$ contributions. Additionally, the quantities from the $\Delta$ sector,
mass $\bf m$ and coupling $h$, have to be considered. In Appendix
\ref{sec:renormalizationrules}, we give the explicit expressions for 
$m_N$, $Z_N$, $g_A$ and $m_\Delta$, $Z_\Delta$, $h_A$ in both chiral
approaches. 
\\\\
After performing all the redundancy shifts and
the renormalization of the leading-order couplings, both discussed
above, the remaining $2\pi\bar N N$-LECs are
renormalized. In particular, we choose to absorb UV divergent pieces
as well as additional finite pieces. The main steps necessary to identify both types of these pieces are:
\begin{itemize}
\item Perform a Taylor series of the full amplitude in powers of small scales while interchanging the loop integration with a power series of the integrand
\item After performing this power series, absorb all remaining parts of the loop amplitude order-by-order into the corresponding LECs
\item Return to the full amplitude and redefine the LECs as constrained in the previous step
\item Absorb all remaining UV divergent pieces in the full amplitude into corresponding LECs
\end{itemize}
In the following, we discuss the results for both chiral approaches after performing the above steps.
Considering the HB framework, we have
\begin{equation}
  \label{eq2:23}
  \begin{aligned}
    c_i &= \bar c_i +\delta c^{(3,\Delta)}_i+\delta c^{(4,\Delta)}_i\komma\\
    d_i &=  \bar d_i +\delta d_i+\delta d^{(3,\Delta)}_i+\delta d^{(4,\Delta)}_i\komma\\
    e_i &=  \bar e_i +\delta e_i +\delta e^{(4,\Delta)}_i\komma
  \end{aligned}
\end{equation}
where for $x\in\{c,d,e\}$ one has
\begin{equation}
  \begin{aligned}
    \delta x_i &=
    \frac{\beta_{x_i}+\beta^\Delta_{x_i}}{F_\pi^2}\left(\bar
      \lambda+\frac{1}{32\pi^2}\ln\left(\frac{M_\pi^2}{\mu^2}\right)\right)\komma\\
    \delta x_i^{(n,\Delta)} &= \frac{\delta\bar x^{(n,\Delta)}_{i,f}}{F_\pi^2}+
    \frac{\beta^{(n,\Delta)}_{x_i}}{F_\pi^2}\left(\bar
      \lambda+\frac{1}{16\pi^2}\ln\left(\frac{2\Delta}{\mu}\right)\right)
  \end{aligned}
\label{eq2:25}
  \end{equation}
with 
\begin{equation}
  \begin{aligned}
    \bar \lambda=\frac{\mu^{d-4}}{16\pi^2}\left(
      \frac{1}{d-4}+\frac{1}{2}(\gamma_E-1-\ln 4\pi) \right)\punkt
  \end{aligned}
\end{equation}
The additional finite pieces are denoted by $\delta\bar
x^{(n,\Delta)}_{i,f}$. They are needed to absorb decoupling breaking
pieces, which are generated by loop functions. 
According to the decoupling theorem, all $\Delta$ contributions
in the amplitude have to vanish in the decoupling limit $\Delta\to\infty$.
As already pointed out in Ref.~\cite{Siemens:2017opr}, 
this theorem is explicitly satisfied for tree-level
contributions of resonances like the $\Delta$ or Roper resonances.
At loop-level, however, loop functions generate terms with positive
power of the mass splitting $\Delta$, which do not vanish explicitly
in the decoupling limit. We emphasize that these decouplings breaking
terms (DBTs) are very similar to the PCBTs in the covariant
formalism of baryon $\chi$PT \cite{Gegelia:1999gf,Fuchs:2003qc}. 
The difference to the PCBTs is that the DBTs in the HB framework obey the naive
power counting, simply because the mass splitting $\Delta$ is regarded as
a small scale and not as a large scale like the nucleon mass $m_N$. 
However, if one chose a counting scheme where the mass splitting is a large
scale $\Delta\sim O(\varepsilon^0)$, then the DBTs would violate
the naive power counting. Clearly, this counting scheme coincides
with the decoupling limit.
At this point, we also would like to point out a feature of the HB framework.
In the construction of the Lagrangian, we employ an expansion around
$m_N\sim O(\varepsilon^0)$ and $\Delta\sim O(\varepsilon^1)$, but then
for the decoupling limit we need to take $\Delta\to\infty$ in the amplitudes. Thus, terms
proportional to e.g., $\Delta/m_N$ which seem to break decoupling, 
actually vanish in the decoupling limit
\begin{equation}
\label{eq2:26}
\Delta/m_N \xrightarrow{\Delta\to\infty} 0 \komma
\end{equation}
because $\Delta\ll m_N$ by construction.
In summary, the DBTs in the $d$-dimensional HB amplitude are cancelled
by absorbing all IR regular contributions from loops into the
$2\pi\bar N N$-LECs. To be precise, we expand in the following scales
\begin{equation}
  \label{eq2:21}
  \omega\sim M_\pi\sim O(\varepsilon^1)\komma\qquad
  t \sim O(\varepsilon^2)\komma\qquad
m_N\sim\Delta\sim \mathcal{O}(\varepsilon^0)\punkt
\end{equation}
Note that this is in evident analogy to EOMS in the covariant formalism.
To calculate the IR regular part of a loop function, the loop momentum
integration is interchanged with a Taylor expansion of the integrand
in powers of the small scales in Eq.~\eqref{eq2:21} and the loop
momentum $l\sim \mathcal{O}(\varepsilon^0)$, see Ref.~\cite{Schindler:2003xv} for the analogous approach in EOMS.
At this point, we emphasize that the IR regular contributions that are
absorbed into the LECs $e_i$ do not violate decoupling. 
Instead, the motivation at this point is that loop contributions should
only saturate the loop-level LECs and none of the tree-level LECs.
Finally, the renormalized HB amplitude fulfills
\begin{itemize}
\item UV finiteness: amplitude is free from UV divergencies
\item decoupling theorem: amplitude is free from DBTs
\item natural saturation: tree-level (loop-level) amplitudes with $\Delta$
saturate the $2\pi \bar N N$-LECs at tree-level (loop-level).
\end{itemize}
Considering the covariant framework, we have
\begin{equation}
  \label{eq2:23Cov}
  \begin{aligned}
    c_i &= \bar c_i +\delta c^{(3)}_i+\delta c^{(3,\Delta)}_i+\delta c^{(4)}_i+\delta c^{(4,\Delta)}_i\komma\\
    d_i &=  \bar d_i +\delta d_i+\delta d^{(3)}_i+\delta d^{(3,\Delta)}_i+\delta d^{(4)}_i+\delta d^{(4,\Delta)}_i\komma\\
    e_i &=  \bar e_i +\delta e_i +\delta e^{(4)}_i+\delta e^{(4,\Delta)}_i
  \end{aligned}
\end{equation}
with 
\begin{equation}
  \label{eq2:34}
  \begin{aligned}
 \delta x_i &=
    \frac{\beta_{x_i}+\beta^\Delta_{x_i}}{F_\pi^2}\left(\bar
      \lambda+\frac{1}{32\pi^2}\ln\left(\frac{M_\pi^2}{\mu^2}\right)\right)\komma\\
  F_\pi^2\;  \delta x^{(n)} &=a_0+a_1 A_0(m_N^2) \komma\\
  F_\pi^2\;  \delta x^{(n,\Delta)} &=a_0+a_1 A_0(m_N^2) +a_2 A_0(m_\Delta^2)
+b_1 B_0(m_N^2,0,m_\Delta^2) +b_2 B_0(m_\Delta^2,0,m_N^2) \\&
+c_1 C_0(m_N^2,0,m_\Delta^2,0,m_N^2,m_N^2) +c_2 C_0(m_N^2,0,m_\Delta^2,0,m_\Delta^2,m_\Delta^2)
\\&+c_3 C_0(m_\Delta^2,0,m_N^2,0,m_N^2,m_\Delta^2) 
+c_4 C_0(m_N^2,0,m_\Delta^2,0,m_N^2,m_\Delta^2) \komma
  \end{aligned}
\end{equation}
where for the only IR divergent $C_0$-function we get in
dimensional regularization
\begin{align}
  \label{eq2:47}
C_0(m_N^2,0,m_\Delta^2,0,m_N^2,m_\Delta^2) &=
\frac{(d-2) \Big(m_\Delta^2 A_0\left(m_N^2\right)-m_N^2 A_0\left(m_\Delta^2\right)\Big)   }{2 (-4+d) m_N^2 m_\Delta^2 \left(m_N^2-m_\Delta^2\right)}\\
&\simeq -\frac{\text{Log}\left(\frac{m_N^2}{m_\Delta^2}\right) \left(2+64 \pi ^2 \bar\lambda+\ln\left(\frac{m_N^2 m_\Delta^2}{\mu ^4}\right)\right)}{4 \left(m_N^2-m_\Delta^2\right)}+O(d-4)\punkt
\end{align}
In Eq. \eqref{eq2:23Cov}, the terms $\delta x^{(n)}$ correspond to the
renormalization of the LECs in the $\Delta$-less
theory, which is discussed extensively in Ref.~\cite{Siemens:2016hdi}. 
However, due to explicit $\Delta$ degrees of freedom, the
order-by-order renormalization of the LECs has to be extended.
Those additional redefinitions of couplings are denoted by $\delta x^{(n,\Delta)}$.
It has to be stressed that both set of terms are calculated in
complete analogy. 
The idea is to expand the $d$-dimensional amplitudes $D$ and $B$ in small parameters
\begin{equation}
  \label{eq2:27}
  M_\pi\sim \mathcal{O}(\varepsilon^1)\komma\quad s-m_N^2\sim \mathcal{O}(\varepsilon^1)\komma\quad 
  u-m_N^2\sim \mathcal{O}(\varepsilon^1)\komma\quad t\sim \mathcal{O}(\varepsilon^2)
\end{equation}
and additionally in the $\Delta$-ful theory 
\begin{equation}
  \label{eq2:28}
  m_N\sim m_\Delta\sim m_\Delta-m_N\sim\mathcal{O}(\varepsilon^0)
\end{equation}
such that the IR regular parts of the amplitudes are identified. 
Note again that the IR regular part of a loop function can be extracted by 
interchanging loop integration with a Taylor series in powers of the small parameters \cite{Schindler:2003xv}.
At this point we have to emphasize that the
nucleon-$\Delta$ mass splitting is a small scale in the power counting
used in this analysis, see Eq.\eqref{eq2:3}. However, in our 
renormalization procedure the mass splitting is treated as a large scale. 
This apparent contradiction is easily resolved.
In the covariant approach, the motivation is to absorb PCBTs and DBTs.
Both set of terms only appear in the IR regular part of loop functions
given by an expansion as specified by Eqs.~\eqref{eq2:27} and \eqref{eq2:28}.
Thus, one has to treat the mass splitting as a large scale.
Furthermore, note the employed notation for the terms
$\delta x$, $\delta x^{(n)}$ and $\delta x^{(n,\Delta)}$ in Eq.~\eqref{eq2:34}. In
these expressions, the loop functions $A_0$, $B_0$ and the IR divergent function
$C_0$ in Eq.~\eqref{eq2:47} include finite and divergent
pieces. The parameters $a_i$, $b_i$ and $c_i$ are functions of
baryonic masses and lower-order coupling constants.
It has to be stressed that only UV and no IR divergencies are included
in the renormalization procedure. In particular, the IR divergent
piece in the $C_0$ function is exactly cancelled by 
the divergence proportional to $\beta_{x_i}^\Delta$ in $\delta x_i$. 
To be precise, the $\beta$-functions $\beta_{x_i}^\Delta$ are
determined after performing the shifts $\delta x^{(n)}$ and
$\delta x^{(n,\Delta)}$ in the pion-nucleon scattering amplitude
by demanding that the amplitude is free of divergencies.
Finally, the renormalized covariant amplitude fulfills (up to $\mathcal{O}(\varepsilon^4)$)
\begin{itemize}
\item UV finiteness: amplitude is free from UV divergencies 
\item proper power counting: amplitude is free from PCBTs 
\item decoupling theorem: amplitude is free from DBTs
\item natural saturation: tree-level (loop-level) amplitudes with $\Delta$
saturate the $2\pi \bar N N$-LECs at tree-level (loop-level).
\end{itemize}
In Appendix \ref{sec:renormalizationrules} and in the supplementary material, we give the explicit
expressions for the renormalization of the LECs $c_i$, $d_i$ and $e_i$
in both chiral approaches. We performed the following check on our results.
The covariant $d$-dimensional expressions were expanded
in the mass splitting $\Delta\sim O(\varepsilon^1)$
and the corresponding $\Delta$ singular parts were successfully matched to the HB
expressions. Note that we define the $\Delta$ singular part of
a loop function in Eq.~\eqref{eq2:34} in analogy to the IR singular
part used in EOMS. In particular, it is the contribution, which is non-analytic in
the mass splitting $\Delta$ in $d$ dimensions.
The $\Delta$ singular part can be extracted from a loop function by
interchanging loop integration with an expansion of the integrand in
$l\sim\Delta\sim\mathcal{O}(\varepsilon^1)$ and
$ m_N\sim\mathcal{O}(\varepsilon^0)$. Note that $l$ is the loop momentum.

\subsection{Complex and Real Counter Terms}
\begin{figure}[ht]
  \centering
\includegraphics[width=0.8\textwidth]{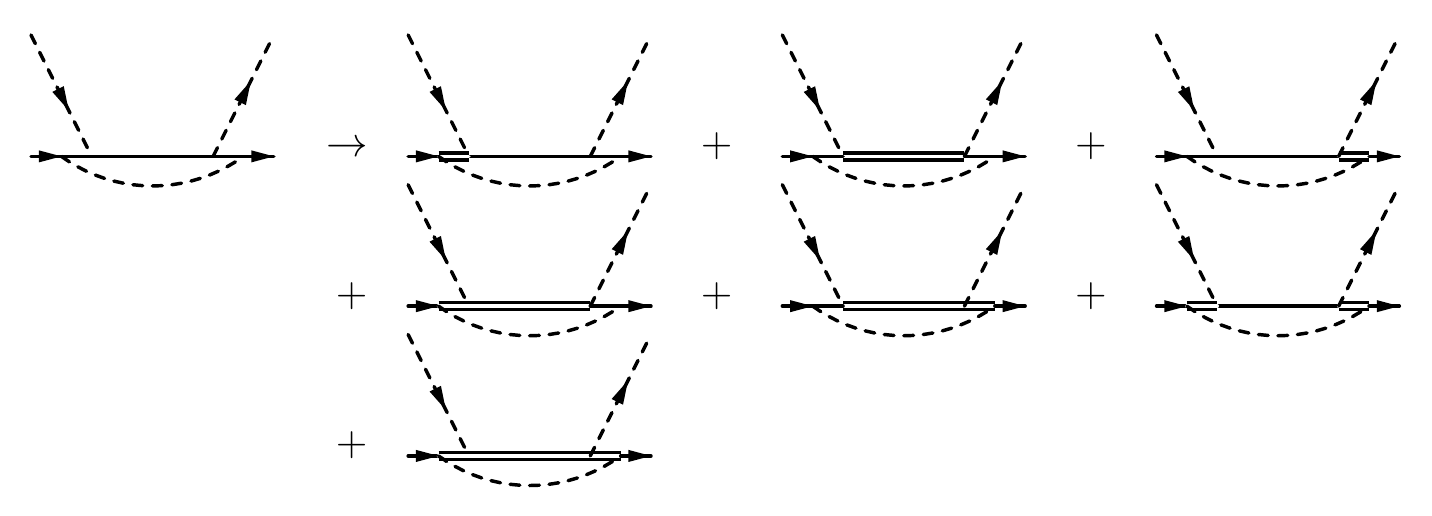}
\includegraphics[width=0.8\textwidth]{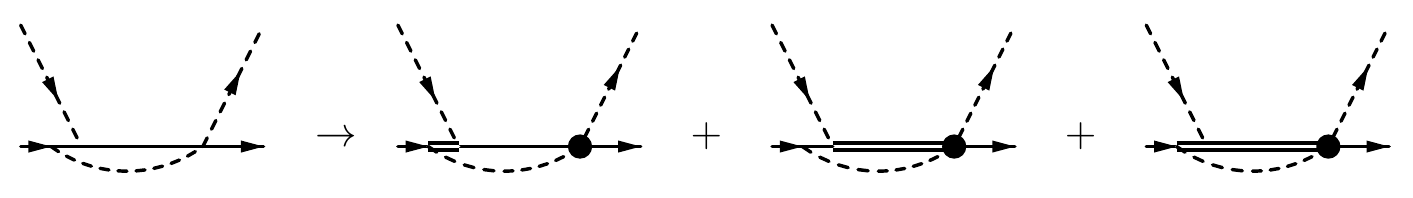}
  \caption{Examples of transitions from $\Delta$-less to $\Delta$-ful graphs at leading and
    next-to-leading order. An insertion of the $c_i$ or $b_i$
    vertices is denoted by a black blob. Dashed, solid, and double solid lines refer to pions,
    nucleons, and $\Delta$ resonances,
    respectively. Crossed and redundant
    diagrams are not shown.}
  \label{fig:LoopEx}
\end{figure}

In Ref.~\cite{Siemens:2016hdi}, we already discussed the pertinent
tree- and loop-level graphs for $\pi N\to\pi N$ up to order $Q^4$.
As shown exemplarily in Fig.~\ref{fig:LoopEx}, the additional graphs
including $\Delta$ are generated by substituting intermediate
nucleon propagators by $\Delta$ propagators. 
Note that this example does not show graphs with redundant contributions.
Finally, the renormalized pion-nucleon scattering amplitude depends on
the following masses and couplings:
\begin{itemize}
\item pion sector: $M_\pi$ and $F_\pi$ from $\mathcal{L}_{\pi\pi}^{(2)}$
\item nucleon sector: $m_N$ and  $g_A$ from $\mathcal{L}_{\pi N}^{(1)}$,
$\bar c_i$ from $\mathcal{L}_{\pi N}^{(2)}$, $\bar d_i$ from
$\mathcal{L}_{\pi N}^{(3)}$ and $\bar e_i$ from $\mathcal{L}_{\pi N}^{(4)}$
\item $\Delta$ sector:
$h_A$ from $\mathcal{L}_{\pi N\Delta}^{(1)}$,
$m_\Delta$ and $g_1$ from $\mathcal{L}_{\pi \Delta}^{(1)}$ and $b_{4,5}$ from $\mathcal{L}_{\pi N\Delta}^{(2)}$
\end{itemize}
At this point, we have to emphasize that in our renormalization procedure the 
real-valued bare quantities in the effective Lagrangian are
renormalized complex. This is a general
consequence when decaying particles are considered explicitly.
A prominent example is the $\Delta$ mass which can be parametrized via
\begin{equation}
  \label{eq2:29}
  m_\Delta=m_\Delta^{\mathrm{Re}}-\frac{\i\Gamma}{2}\komma
\end{equation}
where $\Gamma$ denotes the strong decay width of the $\Delta$. 
Note that $\Gamma\sim \mathcal{O}(\varepsilon^3)$, because 
$\Gamma$ receives its leading contribution at loop-level.
For our calculation, this means that in a loop contribution
the bare $\Delta$ mass $\bf m$
does not need to be replaced by the complex-valued quantity $m_\Delta$.
In a loop contribution, the imaginary part proportional to $\Gamma$ 
counts as $\mathcal{O}(\varepsilon^5)$, such that
for our purposes it is already sufficient to replace $\bf m$ by $m_\Delta^\mathrm{Re}$.
Of course, one can also take the full complex-valued mass $m_\Delta$ in
all loop amplitudes. However, the trick above makes loop calculations
easier and is justified based on the power counting.
If we now consider only the real part $m_\Delta^\mathrm{Re}$ in loop contributions,
we see that the renormalized quantities
$m_\Delta$, $h_A$, $\bar c_i$, $\bar d_i$ and $\bar e_i$
in Appendix \ref{sec:renormalizationrules} still remain complex.
In contrast, the quantities $m_N$ and $g_A$ are real-valued now.
We emphasize that the given renormalization rules
completely determine the imaginary parts of the complex-valued
quantities, because the initial bare quantities are real-valued. 
Clearly, such an imaginary part is generated in the renormalization
procedure, in particular, by a splitting between the renormalized  
quantity and the corresponding counter term, e.g., for the $\Delta$
mass up to order $\varepsilon^4$ one gets the condition
\begin{equation}
  \label{eq2:32}
  \mathrm{Im}\; m_\Delta = -\mathrm{Im}\; \delta {\bf m^{(3)}} - \mathrm{Im}\; \delta {\bf m^{(4)}}\punkt
\end{equation}
This means that these imaginary parts do not need to be constrained by data,
because they are fixed by the theory and, in particular, 
do not represent additional degrees of freedom.
\\\\
As mentioned in section \ref{sec:basicdef-Delta}, two
different unitarization schemes are employed in this work.
To be more precise, in the $K$-matrix untitarization, which is closely
related to the extraction of the Breit-Wigner masses, 
we set the $\Delta$ mass $m_\Delta$ in the leading-order amplitudes
to its real-valued Breit-Wigner mass $m_\Delta^\mathrm{BW}$. 
This means that the imaginary parts of the renormalized mass $m_\Delta$ and
of the counterterms $\delta {\bf m^{(3)}}$ and $\delta {\bf m^{(4)}}$
are neglected in our amplitudes. As can be seen in Eq.~\eqref{eq:1a},
only the real part of the partial wave amplitude is
considered in the $K$-matrix unitarization anyway.
Thus, within this unitarization scheme, only the real parts of the
renormalized quantities $m_\Delta$, $h_A$ and the LECs $\bar c_i$,
$\bar d_i$ and $\bar e_i$ have to be taken into account in our
calculation of the $\pi N$ scattering amplitudes up to order $\varepsilon^4$. 
Note that we could also keep all imaginary parts in the
amplitude and instead fix them to the negative of 
their corresponding counter terms, e.g., as for the $\Delta$ mass in
Eq.~\eqref{eq2:32}. This would be an equivalent procedure, however,
the former one might be more intuitive to the reader.
In the complex mass approach, the $\Delta$ mass in the leading-order amplitudes
is set to its complex-valued pole mass $m_\Delta^\mathrm{Pole}$, where
the imaginary part is fixed by its experimental value. 
As mentioned before, the complex mass unitarization scheme 
is employed in this work to extend the applicability of the chiral EFT 
to the $\Delta$ pole region, where a dressing of the $\Delta$ is inevitable. 
In particular, the real-valued $\Delta$ propagator is divergent in the
$\Delta$ pole region
\begin{equation}
  \label{eq2:33}
  \frac{1}{\slashed{p}-m^\mathrm{Re}_\Delta}\xrightarrow{\slashed{p}\to m^\mathrm{Re}_\Delta}\infty\punkt
\end{equation}
However, as we are going to demonstrate, 
this is only an artifact of the employed power counting.
The resummed or dressed $\Delta$ propagator is the initial starting
point. This propagator is expanded around the renormalization point
$\slashed{p}=m_\Delta^\mathrm{Re}$, which gives
\begin{equation}
  \label{eq2:35}
  \frac{1}{\slashed{p}-m_\Delta^\mathrm{Re}-\Sigma(\slashed{p})}\simeq
  \frac{1}{\slashed{p}-m_\Delta^\mathrm{Re}-\Sigma(m_\Delta^\mathrm{Re})-\Sigma^\prime(m_\Delta^\mathrm{Re}) (\slashed{p}-m_\Delta^\mathrm{Re})+\mathcal{O}((\slashed{p}-m_\Delta^\mathrm{Re})^2)}\punkt
\end{equation}
Here, the self-energy of the $\Delta$ is denoted by 
$-\i\Sigma(\slashed{p})$, which is given by the sum 
of all one-particle-irreducible contributions
to the two-point function of the $\Delta$.
If we employ the following renormalization conditions
\begin{equation}
  \label{eq2:36}
  \mathrm{Re}\;\Sigma(m_\Delta^\mathrm{Re})=0 \komma\qquad
  \mathrm{Im}\;\Sigma(m_\Delta^\mathrm{Re})=-\frac{\Gamma}{2} \komma\qquad
  \Sigma^\prime(m_\Delta^\mathrm{Re})=0\komma
\end{equation}
we are left with
\begin{equation}
  \label{eq2:37}
  \frac{1}{\slashed{p}-m_\Delta^\mathrm{Re}-\Sigma(\slashed{p})}\simeq
  \frac{1}{\slashed{p}-m_\Delta^\mathrm{Re}+\i \frac{\Gamma}{2}}
  +\mathcal{O}((\slashed{p}-m_\Delta^\mathrm{Re})^2)\komma
\end{equation}
which is well-defined in the limit $\slashed{p}\to m_\Delta^\mathrm{Re}$. 
In particular, this is the appropriate behavior in the $\Delta$ pole region. 
Note that a Taylor expansion of Eq. \eqref{eq2:37} in $\Gamma$ is equivalent
to the expression in Eq.~\eqref{eq2:33} to leading order.
To be precise, a strict power counting even demands such an expansion, 
because $\slashed{p}-m_\Delta^\mathrm{Re}\sim
\mathcal{O}(\varepsilon^1)$ and $\Gamma\sim
\mathcal{O}(\varepsilon^3)$ such that $\Gamma\ll
\slashed{p}-m_\Delta^\mathrm{Re}$. 
In contrast, one has $\slashed{p}\simeq m_\Delta^\mathrm{Re}$ 
in the $\Delta$ pole region. This means that $\Gamma\gtrsim
\slashed{p}-m_\Delta^\mathrm{Re}$, which clearly does not justify the expansion of 
Eq. \eqref{eq2:37} in powers of $\Gamma$. 
Thus, for practical calculations, it is advantageous to renormalize
the $\Delta$ mass complex.
In particular, one has in the on mass-shell renormalization scheme
\begin{equation}
  \label{eq2:38}
    \frac{1}{\slashed{p}-m_\Delta-\Sigma(\slashed{p})}\simeq
  \frac{1}{\slashed{p}-m_\Delta}
  +\mathcal{O}((\slashed{p}-m_\Delta)^2)\komma
\end{equation}
where the conditions
\begin{equation}
  \label{eq2:39a}
    \Sigma(m_\Delta)=0 \komma\qquad
  \Sigma^\prime(m_\Delta)=0
\end{equation}
give the desired dressed $\Delta$ propagator.
At one-loop order, we can simplify the above conditions to
\begin{equation}
  \label{eq2:39}
  \Sigma(m_\Delta^\mathrm{Re})=0 \komma\qquad
  \Sigma^\prime(m_\Delta^\mathrm{Re})=0\komma
\end{equation}
where $m_\Delta=m_\Delta^\mathrm{Re}-\i\Gamma/2$. In particular, 
$m_\Delta^\mathrm{Re}\sim\mathcal{O}(\varepsilon^0)$ and
$\Gamma\sim\mathcal{O}(\varepsilon^3)$ such that the higher-order
effects of $\Gamma$, namely $\mathcal{O}(\varepsilon^5)$, are neglected..
Note that in the discussion above, the tensorial structure of the
numerator of the covariant $\Delta$ propagator was suppressed, 
see Eq.~\eqref{eq0:169}.
In the HB approach, we proceed in very close analogy to the covariant case. 
Here, we expand the dressed $\Delta$
propagator around $v\cdot p=\Delta$ such that
\begin{equation}
  \label{eq2:25x}
  \frac{1}{v \cdot p-\Delta-\Sigma(v\cdot p)}\simeq\frac{1}{v \cdot p-\Delta-\Sigma(\Delta)-\Sigma^\prime(\Delta)(v \cdot p-\Delta)+\mathcal{O}((v \cdot p-\Delta)^2)}
\end{equation}
with the conditions at one-loop order
\begin{equation}
  \label{eq2:26x}
   \Sigma(\Delta^\mathrm{Re})=0 \komma\qquad
  \Sigma^\prime(\Delta^\mathrm{Re})=0\komma
\end{equation}
where $\Delta=\Delta^\mathrm{Re}-\i\Gamma/2$.
\\\\
As mentioned before, in the $K$-matrix approach, 
the $\Delta$ width is generated by the unitarization prescription.
However, this unitarization is not reliable in the $\Delta$ region,
because it is simply an approximation for low energies and small phase shifts. 
Instead, in the complex mass approach, the complex-valued $\Delta$
mass is used in the leading-order amplitudes and the
$P_{33}$ partial wave is unitarized as given in Eq.~\eqref{eq:2a}.
In addition, we also would like to avoid $\Delta$ pole contributions at loop-level.
Thus, we employ a complex-valued coupling $h_A$ with its imaginary
part fixed by its counter term, namely
\begin{equation}
  \label{eq2:40}
  \mathrm{Im}\; h_A= -\mathrm{Im}\;\delta h^{(3)}-\mathrm{Im}\;\delta h^{(4)}\punkt
\end{equation}
We emphasize that both quantities $m_\Delta$ and $h_A$ are only
complex-valued in the leading-order $\Delta$ pole graphs.
At loop-level, the imaginary parts of those quantities are regarded
as higher-order effects, such that only the real parts are used.
E.g., in the $s$-channel, the $\Delta$ pole contributions at tree-level scale as
\begin{equation}
  \label{eq2:42}
  \frac{h_A^2}{s-m_\Delta^2}=
  \frac{(h_A^\mathrm{Re}+\i h_A^\mathrm{Im})^2} {s-m_\Delta^2}
  \simeq  \frac{(h_A^\mathrm{Re})^2+2\i
    h_A^\mathrm{Re} h_A^\mathrm{Im}}{s-m_\Delta^2}+\cdots\komma
\end{equation}
where the ellipsis denotes terms neglected in our calculation,
which are at least two-loop-level contributions.
Furthermore, we fix the imaginary parts of the $2\pi \bar N N$-LECs
in the tree-level diagrams by their counter terms, namely
\begin{equation}
  \label{eq2:41}
  \mathrm{Im}\; \bar x_i= -\mathrm{Im}\;\delta x_i^{(3,\Delta)}-\mathrm{Im}\;\delta x_i^{(4,\Delta)}
\quad\text{with}\quad x \in \{ c, d, e\}\punkt
\end{equation}
Instead of employing Eq.~\eqref{eq2:41}, we can also follow an
equivalent approach, where we neglect from the beginning
the imaginary parts of the LECs in the renormalization procedure.
This means just considering $\mathrm{Re}\;\delta x_i^{(n,\Delta)}$,
see Appendix \ref{sec:renormalizationrules}. 
We emphasize that due to the complex leading-order
$\Delta$ pole the above statement 
holds for the coupling $h_A$ only modulo higher order, as seen in Eq.~\eqref{eq2:42}.
\\\\
Furthermore, we use in the HB formalism
the covariant leading-order $\Delta$ pole amplitudes instead of the HB
ones. This modification is motivated by two arguments. 
First, the strict HB expansion is not justified in the $\Delta$ pole region.
Obviously, this can be corrected for by considering the
resummed expressions in the covariant framework. 
Second, the convergence of the quantities $\Gamma$ and
$\mathrm{Im} \; h_A$ in inverse powers of $m_N$ is very poor,
see Table~\ref{tab:Conv}.
Thus, the HB framework does not seem to work well
if we consider those imaginary parts explicitly
like in the complex mass approach. 
Note that the two breakdowns discussed above, the breakdown 
of the HB expansion in the vicinity of the $\Delta$ pole
and the breakdown of the expansion of the covariant propagator
in the $\Delta$ width, are closely related. 
In particular, they are both artifacts of the employed power counting. 
For a different treatment of the resummations on the pole region see Ref.~\cite{Long:2009wq}.

\subsection{Final Amplitudes and Checks}
In this analysis we have calculated the amplitudes of
a total of 60 tree-level and 618 loop-level graphs
(including crossed and redundant ones). 
Due to the enormous size of those amplitudes, we refrain from
showing them explicitly and instead prefer to present them upon
request. In addition, we calculated the 13 leading subthreshold parameters and the 8
leading threshold parameters in both chiral approaches based on those amplitudes.
The explicit expressions are given in Ref. \cite{Siemens:2016jwj}.
\\\\
We emphasize that this is the first full one-loop order 
calculation of $\pi N\to\pi N$ including explicit $\Delta$ degrees of
freedom in the small scale expansion.
We are therefore not in the position to cross check all of our results with
references in the literature. However, some lower-order calculations
are available. 
Up to order $\varepsilon^3$, we performed a cross check with the HB expressions
given in Ref.~\cite{Fettes:2000bb}. Except for obvious typos and the
absent renormalization of the coupling $h_A$, we agree with the
expressions in that work.
In our analysis of Ref.~\cite{Yao:2016vbz}, two groups
calculated the covariant expressions up to order $\varepsilon^3$
independently. Both sets of amplitudes were successfully matched.
Note that the $2\pi\bar N N$-LECs in the above references are renormalized
differently compared to the same LECs in this work.
In addition, our $\varepsilon^4$ results passed the following checks:
\begin{itemize}
\item The explicit independence of the off-shell
  parameters $\alpha$, $z_0$ and of the redundant LECs $b_3$,
  $b_6$, $h_i$ was verified.
\item The renormalized amplitudes exhibit the
  correct analytic structure. To be precise, only the leading-order 
  nucleon and $\Delta$ amplitudes have physical poles corresponding
  to their masses. 
\item The renormalization rules of the $2\pi \bar N N$-LECs in
  the covariant and HB frameworks were successfully matched after an
  appropriate expansion in inverse powers of $m_N$.
\item The subthreshold and threshold parameters in
  the covariant and HB frameworks were successfully matched after an
  appropriate expansion in inverse powers of $m_N$.
\end{itemize}

\section{Fitting Procedure}
\label{sec:fitting-procedure-Delta}
As explained in the previous section, the amplitudes for elastic pion-nucleon
scattering including explicit $\Delta$ degrees of freedom depend on several LECs. 
Throughout this section, we employ the following numerical values for the
leading-order quantities: 
$M_\pi = 139.57$ MeV, $F_\pi =92.2$ MeV, 
$m_N= 938.27$ MeV, $m^{\mathrm{BW}}_\Delta=1232$ MeV or
$m^{\mathrm{Pole}}_\Delta=1210-\i 50$ MeV \cite{Agashe:2014kda} and
$g_A=1.289$ \cite{Siemens:2016hdi}.
Furthermore, the leading- and higher-order LECs are renormalized
as discussed in section \ref{sec:power-count-renorm-Delta}. 
Note that we suppress in the following the bars in the notation of the
renormalized LECs $\bar c_i$, $\bar d_i$ and $\bar e_i$. 
Keep in mind that the numerical values of these LECs are given
in units of GeV$^{-1}$, GeV$^{-2}$ and GeV$^{-3}$, respectively.
\\\\
The fits described below are performed to $\pi N$ scattering
data given in terms of the two observables
$\di\sigma / \di\Omega$ and $P$ of all three channels. Note that we proceed in
very close analogy to the $\Delta$-less fits in Ref.~\cite{Siemens:2016hdi}, where
the reader is referred to for more details. 
The first step is the minimization of the quantity 
\begin{equation}
  \label{eq2:31}
  \begin{aligned}
    \chi^2 =\chi^2_{\pi N}+\chi^2_C\komma
  \end{aligned}
\end{equation}
where the information from the $\pi N$ scattering data
\cite{Workman:2012hx} including
an estimated theoretical uncertainty \cite{Epelbaum:2014efa}
are incorporated in the first term, in particular,
\begin{equation}
  \label{eq:31}
  \begin{aligned}
    \chi_{\pi N}^2 =\sum_i\left( \frac{\mathcal{O}^{exp}_{i}-N_i
      \mathcal{O}^{(n)}_{i}}{\delta \mathcal{O}_i}\right)^2\qquad
  \mathrm{with} \qquad \delta\mathcal{O}_i=\sqrt{(\delta\mathcal{O}^{exp}_i)^2+
(\delta\mathcal{O}^{(n)}_i)^2}\punkt
  \end{aligned}
\end{equation}
The experimental data $\mathcal{O}^{exp}_i$, experimental errors
$\delta\mathcal{O}^{exp}_i$ and normalization factors $N_i$ are
taken from the GWU-SAID data base 
\cite{Workman:2012hx}. The quantity $\mathcal{O}^{(n)}_i$ is the corresponding
observable $\mathcal{O}_i$ but calculated from our chiral
amplitudes at order $n$.
The theoretical error $\delta\mathcal{O}^{(n)}_i$ is based on the
truncation of the chiral expansion and is estimated as
\begin{equation}
  \label{eq:5}
  \delta\mathcal{O}^{(n)}_i =\max( |\mathcal{O}^{(\mathrm{LO})}_i| Q^{n-\mathrm{LO}+1} ,
\{|\mathcal{O}^{(k)}_i-\mathcal{O}^{(j)}_i| Q^{n-j} \})\qquad
\mathrm{with}\qquad j<k\leq n\punkt
\end{equation}
Here, we use $Q=\omega_{CMS}/\Lambda_b$, where the energy
of the incoming pion in the CMS frame is denoted by $\omega_{CMS}$. 
The acronym LO refers to leading nonvanishing chiral order of the
observable $\mathcal{O}_i$.
Furthermore, the second term in Eq. \eqref{eq2:31} is defined as
\begin{equation}
  \label{eq:2}
  \qquad
  \chi^2_C=\sum_i\left( \frac{a^2_i-\bar a^2_i}{\delta a_i^2}\right)^2\komma
\end{equation}
where we employ $\vec a=\{ g_1,b_4,b_5 \}$, $ \bar {\vec a}=\{
9/5 g_A,1,1 \}$ and $\delta a_i=1$. In these conditions we assume
the large $N_c$ prediction for $g_1$, naturalness for $b_4$ and $b_5$
and altogether rather conservative errors $\delta a_i$.
In particular, we use the quantity $\chi^2_C$ to enforce additional
constraints on the LECs $g_1$ and $b_4$, $b_5$ due to their first
appearance in loops only at order $\varepsilon^3$ and $\varepsilon^4$, respectively.
Unfortunately, we were not able to reliably constrain these LECs by mere scattering
data such that additional constraints were needed in the fitting procedure.
In contrast to the $\Delta$-less analysis in
Ref.~\cite{Siemens:2016hdi}, where we used a conservative estimate for
the breakdown scale $\Lambda_b$, we adopted in this analysis
a more aggressive estimate for the breakdown scale, namely $\Lambda_b = 700$~MeV. 
This value is justified due to explicit $\Delta$ resonances in our calculation. 
The second step, after minimizing the quantity in Eq.~\eqref{eq2:31}
and extracting a preferred set of LECs, we included additional information from the
subthreshold region and minimized the following quantity
\begin{equation}
  \label{eq2:48}
    \hat \chi^2 =\chi^2_{\pi N}+\chi^2_\mathrm{RS}+\chi^2_C\punkt
\end{equation}
The quantities $\chi^2_{\pi N}$ and $\chi^2_C$ are defined above and
the quantity $\chi^2_{\mathrm{RS}}$ is defined in close analogy to $\chi^2_{\pi N}$ as 
\begin{equation}
  \label{eq:3}
  \begin{aligned}
    \chi_C^2 =\sum_i\left( \frac{\mathcal{O}^\mathrm{RS}_{i}-
      \mathcal{O}^{(n)}_{i}}{\delta \mathcal{O}_i}\right)^2\qquad
  \mathrm{with} \qquad \delta\mathcal{O}_i=\sqrt{(\delta\mathcal{O}^\mathrm{RS}_i)^2+
(\delta\mathcal{O}^{(n)}_i)^2}\komma
  \end{aligned}
\end{equation}
where $\mathcal{O}^\mathrm{RS} = \{d_{00}^\pm, d_{10}^\pm, d_{01}^\pm, b_{00}^\pm\}$
denotes the set of the 8 leading subthreshold parameters taken 
from Ref.~\cite{Hoferichter:2015hva}. 
Note that we took the set of LECs extracted from
minimizing $\chi^2$ in Eq.~\eqref{eq2:31} as
the starting point in the iterative minimization procedure
of $\hat \chi^2$ in Eq.~\eqref{eq2:48}. 
\\
In the following section, we study pion-nucleon scattering data in three chiral approaches employing two
different unitarization prescriptions, $K$-matrix and complex mass approach. 
The chiral approaches are the two HB formalisms called HB-NN and HB-$\pi$N in the following, 
which differ by the treatment of the $1/m$ corrections, and the fully covariant formalism denoted by Cov.
See Ref.~\cite{Siemens:2016hdi} for more details.

\section{Fit Results, Predictions, and Discussion}
\label{sec:predictions-Delta}
\subsection{$K$-Matrix Approach}
\label{sec:k-matrix-approach}
\begin{figure}[ht]
  \centering
\includegraphics[width=\textwidth]{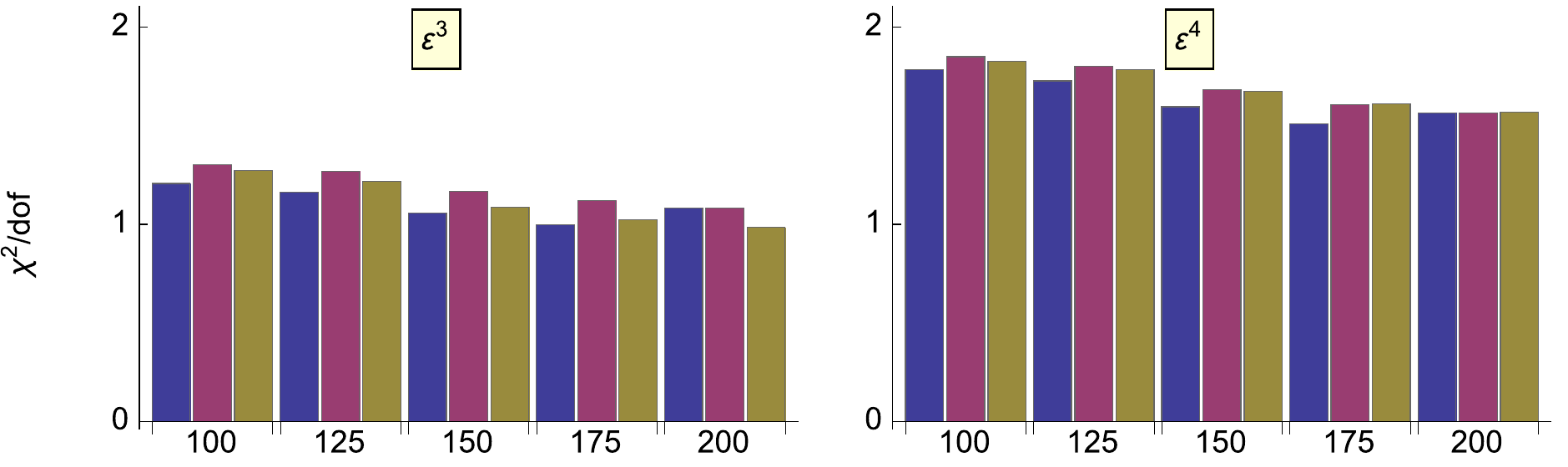}
\includegraphics[width=\textwidth]{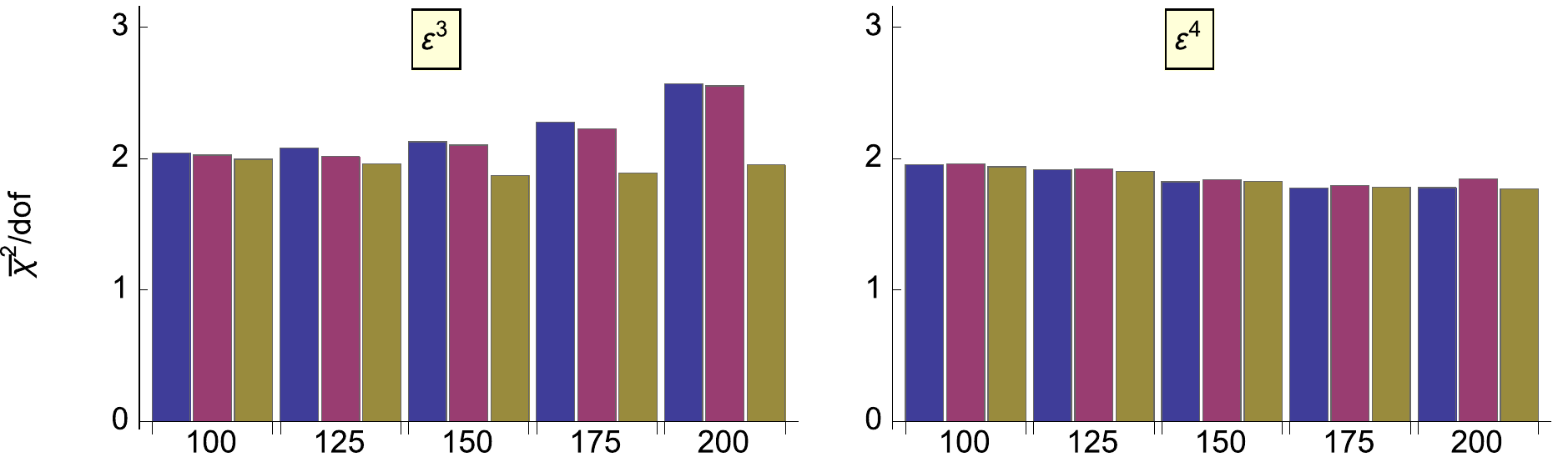}
\includegraphics[width=\textwidth]{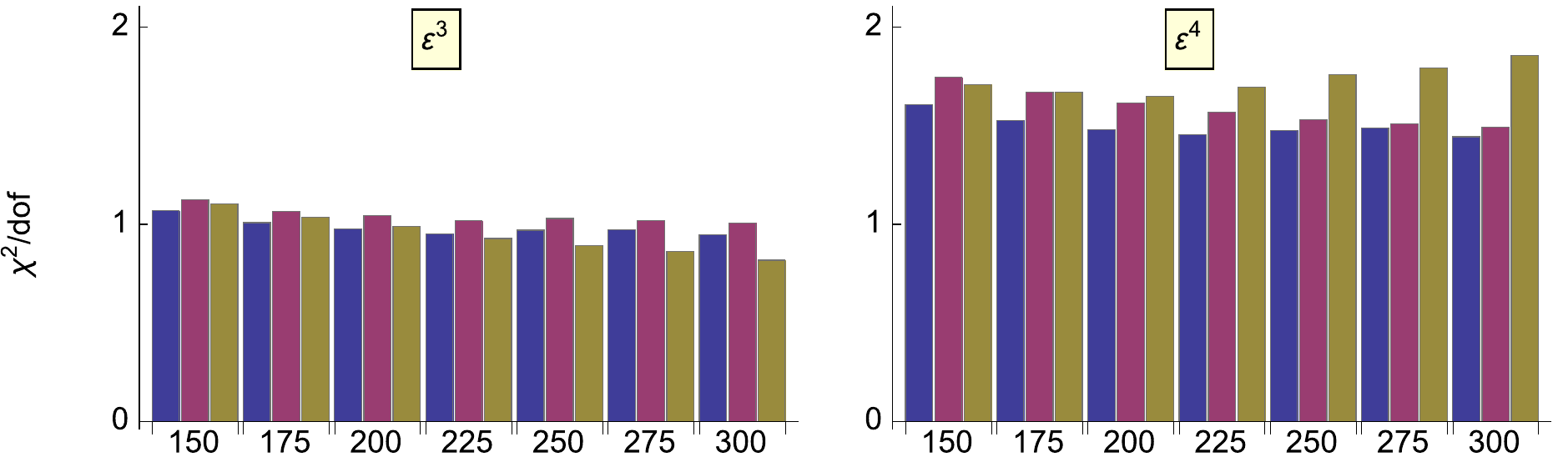}
\includegraphics[width=\textwidth]{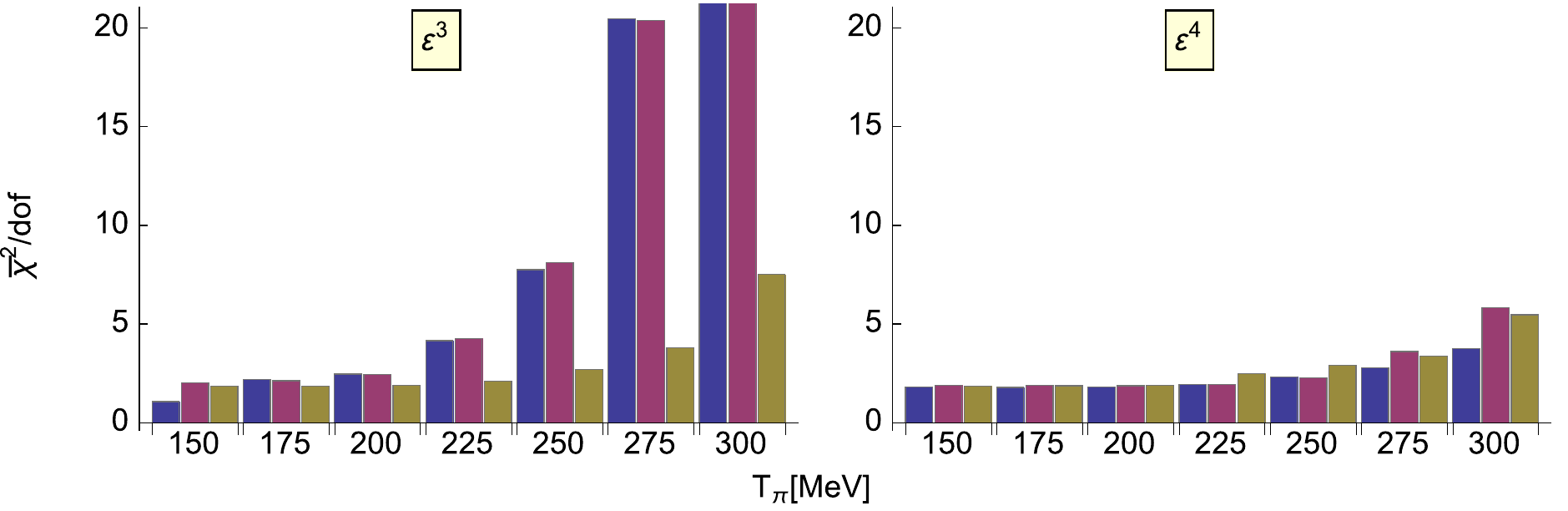}
  \caption{Reduced $\chi^2$ (with theoretical error) and $\bar\chi^2$
    (without theoretical error) as functions of the maximum fit energy
    $T_\pi$, see Eq.~\eqref{eq2:31}. The results for the HB-NN,
    HB-$\pi$N, and covariant counting
    are denoted by blue, red, and green bars. The upper two rows refer to fits
    in the $K$-matrix approach, whereas the lower two rows refer to fits
    in the complex mass approach.}
  \label{fig:RedChiSqK}
\end{figure}

\begin{figure}[ht]
  \centering
\includegraphics[width=0.7\textwidth]{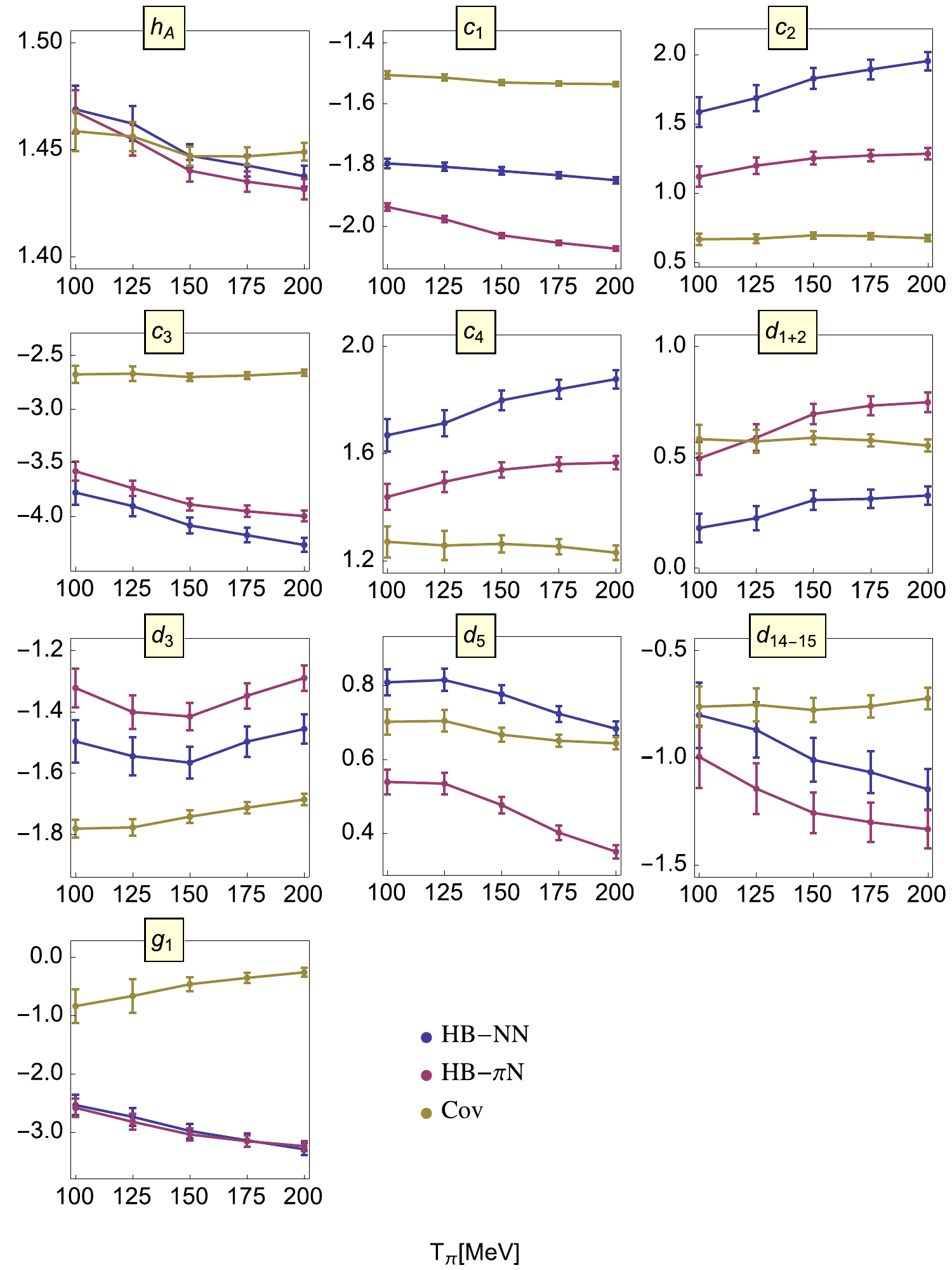}
  \caption{$K$-matrix approach: LECs extracted at order
    $\varepsilon^3$ as functions of the maximum fit energy
    $T_\pi$, see Eq.~\eqref{eq2:31}. The labels
 HB-NN, HB-$\pi$N, and Cov (covariant) denote the different counting schemes
 of $1/m_N$ contributions, see section \ref{sec:power-count-renorm-Delta}.}
  \label{fig:LECsQ3K}
\end{figure}

\begin{figure}[ht]
  \centering
\includegraphics[width=0.65\textwidth]{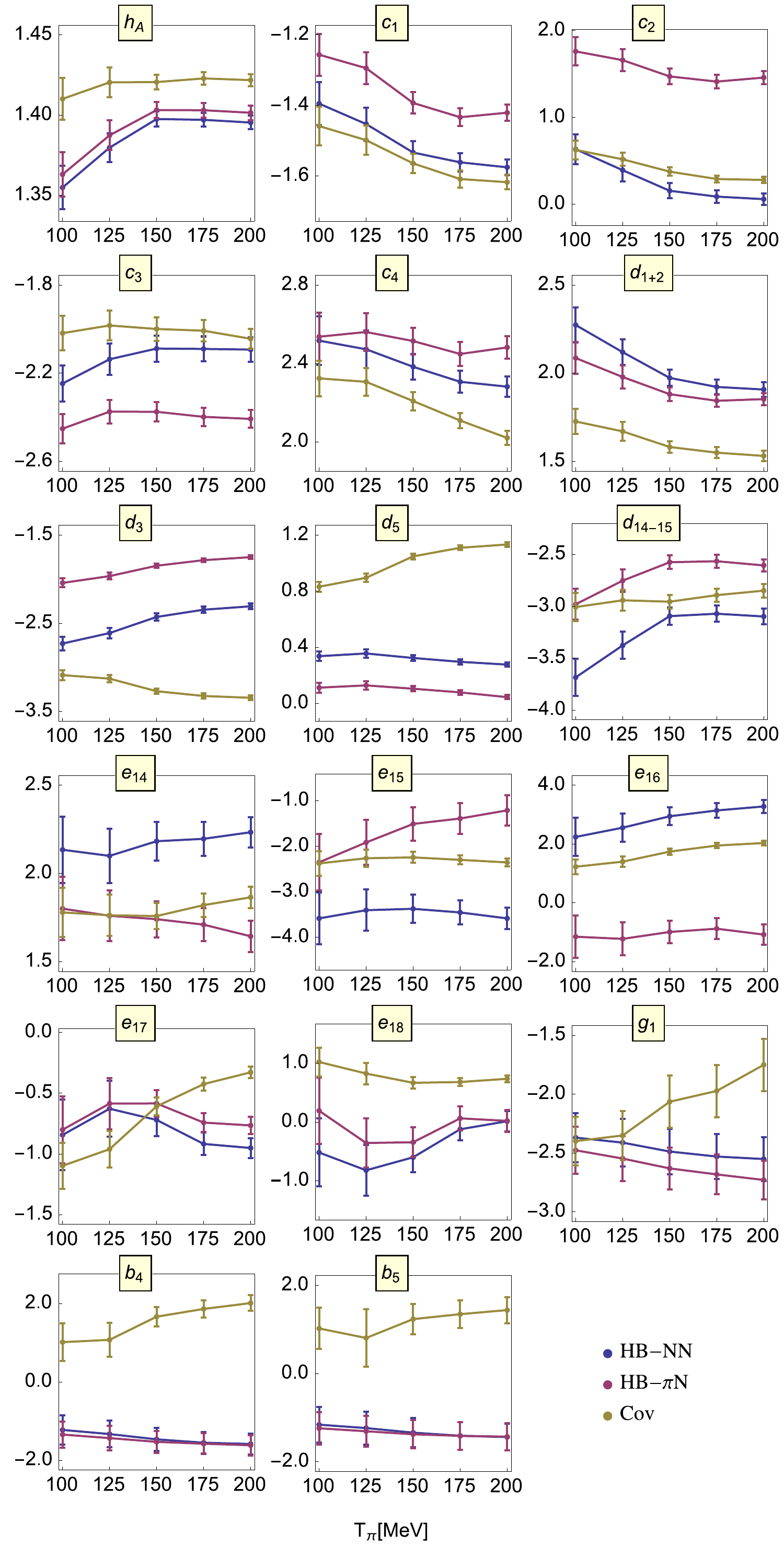}
  \caption{$K$-matrix approach: LECs extracted at order
    $\varepsilon^4$ as functions of the maximum fit energy
    $T_\pi$, see Eq.~\eqref{eq2:31}. The labels
 HB-NN, HB-$\pi$N, and Cov (covariant) denote the different counting schemes
 of $1/m_N$ contributions, see section \ref{sec:power-count-renorm-Delta}.}
  \label{fig:LECsQ4K}
\end{figure}

\begin{figure}[ht!]
  \centering
  \includegraphics[width=0.78\textwidth]{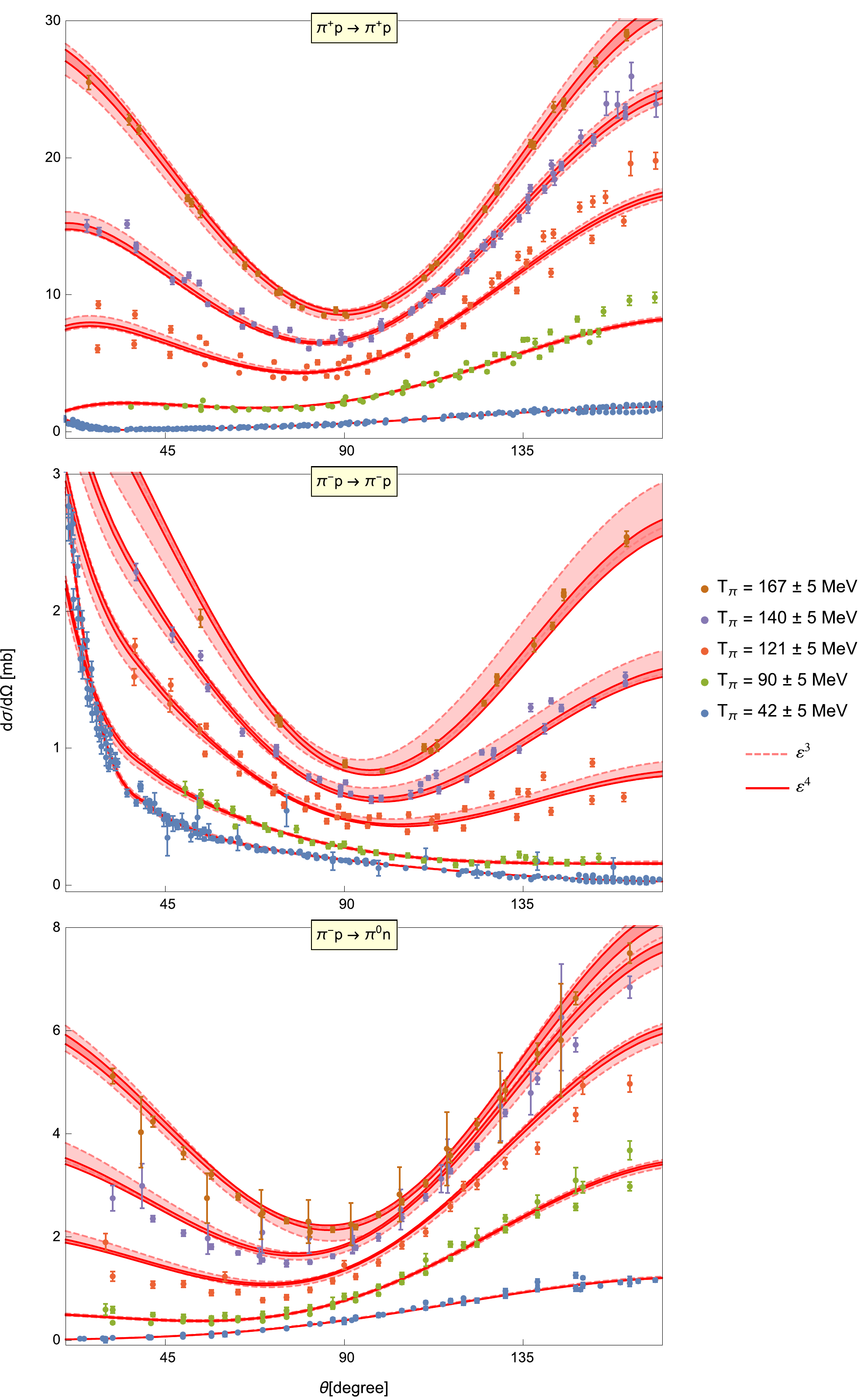}
\caption{$K$-matrix approach: Covariant predictions for the
  differential cross sections $\di\sigma/\di\Omega$ up to pion energies
  $T_\pi=170$ MeV.   The pink and red (dashed and solid) bands refer to $\varepsilon^3$ and $\varepsilon^4$ results
  including theoretical uncertainties,
  respectively. The experimental data are taken from the GWU-SAID data
  base \cite{Workman:2012hx}.}
\label{fig:DataPlotK}
\end{figure}

\begin{figure}[ht!]
  \centering
  \includegraphics[width=0.8\textwidth]{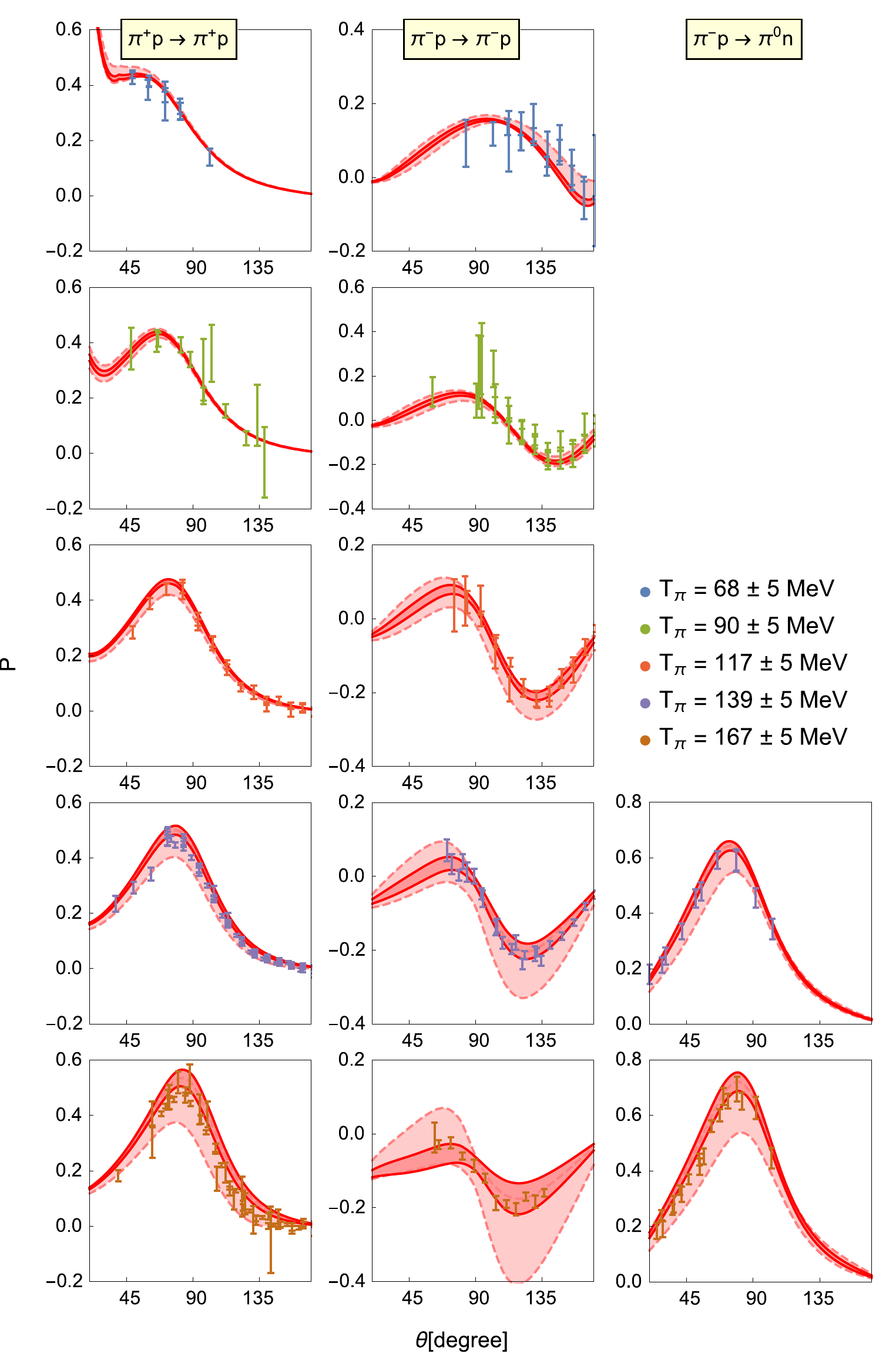}
\caption{$K$-matrix approach: Covariant predictions for the
  polarizations $P$ up to pion energies
  $T_\pi=170$ MeV.   The pink and red (dashed  and solid) bands refer to $\varepsilon^3$ and $\varepsilon^4$ results
  including theoretical uncertainties,
  respectively. The experimental data are taken from the GWU-SAID data
  base \cite{Workman:2012hx}.}
\label{fig:DataPlotPK}
\end{figure}

\begin{figure}[ht!]
  \centering
  \includegraphics[width=0.6\textwidth]{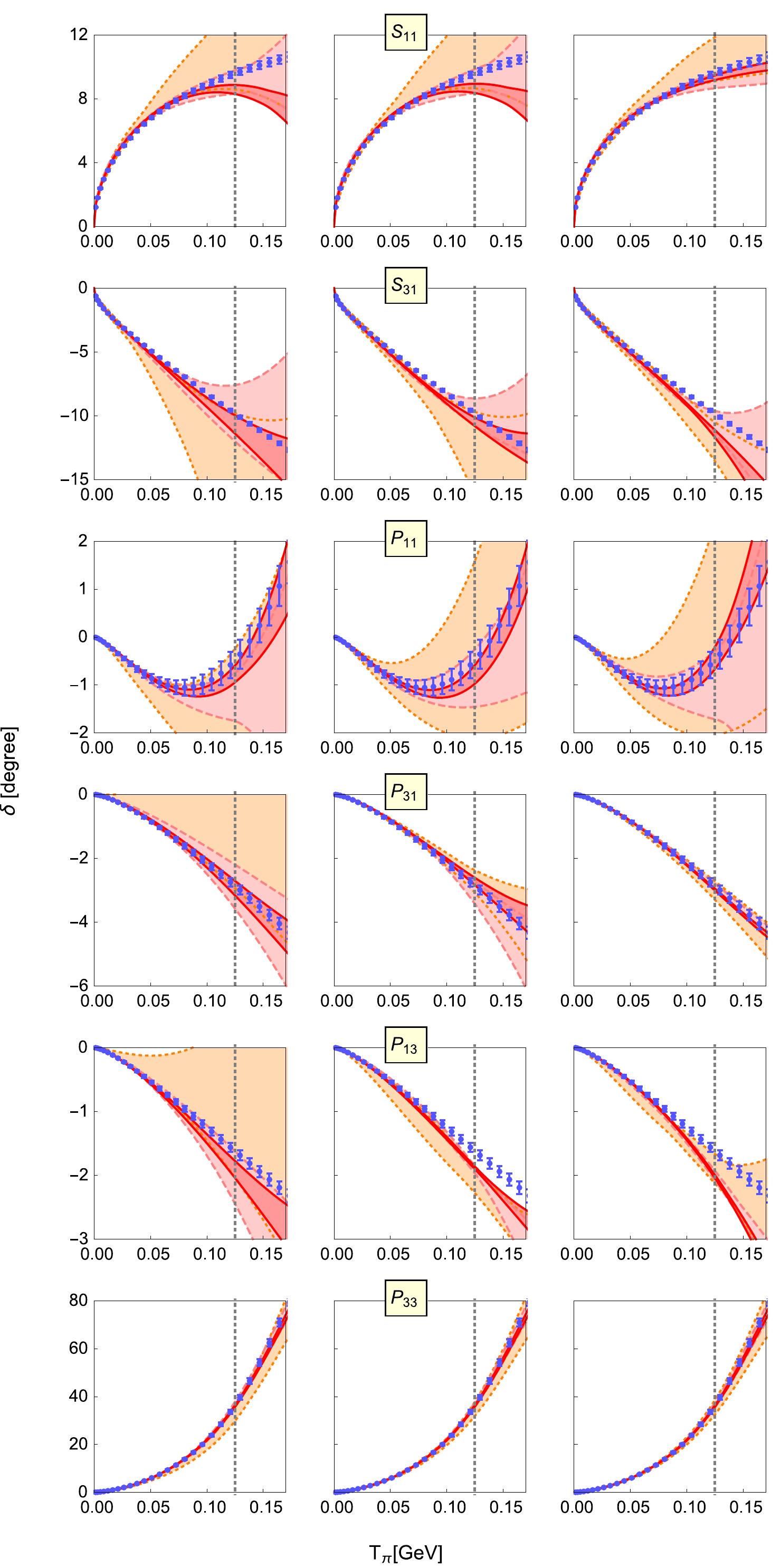}
\caption{$K$-matrix approach: Predicted $S$- and $P$-wave phase shifts up to pion energies
  $T_\pi=170$~MeV. The predictions in the HB-NN, HB-$\pi$N, and
  covariant counting are given in the columns from left to right, respectively.
  The orange, pink, and red (dotted, dashed, and solid) bands refer to
  $\varepsilon^2$, $\varepsilon^3$, and $\varepsilon^4$ results
  including theoretical uncertainties, respectively. The gray dotted
  vertical line marks the fitting limit. The data are taken from the RS analysis \cite{Hoferichter:2015hva}.}
\label{fig:SnPwavesK}
\end{figure}

We first start our discussion with the more commonly used $K$-matrix approach.
In this approach, we performed fits to all available data for all scattering angles
and incoming pion kinetic energy of $T_\pi<\{100, 125, 150, 175, 200\}$~MeV, which corresponds
to $\{1704, 1854, 2176, 2399, 2564\}$ data points, respectively.
In Fig.~\ref{fig:RedChiSqK}, the reduced $\chi^2$ and $\bar\chi^2$ values for these set of fits are presented 
as functions of the maximum fit energy $T_\pi$.
In Figs.~\ref{fig:LECsQ3K} and \ref{fig:LECsQ4K}, the extracted LECs as functions of $T_\pi$ are plotted as well.
As can be seen in these figures, we get a plateau-like behavior of the
extracted LECs and the reduced $\chi^2$ is close to $1$ i in the energy range between $100$~MeV and $150$~MeV. 
Interestingly, the reduced $\chi^2$ and $\bar\chi^2$ slightly decrease for fits including data at higher energies,
however, the extracted LECs start to deviate from their plateaus.
This behavior is clearly attributed to the $K$-matrix unitarization, which is used in this analysis
in the calculation of the phase shifts. As can be seen in Eq.~\eqref{eq:1a}, the employed unitarization prescription
modifies the perturbative amplitude, where the modification is getting stronger with an increase in the magnitude of the phase shift.
As mentioned before, the reliability of the $K$-matrix unitarization is only ensured for small phase shifts ($|\delta|<\pi/6$).
Thus, guided by arguments above, we choose the fits with $T_\pi<125$~MeV as representative results for the following discussion.
\\\\
In Table~\ref{tab:FitK}, the extracted LECs at different chiral orders are compared
for the three chiral approaches. Every set of LECs is also supplemented by 
values of the reduced $\chi_{\pi N}^2$ and $\bar\chi_{\pi N}^2$ of the corresponding fit.
Additionally, the same set of values but including the constraints $\chi^2_\mathrm{RS}$ are given as comparison.
Fortunately, we can observe that an increase in the chiral order leads to a decrease of the reduced $\bar\chi_{\pi N}^2$.
This clearly demonstrates an improved description of the scattering data when higher chiral orders are taken into account.
Due to large theoretical uncertainties at lower orders, we can see that the reduced $\chi_{\pi N}^2$, unfortunately, has the opposite rising behavior
Note that this was also observed in the previous $\Delta$-less analysis. Furthermore, the values for $\bar\chi_{\pi N}^2$/dof at fourth order
show that all three considered chiral approaches give a similar description of the scattering data.
When considering the additional constraints  $\chi^2_\mathrm{RS}$ from the subthreshold region, we observe
on the one hand a negligible increase of $\bar \chi^2_{\pi N}$/dof. Meaning that the description quality of the scattering data is
only slightly degraded. But, on the other hand, as seen in
Tables~\ref{tab:SubThrParaKvsC} and \ref{tab:ThrParaKvsC}, that the determined values for the
subthreshold and threshold parameters are substantially closer to the RS results.
Given these findings, we decided to select fits including the constraints from the subthreshold region as
as representative results of the approach employing the $K$-matrix unitarization. In what follows,
we only discuss predictions based on the LECs extracted from Eq.~\eqref{eq2:48} with $T_\pi<125$~MeV.
In Tables~\ref{tab:Q4piNCorrCov1} - \ref{tab:Q4piNCorrCov3}, we provide for these selected fits the corresponding 
correlation and covariance matrices. Although some stronger correlations occur in the two HB approaches, it is 
comforting that in the covariant approach these stronger correlations are visibly reduced.
\\\\
Next, we focus on the predictions of the numerous $\pi N$ scattering observables. See Ref.~\cite{Siemens:2016hdi}
for an extensive summary of these observables. All predictions are based on the values of LECs taken from Table~\ref{tab:FitK} 
in the columns denoted by $\pi$N+RS. Furthermore, we present all results with a statistical error and a theoretical error.
\\\\
In the columns denoted by $\pi$N+RS in Tables~\ref{tab:SubThrParaKvsC} and \ref{tab:ThrParaKvsC},
the determined values for the subthreshold und threshold parameters are given at order $\varepsilon^4$ .
Unfortunately, both HB counting schemes give rather disappointing results, where the HB-NN counting seems to be
slightly superior to the HB-$\pi$N counting. In addition, a majority of these parameters have rather large theoretical errors,
which is directly contributed to strong changes in between the determined values at lower chiral orders.
Note, however, that many parameters only coincide with the RS values when considering more than one standard deviation.
We only get a good agreement with the RS values for the four leading parameters $d^+_{00}$, $d^+_{10}$,
$d^+_{01}$ and $b^-_{00}$. Fortunately, for the covariant approach we get a better results, where 
most of the 8 leading parameters agree closely with the RS analysis. Only $b^+_{00}$
turns out too small. Additionally, the theoretical errors estimated for these parameters are relatively small,
which is an indication for a superior convergence in the covariant approach.
Finally, we turn to the the predictions for the threshold parameters. 
As can be seen, these predicted parameters are in good agreement with the RS results.
This can be observed for all three considered chiral approaches, where the covariant one
gives slightly better predictions. Only $a_{0+}^-$ turns out too large.
\\\\
Next, we turn to our results for the scattering observables, in particular, 
the differential cross sections $\di\sigma/\di\Omega$ and polarizations $P$.
In Figs.~\ref{fig:DataPlotK} and \ref{fig:DataPlotPK}, we show our results for these
observables up to pion energies $T_\pi=170$~MeV. Note that we only visualize the
dominant theoretical error in these figures. Furthermore, the theoretical curves
for $\di\sigma/\di\Omega$ and $P$ are calculated at
the mean energies $T_\pi=\{42, 90, 121, 140, 167\}$~MeV and $T_\pi=\{68,
90, 117, 139, 167\}$~MeV, respectively. The experimental data, however are shown
in an additional energy range of $\pm 5$ MeV around the above mentioned mean values.
As can be seen, we get an excellent agreement between theory and experiment
even above the fitted data, namely for energies higher than $T_\pi=125$~MeV.
Note that we only show the results calculated based on the covariant framework.
These results should, however, be understood as representative for all three chiral approaches.
\\\\
Finally, he predictions for the $S$-, $P$-, $D$- and $F$-wave phase shift are shown
in Figs.~\ref{fig:SnPwavesK} - \ref{fig:FwavesK}, where we see an 
excellent agreement with the RS values for the $S$- and $P$-waves up to energies
$T_\pi=100$~MeV. For higher energies, however, the phase shifts of the $S_{11}$-wave
in the HB approaches, the  $S_{31}$-wave in the covariant approach and of the $P_{13}$-wave
in all approaches behave rather different compared to the RS results.
Due to the relatively small magnitude of the phase shifts of the $P_{13}$-wave, the observed deviation
should be taken with care. In contrast, the strong deviations for the leading $S$-waves is quite surprising.
The higher order phase shifts for the $D$- and $F$-wave are compared with the results from 
GWU-SAID. We are able to reproduce these results with some exceptions. 
In particular, we get stronger discrepancies for the $D_{33}$-wave in the
covariant approach, the $D_{35}$-wave in the HB-$\pi$N approach and 
the $F_{35}$-, $F_{17}$-waves in both HB approaches.
Note, however, that the contributions from the $D$, $F$ (and higher) partial waves
are negligible at the considered low energies. This is also explicitly demonstrated in the RS analysis.
In contrast, the $D$- and $F$-waves in the GWU-SAID analysis are constrained by scattering data 
at higher energies and then continuously extended to lower energies, in particular, the threshold region.
Unfortunately, they do not provide any error estimates of their approach.
Thus, minor deviations for higher partial waves should be taken with caution.

\subsection{Complex Mass Approach}
\label{sec:compl-mass-appr}
\begin{figure}[ht!]
  \centering
  \includegraphics[width=0.78\textwidth]{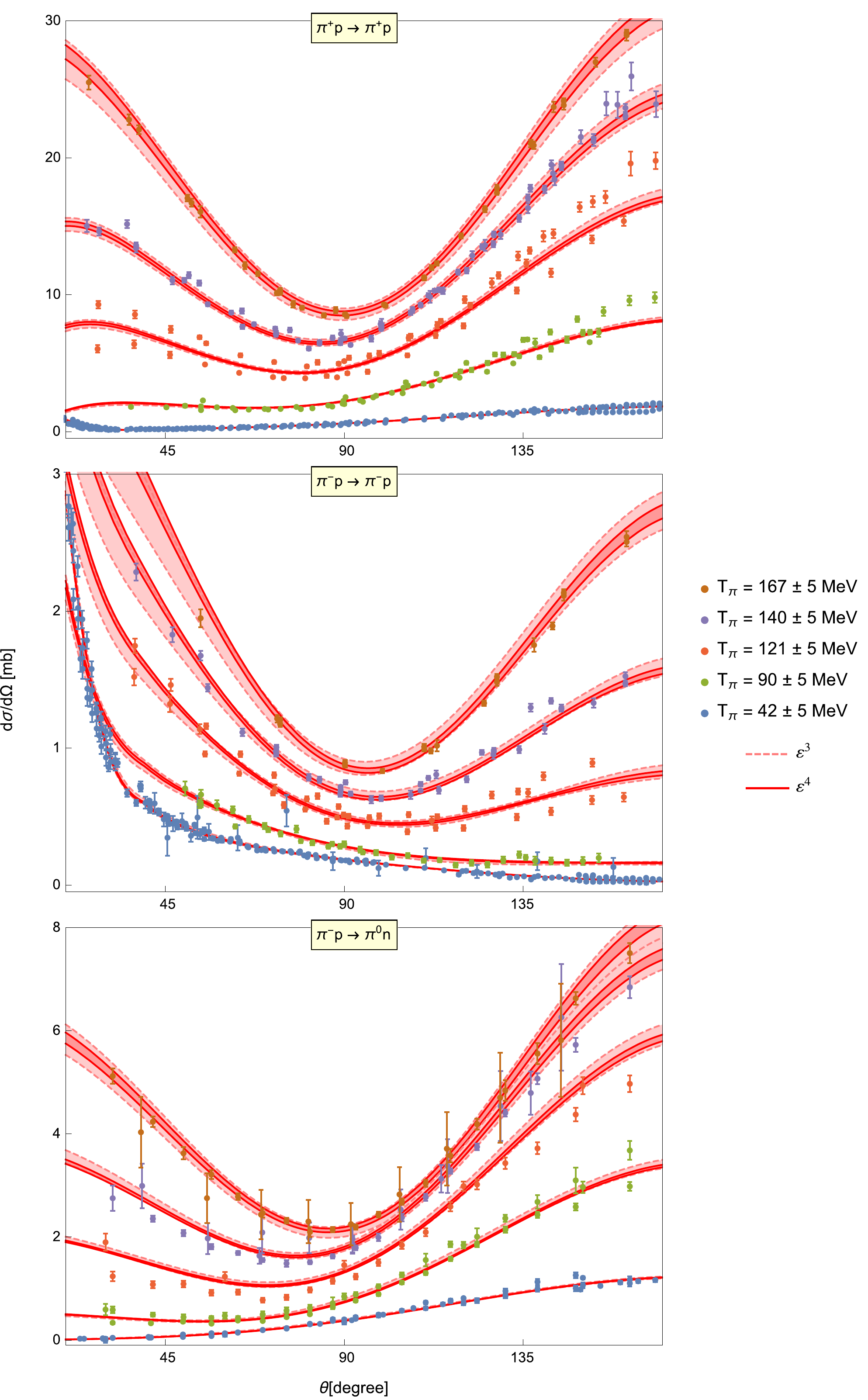}
\caption{Complex mass approach: Covariant predictions for the
  differential cross sections $\di\sigma/\di\Omega$ up to pion energies
  $T_\pi=170$ MeV.   The pink and red (dashed  and solid) bands refer to $\varepsilon^3$ and $\varepsilon^4$ results
  including theoretical uncertainties,
  respectively. The experimental data are taken from the GWU-SAID data
  base \cite{Workman:2012hx}.}
\label{fig:DataPlotC}
\end{figure}

\begin{figure}[ht!]
  \centering
  \includegraphics[width=0.8\textwidth]{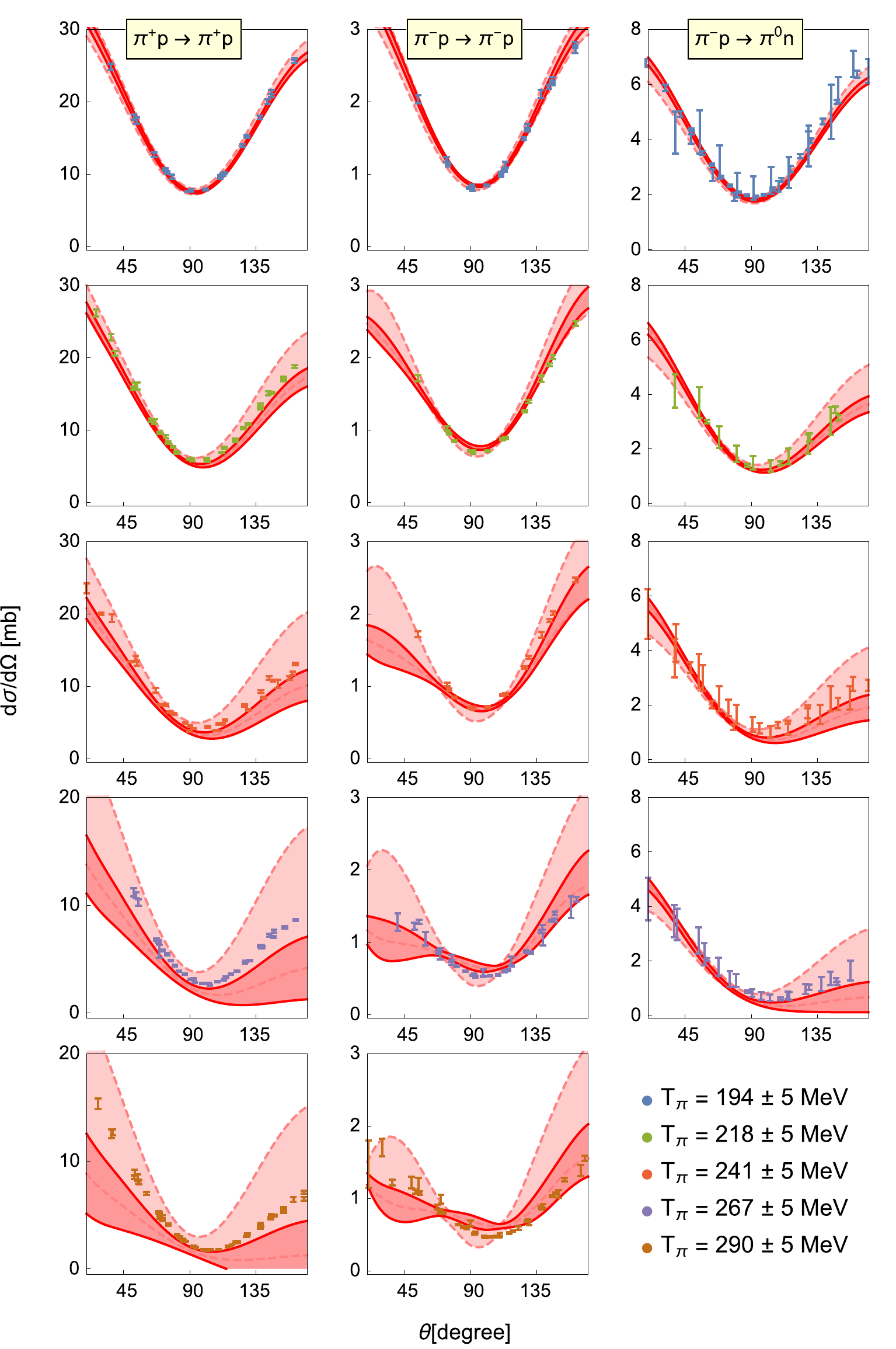}
\caption{Complex mass approach: Covariant predictions for the
  differential cross sections $\di\sigma/\di\Omega$ up to pion energies
  $T_\pi=300$ MeV.   The pink and red (dashed  and solid) bands refer to $\varepsilon^3$ and $\varepsilon^4$ results
  including theoretical uncertainties,
  respectively. The experimental data are taken from the GWU-SAID data
  base \cite{Workman:2012hx}.}
\label{fig:DataPlotC2}
\end{figure}

\begin{figure}[ht!]
  \centering
  \includegraphics[width=0.8\textwidth]{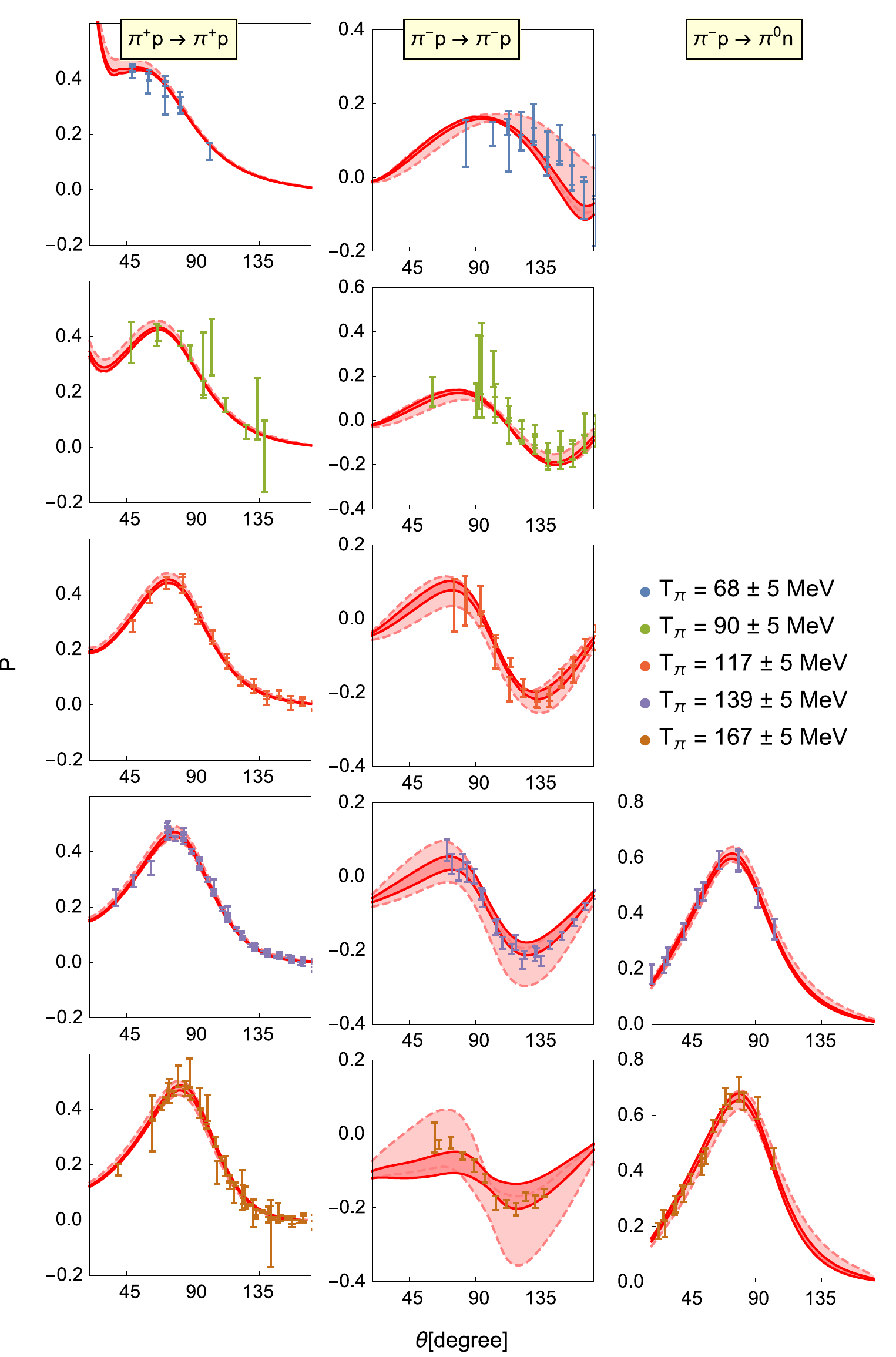}
\caption{Complex mass approach: Covariant predictions for the
  polarizations $P$ up to pion energies
  $T_\pi=170$ MeV.   The pink and red (dashed  and solid) bands refer to $\varepsilon^3$ and $\varepsilon^4$ results
  including theoretical uncertainties,
  respectively. The experimental data are taken from the GWU-SAID data
  base \cite{Workman:2012hx}.}
\label{fig:DataPlotPC}
\end{figure}

\begin{figure}[ht]
  \centering
  \includegraphics[width=0.8\textwidth]{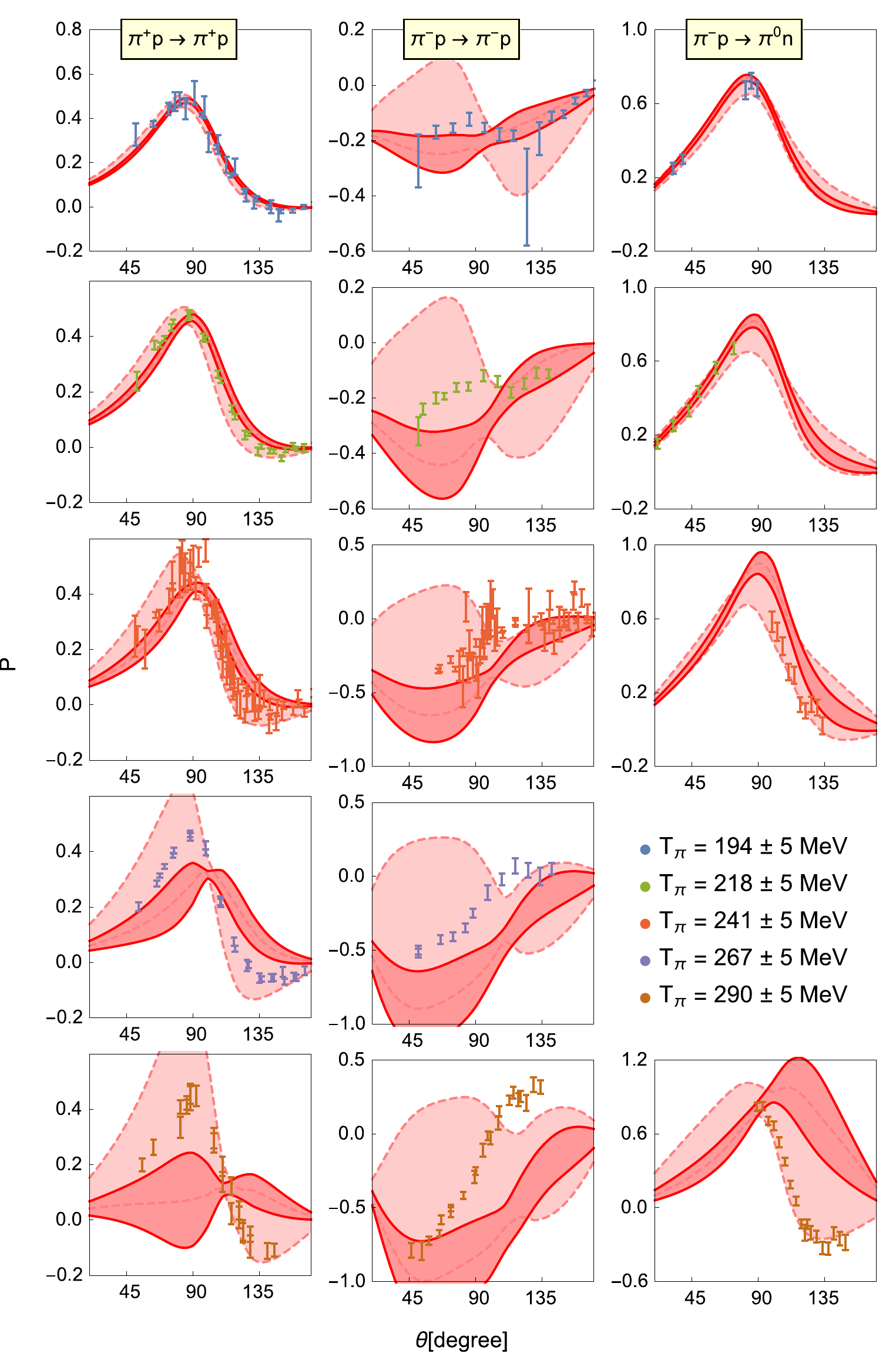}
\caption{Complex mass approach: Covariant predictions for the
  polarizations $P$ up to pion energies
  $T_\pi=300$ MeV.   The pink and red (dashed  and solid) bands refer to $\varepsilon^3$ and $\varepsilon^4$ results
  including theoretical uncertainties,
  respectively. The experimental data are taken from the GWU-SAID data
  base \cite{Workman:2012hx}.}
\label{fig:DataPlotPC2}
\end{figure}

\begin{figure}[ht]
  \centering
  \includegraphics[width=0.6\textwidth]{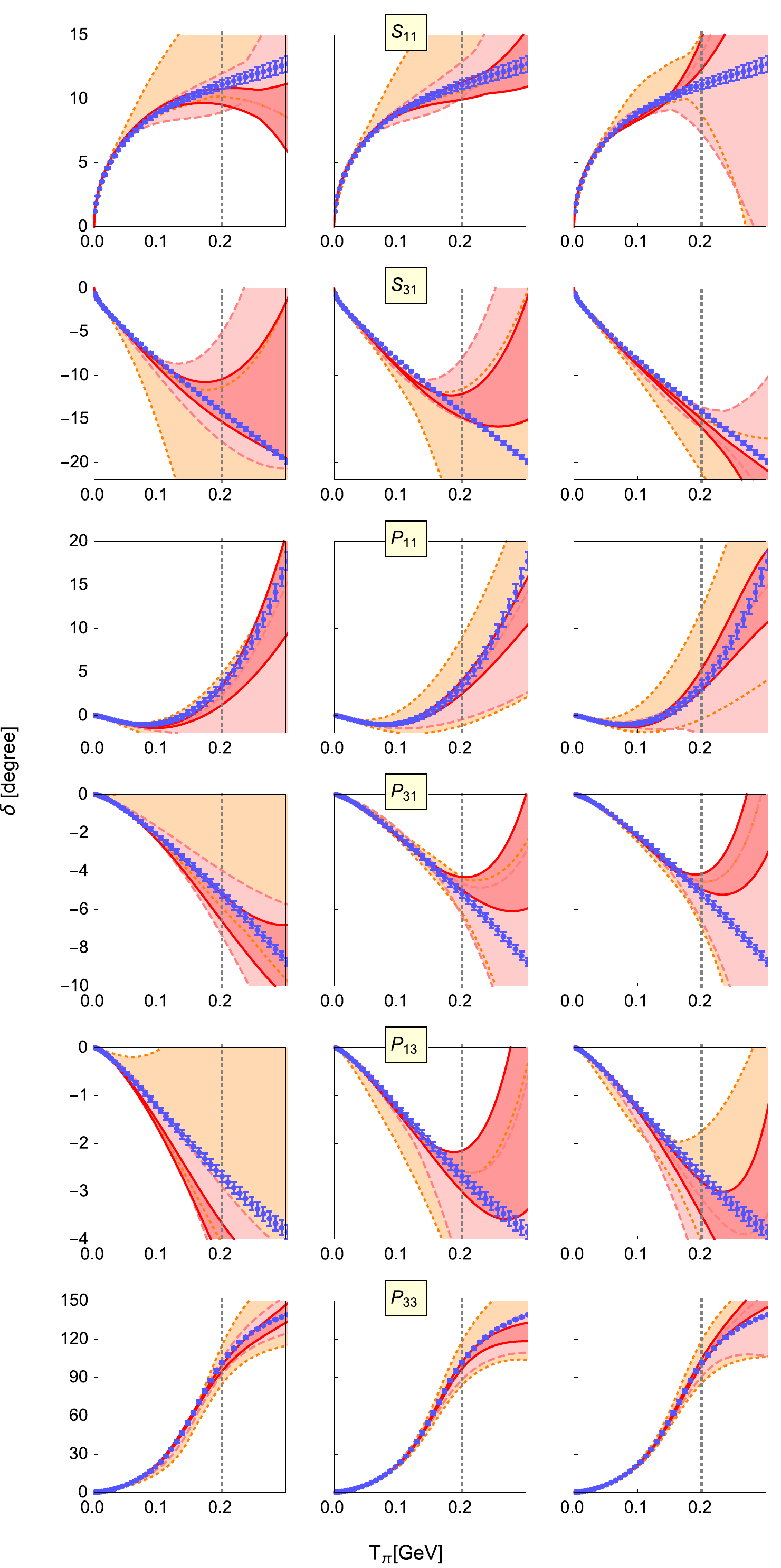}
\caption{Complex mass approach: Predicted $S$- and $P$-wave phase shifts up to pion energies
  $T_\pi=300$~MeV. The predictions in the HB-NN, HB-$\pi$N, and
  covariant counting are given in the columns from left to right, respectively.
  The orange, pink, and red (dotted, dashed, and solid) bands refer to
  $\varepsilon^2$, $\varepsilon^3$, and $\varepsilon^4$ results
  including theoretical uncertainties, respectively. The gray dotted
  vertical line marks the fitting limit. The data are taken from the RS analysis \cite{Hoferichter:2015hva}.}
\label{fig:SnPwavesC}
\end{figure}

As discussed in the previous section, the $K$-matrix approach has only a limited energy range of applicability.
In contrast, the complex mass approach should be applicable at higher energies as well. 
Thus, in the complex mass approach, we extend the energy region by performing fits to all available scattering
data with energy $T_\pi<\{150, 175, 200, 225, 250, 275, 300\}$~MeV corresponding to 
$\{2176, 2399, 2564, 2727, 3004, 3147, 3413\}$~data points,
respectively. In Figs.~\ref{fig:RedChiSqK}, \ref{fig:LECsQ3C} and \ref{fig:LECsQ4C},
we present the reduced $\chi^2$, $\bar\chi^2$ and the extracted LECs as functions of the maximum energy $T_\pi$, respectively.
As can be seen in these figures, by employing the complex mass approach we are able to extend the
plateau-like behavior of the extracted LECs to higher energies of up to $T_\pi\simeq 225$~MeV. 
We also observe that the reduced $\chi^2$ slightly decreases with increasing energy in the two HB approaches.
In the covariant counting scheme the reduced $\chi^2$ is minimal at $T_\pi\simeq200$~MeV. 
At this point, we remind the reader of the discussion in the end of section \ref{sec:power-count-renorm-Delta},
where we pointed out that we had to modify the amplitudes in both HB approaches
by replacing the $\Delta$ pole contribution by the covariant expression.
Thus, these result should be taken with care and we rather focus on the results in the covariant formalism from now on.
In particular, as representative results of the first step of the fitting procedure we choose the fits with $T_\pi<200$ MeV.
\\\\
Next, we turn to the results of the fitting procedure. The extracted LECs at different chiral orders
are shown in Table~\ref{tab:FitC} for all three chiral approaches. Additionally, we give the
corresponding reduced $\chi_{\pi N}^2$ and $\bar\chi_{\pi N}^2$.
As in the $K$-matrix approach, additional constraints from the subthreshold region ($\chi^2_\mathrm{RS}$)
are considered in the fitting routine and the results without and with these constraints are compared.
As can be seen, the reduced $\bar\chi^2$ decreases with an increasing chiral order, which 
is a clear indicator of an improving description of the scattering data.
Furthermore, we observe rather large theoretical uncertainties at lower orders, which
explains the increasing values of the reduced $\chi^2$ with increasing chiral order.
In summary, the three chiral approaches give a similar fit quality. However, the 
convergence pattern in the covariant framework, seen in the small changes
of the reduced $\chi^2_{\pi N}$ between chiral orders, seems to be superior to the other approaches.
Note the small difference between the values at order $\varepsilon^3$ and $\varepsilon^4$.
Including the constraints $\chi^2_\mathrm{RS}$ visibly impacts the values of the LECs
but only slightly effects the value of the reduced $\chi^2_{\pi N}$.
In addition, we also get an improved description of the subthreshold and threshold parameters
given in Tables~\ref{tab:SubThrParaKvsC} and \ref{tab:ThrParaKvsC}. Unfortunately,
most of the subthreshold parameters still exhibit rather strong deviations from the RS values.
It is surprising that not even the leading subthreshold parameters are properly reproduced.
Even the covariant result turn out poorly, although this approach worked quite well within the
$K$-matrix unitarization studied in the previous section.
Similar conclusions can also be drawn for the threshold parameters.
The observed inability to describe the threshold and subthreshold region properly is due to
the employed complex mass approach. By construction, we use a constant $\Delta$ width in the
amplitudes. But this is only a good approximation in the vicinity of the renormalization point $\slashed{p}=m_\Delta^\mathrm{Re}$,
which corresponds to what we labeled as the $\Delta$ region. Clearly, if we move away from the renormalization point,
this approximation becomes less reliable. In particular, the energy-dependent width vanishes at the threshold
and subthreshold point, which is not the case for the constant width employed in our amplitudes.
Thus, in the analysis of the subthreshold and threshold parameters, it was necessary to perform an additional expansion
of the amplitudes in powers of the $\Delta$ width. This clearly modifies the amplitudes constrained by fits to experimental data.
Note that these modifications are strictly speaking of higher chiral order.
We have to stress that it was not necessary to perform such kind of modification in the $K$-matrix approach.
Instead, we were able to use the same amplitudes as input for the fits to the experimental data and for the 
subthreshold and threshold parameters. Another difference between both approaches, is that in the
the $K$-matrix unitarization only the real part of the partial wave amplitude is considered, whereas in the 
complex mass approach the full amplitude including the imaginary parts is taken into account.
In particular, the imaginary part of the amplitude, not considering the $\Delta$ pole dressing for a moment,
is only calculated up to its next-to-leading order in the $\varepsilon^4$ $\pi$N-amplitudes.
Taken our observations for the real part contributions at next-to-leading order into account, it is highly unlikely
that the we have reasonably well converged next-to-leading order expressions for the imaginary parts.
Additionally, in the $\Delta$ pole region, the dressed $\Delta$ pole amplitude are enhanced to order $\varepsilon^{-1}$ 
and other parts of the amplitude keep their original chiral order.\footnote{The enhancement is due to the different 
countings of the denominator of the $\Delta$ propagator in both kinematical regions. 
In the threshold region, one has $\slashed{p}-m_\Delta\sim \mathcal{O}(\varepsilon^1)$ and in the vicinity of the $\Delta$ pole, 
one has $\slashed{p}-m_\Delta\sim\Gamma\sim \mathcal{O}(\varepsilon^3)$.}
Thus, the $\Delta$ pole contributions become extremely important and the specific treatment of these contributions is crucial, see Eq.~\eqref{eq2:42}.
To sum up, it should not be surprising that we observe superior results in the $K$-matrix approach when connecting threshold 
and subthreshold region. The motivation of the complex mass approach is to extend the applicability of the theory to higher energies and the
$\Delta$ pole region. The cost of this extension is the reduced accuracy of describing subthreshold and threshold parameters.
Based on this discussion, we prefer the fits without additional constraints from the subthreshold region as representative results.
In particular, we only discuss in what follows the results from the complex mass approach based on the LECs extracted 
from Eq.~\eqref{eq2:31} with $T_\pi<200$ MeV. In Tables~\ref{tab:Q4piNCorrCov1} - \ref{tab:Q4piNCorrCov3}, the corresponding
correlation and covariance matrices can be found. Note the rather small correlations between the extracted
LECs in the covariant approach.
\\\\
Like in the previous section on the $K$-matrix approach, we discuss the predicted observables once again.
In particular, the LECs collected in Table~\ref{tab:FitC} in the columns denoted by $\pi$N are employed.
In Tables~\ref{tab:SubThrParaKvsC} and \ref{tab:ThrParaKvsC}, we show the
predictions for the subthreshold and threshold parameters. 
The results are quite unsatisfactory which is, as we already mentioned, a consequence of the employed scheme.
We refer the reader to the previous section for a more consistent determination of these parameters.
\\\\
Next, we turn to the predictions for the differential cross sections
$\di\sigma/\di\Omega$ and polarizations $P$. In Figs.~\ref{fig:DataPlotC} and \ref{fig:DataPlotPC}
we show the predicted values up to $T_\pi=170$~MeV. In Figs.~\ref{fig:DataPlotC2} and \ref{fig:DataPlotPC2}
we extend the energy range up to $T_\pi=300$ MeV.
Note that we calculate the theoretical curves based on mean values, whereas the experimental data
are shown in an additional energy range of $\pm 5$ MeV around those means. For $\di\sigma/\di\Omega$
the mean energies are $T_\pi=\{42, 90, 121, 140, 167, 194,218,241,267,290\}$~MeV and for $P$
we have chosen $T_\pi=\{68, 90, 117, 139, 167,194,218,241,267,290\}$~MeV.
In these figures we only show the results in the covariant framework, which should be 
taken as representative examples. As can be seen, the theoretical curves match the 
experimental data perfectly up to energies $T_\pi=220$~MeV, whereas for 
higher energies the predictions deviate from the experimental values.
Note the stronger deviations for the observable $P$.
\\\\
Finally, we present the predictions for the phase shifts in all three chiral approaches. 
The predicted $S$-, $P$-, $D$- and $F$-wave phase shifts are shown in Figs.~\ref{fig:SnPwavesC} - \ref{fig:FwavesC}.
Considering the fitting region, the $S$- and $P$-waves in all approaches agree very well with the RS results,
except the $P_{13}$-wave in the HB-NN counting.
Going to higher energies, we observe stronger deviations for both $S$-waves and the
$P_{31}$-wave. In the covariant approach, we see an agreement of the $D$- and $F$-waves with the GWU-SAID mean
values, in particular up to $T_\pi=200$~MeV. The only exception is the $F_{17}$-wave, which exhibits the 
opposite behavior as indicated by the data. Except for the $D_{33}$-wave, the predicted values for higher energies are reasonable as well.
The HB approaches are not able to describe the $D_{35}$- and $F_{35}$-wave properly and the HB-NN counting is not able
to predict $F_{17}$-wave. For the rest of the partial waves, both countings give a fair description.

\section{Summary and Outlook}
\label{sec:sum-Delta}
The main results of this paper are summarized as follows:
\begin{itemize}
\item
We calculated the pion-nucleon scattering amplitudes in two heavy baryon and in the covariant
framework of baryon $\chi$PT with explicit $\Delta$(1232) degrees of freedom. In particular,
the calculation was performed up to order $\varepsilon^4$ in the small scale expansion.
Additionally, we discussed in detail the renormalization procedure including $\Delta$ contributions of the pion-nucleon LECs.
The explicit expressions for the UV divergent parts and additional finite shifts are given 
in the appendix and/or in the supplementary material.
\item
We performed fits to low-energy pion-nucleon scattering observables employing three
different countings of relativistic corrections, labeled as HB-NN, HB-$\pi$N and Cov.
Furthermore, we used a systematical approach to account for the
theoretical uncertainties due to the truncation of the chiral series.
The LECs were extracted from two kinematical regions, the threshold
and the $\Delta$ pole region. In particular, for each region we employed 
different unitarization procedures, the $K$-matrix and the complex mass approach, respectively.
Additionally, we studied the impact of the 8 leading subthreshold parameters when included
as additional constraints in the fitting procedure. All in all, the extracted LECs turned out to be of natural size.
\item
In the $K$-matrix approach, we chose the LECs extracted from fits to experimental data with $T_\pi<125$~MeV
including the constraints from the subthreshold region as the representative and most reliable values.
With these LECs we get an excellent description of the low-energy experimental data up to $T_\pi=170$~MeV 
and a very good agreement with the threshold and subthreshold parameters of the RS analysis.
In addition, our predictions for the phase shifts of the $S$- and $P$-waves are in agreement with the RS values near threshold. 
However, some of the higher-order partial waves show a different behavior compared to the GWU results.
Comparing the different chiral approaches, we were able to observe that the covariant approach is far more
superior in connecting the threshold and subthreshold region.
\item
In the complex mass approach, we chose the LECs extracted from fits to experimental data with $T_\pi<200$~MeV
and without the constraints from the subthreshold region as the representative and most reliable values.
Given these LECs we get an excellent agreement of the theoretical values with the low-energy data up to $T_\pi=220$~MeV.
Thus, employing this approach we were able to extend the applicability of $\chi$PT to higher energies up to the
$\Delta$ pole region. However, we had to pay a price for this extension. The theory fails to predict the 
threshold and subthreshold parameters. Like in the $K$-matrix approach, we observed superior results
in the covariant framework.
\end{itemize}
The extensive analysis in this paper is a fundamental step in the study of pion-nucleon-$\Delta$ physics.
In particular, this analysis should be extended by performing a combined analysis of $\pi N \to \pi N$ and $\pi N \to
\pi \pi N$.


\clearpage
\appendix

\newpage
\section{Effective Lagrangian}
\label{sec:effective-lagrangian}
The effective pion-nucleon Lagrangian with explicit $\Delta$ degrees of freedom consists of four parts: 
pion-pion $(\pi\pi)$, pion-nucleon~($\pi N$), pion-$\Delta$ ($\pi\Delta$), 
and pion-nucleon-$\Delta$ ($\pi N\Delta$) interactions. Thus, at a given chiral order $n$ one has
\begin{equation}
  \label{eq0:137}
  \mathcal{L}^{(n)}=\mathcal{L}^{(n)}_{\pi \pi}+\mathcal{L}^{(n)}_{\pi N}+  \mathcal{L}_{\pi
    \Delta}^{(n)}+
\mathcal{L}^{(n)}_{\pi
    N\Delta}\komma
\end{equation}
where the chiral order of an interaction term is determined by the small scale expansion with
the expansion parameter \cite{Hemmert:1997ye}
\begin{equation}
  \label{eq0:140}
  \varepsilon=\left\{\frac{q}{\Lambda_b},\frac{M}{\Lambda_b},\frac{\Delta_0}{\Lambda_b}\right\}\qquad\mathrm{with}\qquad \Lambda_b\in\{\Lambda_\chi,4\pi F,m\}\punkt
\end{equation}
Furthermore, the chiral order $D$ of a Feynman diagram 
with $L$ loops, $V^M_d$ mesonic vertices and $V^B_d$ baryonic vertices
is given by 
\begin{equation}
  \label{eq0:131}
  D= 2L + 1+ \sum_d (d-2)V_d^M +\sum_d (d-1)V_d^B\komma
\end{equation}
where $\Delta$ and nucleonic vertices are treated on the same footing. 
In the following, we only discuss terms contributing to pion nucleon scattering up to chiral order $n=4$
and in the case of $\mathrm{SU}(2)$. Thus, we set the external sources to $p=a_\mu=v_\mu=0$ and $s=\mathcal{M}$.

\subsection{Chiral Perturbation Theory for Pions}
\label{sec:chipt-mesons}
The pion fields are conveniently collected in a matrix $U$, whose most general parametrization
constrained by unitarity is given by
\begin{equation}
  \label{eq0:73}
  \begin{aligned}
    U &= 1 + \i \frac{\vec{\tau} \cdot \vec{\pi}}{F} -
    \frac{\vec{\pi}^2}{2 F^2} - \i \alpha \frac{ \vec{ \pi}^2
      \vec{\tau} \cdot \vec{\pi} }{F^3} + \frac{(8 \alpha - 1)}{8
      F^4} \vec{\pi}^4 \\&- \i \beta \frac{\vec{ \pi}^4 \vec{\tau}
      \cdot \vec{\pi} }{F^5} +
    \frac{-1+8\alpha-8\alpha^2+16\beta}{16F^6} \vec{\pi}^6 +
    \ldots 
  \end{aligned}
\end{equation}
with
\begin{equation}
  \label{eq0:74}
\vec{\tau}\cdot\vec{\pi}=
  \begin{pmatrix}
        \pi^3& \pi^1-\i\pi^2\\
    \pi^1+\i\pi^2 & -\pi^3
  \end{pmatrix}
\equiv
  \begin{pmatrix}
    \pi^0 & \sqrt{2}\pi^+\\
    \sqrt{2}\pi^- & - \pi^0
  \end{pmatrix}\komma
\end{equation}
where we used the Pauli matrices $\tau_i$ and denoted the unphysical off-shell parameters by $\alpha$, $\beta$.
Employing the above matrix notation for the pion fields, the effective pion Lagrangian up to next-to-leading order reads 
\cite{Gasser:1983yg,Weinberg:1978kz}
\begin{align*}
    \mathcal{L}_{\pi\pi}^{(2)}&=\frac{F^2}{4}\braket{\partial_\mu
    U^\dagger \partial^\mu U}+\frac{F^2}{4}\braket{\chi_+}\komma\\
  \mathcal{L}_{\pi\pi}^{(4)}&=
      \frac{l_1}{4} \Braket{\partial_\mu U\partial^\mu U^\dagger}^2
      +\frac{l_2}{4} \Braket{\partial_\mu U\partial_\nu U^\dagger}\Braket{\partial^\mu U\partial^\nu U^\dagger}    
      +\frac{l_3}{16} \Braket{\chi_+}^2 \numberthis\label{eq0:L1}\\& 
     +\frac{l_4}{16}\Big(2 \Braket{\partial_\mu U \partial^\mu
      U^\dagger}\Braket{\chi_+}
    +2 \Braket{\chi^\dagger U \chi^\dagger U + U^\dagger \chi
      U^\dagger\chi}-4\Braket{\chi^\dagger \chi}-\Braket{\chi_-}^2\Big)
\end{align*}
where
\begin{equation}
  \label{eq0:76}
\chi = 2B_0 \mathcal{M} \qquad\text{with} \qquad \mathcal{M}=\textrm{diag}(m_u, m_d)
\end{equation}
and
\begin{equation}
  \label{eq0:23}
  \begin{aligned}
    \chi_\pm &= u^\dagger \chi u^\dagger \pm u\chi^\dagger u\komma
  \end{aligned}
\end{equation}
where we used the definition $u(\vec{\pi})=\sqrt{U(\vec{\pi})}$. 
Note the explicit chiral symmetry breaking by the non-zero quark masses $m_u$ and $m_d$.

\subsection{Chiral Perturbation Theory for Pions and Nucleons}
\label{sec:chipthbchipt-nucleons}
The proton $p$ and neutron $n$ fields are conveniently written in the isodoublet representation
\begin{equation}
  \label{eq0:203}
  \Psi=
  \begin{pmatrix}
    p\\n 
  \end{pmatrix}
  \punkt
\end{equation}
Given the introduced notation for the pion and nucleon fields, the covariant pion-nucleon Lagrangian up to fourth chiral order reads \cite{Fettes:2000gb}
  \begin{align*}
    \mathcal{L}_{\pi N}^{(1)}&=\bar{\Psi} \left[ \i
      \slashed{D}-m+\frac{g}{2}\slashed{u}\gamma_5 \right]\Psi\komma\\
    \mathcal{L}_{\pi N}^{(2)}&= \bar{\Psi} \left[c_1\Braket{\chi_+} +
      \frac{c_2}{8m^2} (-\braket{u_\mu
        u_\nu}D^{\mu\nu}+\mathrm{h.c.} ) +
      \frac{c_3}{2}\Braket{u\cdot u}-\frac{c_4}{2} \sigma^{\mu\nu}[u_\mu, u_\nu]\right]
    \Psi\komma \\
    \mathcal{L}_{\pi N}^{(3)}&=\bar{\Psi} \Big[-\frac{d_1}{2m}[u_\mu,[D_\nu,u^\mu]]D^\nu
      -\frac{d_2}{2m}[u_\mu,[D^\mu,u_\nu]]D^\nu+\frac{d_3}{12m^3}[u_\mu,[D_\nu,u_\rho]]D^{\mu\nu\rho}\\
      &+\frac{d_4}{2m}\epsilon^{\mu\nu\alpha\beta}\braket{u_\mu u_\nu u_\alpha}D_\beta
      +\frac{d_5}{2m}\i [\chi_-,u_\mu]D^\mu
      -\frac{d_{12}}{8m^2}\gamma^\mu \gamma_5 \braket{u_\lambda
        u_\nu}u_\mu D^{\lambda\nu}
      \\&-\frac{d_{13}}{8m^2}\gamma^\mu \gamma_5 \braket{u_\mu
        u_\nu}u_\lambda D^{\lambda\nu}
      -\frac{d_{14}}{2m} \sigma^{\mu\nu}\Braket{[D_\lambda,u_\mu]u_\nu}D^\lambda
       -\frac{d_{15}}{2m} \sigma^{\mu\nu}\Braket{u_\mu[D_\nu,u_\lambda]}D^\lambda 
       +\mathrm{h.c.} \Big] \Psi\\
       &+\bar\Psi \Big[\frac{d_{10}}{2}\gamma^\mu \gamma_5 \braket{u\cdot u}u_\mu
      +\frac{d_{11}}{2}\gamma^\mu \gamma_5 \braket{u_\mu u_\nu}u^\nu+\frac{d_{16}}{2}\gamma^\mu\gamma_5\Braket{\chi_+}u_\mu
       +\frac{d_{18}}{2}\i\gamma^\mu\gamma_5[D_\mu,\chi_-]\Big]\Psi\komma\\
    \mathcal{L}_{\pi N}^{(4)}&=\bar{\Psi}\Big[
       (\frac{e_{10}}{4m^2}\braket{h_{\lambda\mu}u_\nu}u_\rho \epsilon^{\mu\nu\rho}_{\quad\;\tau}D^{\lambda\tau}+\mathrm{h.c.})
+(\frac{e_{11}}{4m}\braket{h_{\lambda\mu}[u^\lambda,u_\nu]}\gamma_5\gamma^\mu D^{\nu}+\mathrm{h.c.}) \numberthis\label{eq0:L1piN}
\\&+(\frac{e_{12}}{4m}\braket{h_{\lambda\mu}[u^\lambda,u_\nu]}\gamma_5\gamma^\nu D^{\mu}+\mathrm{h.c.})
+(-\frac{e_{13}}{24m^3}\braket{h_{\lambda\mu}[u_\nu,u_\rho]}\gamma_5\gamma^\rho D^{\lambda\mu\nu}+\mathrm{h.c.})
      \\&+e_{14}\Braket{h_{\mu\nu}h^{\mu\nu}}
      +(-\frac{e_{15}}{4m^2}\braket{h_{\lambda\mu}h^{\lambda}_{\;\;\nu}}D^{\mu\nu}+\mathrm{h.c.})
      +(\frac{e_{16}}{48m^4}\Braket{h_{\lambda\mu}h_{\nu\rho}}D^{\lambda\mu\nu\rho}+\mathrm{h.c.})\\
      &-e_{17}[h_{\lambda\mu},h^\lambda_{\;\;\nu}]\sigma^{\mu\nu}
      +(\frac{e_{18}}{4m^2}[h_{\lambda\mu},h_{\nu\rho}]\sigma^{\mu\nu}D^{\lambda\rho}+\mathrm{h.c.})
      +e_{19}\braket{\chi_+}\braket{u_\mu u^\mu}\\
      &+(-\frac{e_{20}}{4m^2}\braket{\chi_+}\braket{u_\mu u_\nu}D^{\mu\nu}+\mathrm{h.c.})
      -e_{21}\braket{\chi_+}[u_\mu,u_\nu]\sigma^{\mu\nu}
      +e_{22}[D_\mu,[D^\mu,\braket{\chi_+}]]\\
      &+(\frac{e_{34}}{4m}\i\braket{\tilde\chi_-[u_\mu,u_\nu]}\gamma_5\gamma^\mu
      D^{\nu}+\mathrm{h.c.})
      +(-\frac{e_{35}}{4m^2}\i\braket{\tilde\chi_-h_{\mu\nu}}D^{\mu\nu}+\mathrm{h.c.})
      \\&+e_{36}\i \braket{u_\mu[D^\mu,\tilde\chi_-]}-e_{37}\i [u_\mu,[D_\nu,\tilde\chi_-]]\sigma^{\mu\nu}
      +e_{38}\braket{\chi_+}\braket{\chi_+}\\&+\frac{e_{115}}{4}\braket{\chi_+^2-\chi_-^2}
      -\frac{e_{116}}{4}(\braket{\chi_-^2}-\braket{\chi_-}^2+\braket{\chi_+^2}-\braket{\chi_+})^2
 \Big]\Psi\komma
  \end{align*}
where we employed the chiral vielbein
\begin{equation}
  \label{eq0:99}
  u_\mu= \i ( u^\dagger\partial_\mu u-u\,\partial_\mu u^\dagger )\komma
\end{equation}
the chiral connection
\begin{equation}
  \label{eq0:100}
  \Gamma_\mu=\frac{1}{2}(u^\dagger\partial_\mu u +u \,\partial_\mu u^\dagger)\komma
\end{equation}
the covariant derivative
\begin{equation}
  \label{eq0:101}
  D_\mu = \partial_\mu + \Gamma_\mu\komma
\end{equation}
the totally symmetrized product of covariant derivatives $D_{\mu\nu\dots}$ and additionally
\begin{equation}
  \label{eq0:24}
  \begin{aligned}
    h_{\mu\nu}&=[D_\mu,u_\nu]+[D_\nu,u_\mu]\komma\qquad
\tilde\chi_-=\chi_--\braket{\chi_-}\komma\\
[\tau_a,\tau_b]&=2\i \epsilon_{abc}\tau_c\komma\qquad
\{ \gamma_\mu,\gamma_\nu \}=2 g_{\mu\nu}\komma\qquad\sigma_{\mu\nu}=\frac{1}{4}[\gamma_\mu,\gamma_\nu]\punkt
  \end{aligned}
\end{equation}
We emphasize two differences between Eq.~\eqref{eq0:L1piN} and the effective Lagrangian in Ref.~\cite{Fettes:2000gb}.
In that reference, one defines $\sigma_{\mu\nu}=\frac{\i}{2}[\gamma_\mu,\gamma_\nu]$ and uses the
opposite sign for the interaction terms proportional to the LECs $d_4$, $e_{10}$ and $e_{34}$
\\\\
In the HB approach, the nucleon momentum is split into two parts
\begin{equation}
  \label{eq0:91}
  p_\mu=m\, v_\mu +k_\mu\komma
\end{equation}
where the first summand is a large contribution close to on-shell kinematics, with $v_\mu$ the four-velocity of the nucleon,
and the second summand is a residual contribution $k_\mu$.
Additionally, a decomposition of the nucleon field $\Psi$ into eigenstates of $\slashed{v}$ is employed, where
\begin{equation}
  \label{eq0:94}
  \begin{aligned}
    N &= e^{\i m v\cdot x}P_v^+\Psi\komma\qquad
    h =  e^{\i m v\cdot x}P_v^-\Psi
  \end{aligned}
\end{equation}
are the so-called light and heavy fields, respectively. Note that the projection operators are given by 
\begin{equation}
  \label{eq:28}
  P_v^\pm=\frac{1}{2}(1\pm\slashed{v})\punkt
\end{equation}
\\
In our analysis, we employ the HB Lagrangian only to calculate the loop-level amplitudes, whereas
the tree-level contributions are directly determined by an expansion of the covariant amplitudes in
inverse powers of the nucleon mass. Thus, for our purpose, we only need to consider the HB Lagrangian
up to next-to-leading order \cite{Fettes:2000gb}
\begin{equation}
  \label{eq0:98}
  \begin{aligned}
    \mathcal{\hat{L}}_{\pi N}^{(1)}&=\bar{N}\left[ \i v\cdot D+g
      S\cdot u \right]N\komma\\
    \mathcal{\hat{L}}_{\pi N}^{(2)}&= \bar{N}\left[{c}_1\Braket{\chi_+} +
    \frac{{c}_2}{2}\braket{(v\cdot
      u)^2} + \frac{{c}_3}{2}\Braket{u\cdot u}+\frac{{c}_4}{2} [S^\mu, S^\nu][u_\mu, u_\nu]\right] N\\
&+\frac{1}{2m}\bar N\Big[(v\cdot D)^2-D^2-\i g \{S\cdot D,v\cdot u\}
    -\frac{g^2}{8}\braket{(v\cdot u)^2} 
    +\frac{1}{4}[S^\mu,S^\nu][u_\mu,u_\nu]\Big]N  
\end{aligned}
 \end{equation}
with the Pauli-Lubanski spin vector
\begin{equation}
S_\mu=-\gamma_5\sigma_{\mu\nu}v^\nu=-\frac{1}{2}\gamma_5(\gamma_\mu
\slashed{v}-v_\mu)\komma
\end{equation}
which inherits the Dirac spin structure. This spin vector has the properties
\begin{equation}
  \label{eq0:102}
  S\cdot v= 0\komma\quad S^2=\frac{1-d}{4}\komma\quad 
  \{ S_\mu, S_\nu \}=\frac{1}{2}(v_\mu v_\nu-g_{\mu\nu})\komma\quad
  [S_\mu, S_\nu]=\i\epsilon_{\mu\nu\alpha\beta}v^\alpha S^\beta\komma
\end{equation}
with the space-time dimension $d$ and the Levi-Civita symbol
$\epsilon^{\mu\nu\alpha\beta}$, where $\epsilon^{0123}=-1$.

\subsection{Chiral Perturbation Theory for Pions, Nucleons,
 and ${\Delta}$(1232) Resonances}
\label{sec:chihbchipt-delta-reson}
The $\Delta(1232)$ resonance has four physical states $(\Delta^{++}, \Delta^+, \Delta^0, \Delta^-)$, 
which are conveniently collected in three isospin doublets
\begin{equation}
  \label{eq0:118}
  \Psi_\mu^1=\frac{1}{\sqrt{2}}
  \begin{pmatrix}
    \Delta^{++} -\frac{1}{\sqrt{3}}\Delta^0\\
    \frac{1}{\sqrt{3}}\Delta^0 -\Delta^-
  \end{pmatrix}_\mu
\komma\quad
  \Psi_\mu^2=\frac{\i}{\sqrt{2}}
  \begin{pmatrix}
    \Delta^{++} +\frac{1}{\sqrt{3}}\Delta^0\\
    \frac{1}{\sqrt{3}}\Delta^+ +\Delta^-
  \end{pmatrix}_\mu
\komma\quad
  \Psi_\mu^3=-\sqrt{\frac{2}{3}}
  \begin{pmatrix}
    \Delta^{+}\\
    \Delta^0
  \end{pmatrix}_\mu\komma
\end{equation}
where the field $\Psi_\mu^i$ is a Rarita-Schwinger-like isospurion, a spin-$3/2$
field constructed via coupling of a spin-$1$ to a spin-$1/2$ field.
Additionally, each spin component is an isodoublet with its own isovector index and
with the constraint
\begin{equation}
  \label{eq0:116}
  \tau^i \Psi^i_\mu=0
\end{equation}
to reduce the number of independent states from six to four. 
Note that to ensure the independence of unobservable off-shell parameters, the Lagrangian of the $\Delta$ resonance  
has to be invariant under the so-called point transformation \cite{Nath:1971wp}
\begin{equation}
  \label{eq0:204}
  \begin{aligned}
    \Psi_\mu &\to \Psi_\mu + a\gamma_\mu\gamma_\nu\Psi^\nu\komma\qquad
    A\to \frac{A-2a}{1+4a}\komma
  \end{aligned}
\end{equation}
where the unphysical gauge parameter $A$ appears in the most general leading-order Lagrangian.
A convenient choice for this parameter is $A=-1$, resulting in a more compact form for the $\Delta$ propagator.
However, in the original formulation of the HB Lagrangian \cite{Hemmert:1997ye} the $\Delta$ fields
are redefined to absorb the dependence on the parameter $A$. This is equivalent to the choice $A=0$.
We discuss the matching of coupling constants between these two choices at the end of this section, 
after introducing the HB Lagrangian. 
In the following, we employ the convenient choice $A=-1$. 
Thus, employing the above notation, the covariant pion-$\Delta$ Lagrangian up to fourth order reads \cite{Hemmert:1997ye,Hemmert:1997wz}
  \begin{align*}
  \mathcal{L}_{\pi\Delta}^{(1)} &=- \bar{\Psi}_i^\mu\Big[
    (\i\slashed{D}^{ij}-\mathbf{m} \delta^{ij} )g_{\mu\nu}-\i (\gamma_\mu
    D_\nu^{ij}+\gamma_\nu D_\nu^{ij})+\i \gamma_\mu
    \slashed{D}^{ij}\gamma_\nu+\mathbf{m} \delta^{ij}\gamma_\mu \gamma_\nu \\
    &+\frac{{\bf g_1}}{2}g_{\mu\nu} \slashed{u}^{ij} \gamma_5 +
\frac{{\bf g_2}}{2}(\gamma_\mu
u_\nu^{ij}+u_\mu^{ij}\gamma_\nu)\gamma_5+\frac{{\bf g_3}}{2}\gamma_\mu
\slashed{u}^{ij}\gamma_5 \gamma_\nu
    \Big]\Psi^\nu_j\komma\\
     \mathcal{L}_{\pi\Delta}^{(2)}&=\bar{\Psi}_\mu^i\Theta^{\mu\alpha}(y_1)\Big[
     c_1^\Delta\sigma_{\alpha\beta}\delta^{ij}\braket{\chi_+}
     +\frac{c_2^\Delta}{16\mathbf{m}^2}g_{\alpha\beta}\braket{u^\rho u^\lambda}D^{ij}_{\rho\lambda}
     -\frac{c_3^\Delta}{4}g_{\alpha\beta}\braket{u\cdot u}\delta^{ij}\\
     &+\frac{c_4^\Delta}{4}g_{\alpha\beta}[u_\rho,u_\lambda]\sigma^{\rho\lambda}\delta^{ij}
     -\frac{c_{11}^\Delta}{4}\braket{u^\alpha u^\beta}\delta^{ij}
     -\frac{c_{12}^\Delta}{2}(w^i_\alpha w^j_\beta+w^j_\alpha w^i_\beta)\\
     &-\frac{c_{13}^\Delta}{8\mathbf{m}^2}(w^i_\rho w^k_\lambda+w^i_\lambda w^k_\rho)D^{\rho\lambda}_{kj}
     \Big]\Theta^{\beta\nu}(y_1)\Psi_\nu^j
    +\mathrm{h.c.}  \komma\numberthis\label{eq0:L2}\\
    \mathcal{L}_{\pi \Delta}^{(3)}&=\bar{\Psi}_\mu^i\Theta^{\mu\alpha}
    (y_2)\Big[ 
    d_1^\Delta\i w_{\alpha\beta}^k\tau^k\gamma_5
    +d_2^\Delta \braket{\chi_- \tau^k} \tau^k\gamma_5 \delta^{ij}g_{\alpha\beta}
    +d_3^\Delta\braket{\chi_+} w_\lambda^k\tau^k\gamma_5
    \gamma^\lambda  g_{\alpha\beta}
    \\&+d_4^\Delta w_{\rho\tau \lambda}^kg^{\rho\tau}\tau^k\gamma_5
    \gamma^\lambda  g_{\alpha\beta}
    +d_5^\Delta w_{\alpha\beta\lambda}^k\tau^k\gamma_5
    \gamma^\lambda  
    \Big]\delta^{ij} \Theta^{\beta\nu}
    (y_2)\Psi_\nu^j\komma\\
    \mathcal{L}_{\pi \Delta}^{(4)}&=\bar{\Psi}_\mu^i\Theta^{\mu\alpha}
    (y_3)\Big[ e^\Delta_{38}\braket{\chi_+}\braket{\chi_+}
    +\frac{e^\Delta_{115}}{4}\braket{\chi_+^2-\chi_-^2}\\
    &-\frac{e^\Delta_{116}}{4}(\braket{\chi_-^2}-\braket{\chi_-}^2+\braket{\chi_+^2}-\braket{\chi_+})^2  
    \Big]g_{\alpha\beta}\delta^{ij} \Theta^{\beta\nu}
    (y_3)\Psi_\nu^j\komma
  \end{align*}
with $\Theta^{\mu\alpha}(z)=g^{\mu\alpha}+z\gamma^\mu\gamma^\alpha$
and the off-shell parameters $y_i$, ${\bf g_2}$ and ${\bf g_3}$.
 Additionally, we used the covariant derivative
\begin{equation}
  \label{eq0:114}
  D^{ij}_\mu=\partial_\mu \delta^{ij}+\Gamma_\mu^{ij}
\end{equation}
with the chiral connection
\begin{equation}
  \label{eq0:115}
  \Gamma_\mu^{ij}=\Gamma_\mu\delta^{ij} - \i \epsilon^{ijk} \braket{\tau^k\Gamma_\mu}
\end{equation}
and the following set of definitions
\begin{equation}
  \label{eq0:199}
  \begin{aligned}
    w_\alpha^i &= \frac{1}{2}\mathrm{Tr}\left[ \tau^i u_\alpha
    \right]\komma\quad
    w_{\alpha\beta}^i &=\frac{1}{2}\mathrm{Tr}\left[
      \tau^i\left[D_\alpha,u_\beta\right] \right]
\komma\quad
    w_{\alpha\beta\tau}^i &=\frac{1}{2}\mathrm{Tr}\left[ \tau^i\left[D_\alpha,\left[D_\beta,u_\tau\right]\right] \right]\punkt
  \end{aligned}
\end{equation}
The $\Delta$ propagator constrained by the leading-order Lagrangian $\mathcal{L}_{\pi\Delta}^{(1)}$
is given by \cite{Bernard:2003xf}
\begin{equation}
  \label{eq0:169}
  \begin{aligned}
    \mathcal{G}^{\mu\nu}_{ij ,\Delta} (p)&=
    -\frac{\slashed{p}+{\bf m}}{p^2-{\bf m}^2} \left(
      g^{\mu\nu}-\frac{1}{d-1}\gamma^\mu \gamma^\nu
      +\frac{1}{d-1}\frac{p^\mu \gamma^\nu-p^\nu \gamma^\mu}{{\bf m}}
      -\frac{d-2}{d-1}\frac{p^\mu p^\nu}{{\bf m}^2}\right) \xi^{ij}_{3/2}
\\&=-\frac{\slashed{p}+{\bf m}}{p^2-{\bf m}^2}
\left(P^{3/2}_{33}\right)^{\mu\nu}\xi^{ij}_{3/2}
-\frac{1}{\sqrt{d-1}{\bf m}}\left(
  \left(P^{1/2}_{12}\right)^{\mu\nu}+
  \left(P^{1/2}_{21}\right)^{\mu\nu} \right) \xi^{ij}_{3/2}
\\&+\frac{d-2}{d-1}\frac{\slashed{p}+{\bf m}}{{\bf m}^2}\left(P^{1/2}_{22}\right)^{\mu\nu}\xi^{ij}_{3/2}
  \end{aligned}
\end{equation}
with the spin projection operators
\begin{equation}
  \label{eq0:177}
  \begin{aligned}
    \left(P^{3/2}_{33}\right)_{\mu\nu}&=g_{\mu\nu}-\frac{1}{d-1}\gamma_\mu
    \gamma_\nu-\frac{1}{(d-1)p^2}(\slashed{p}\gamma_\mu p_\nu+p_\mu
    \gamma_\nu \slashed{p})-\frac{d-4}{d-1}\frac{p_\mu p_\nu}{p^2}\komma\\
    \left(P^{1/2}_{12}\right)_{\mu\nu}&=\frac{1}{\sqrt{d-1}p^2}(p_\mu
    p_\nu -\slashed{p}p_\nu \gamma_\mu) \komma\\
    \left(P^{1/2}_{21}\right)_{\mu\nu}&=\frac{1}{\sqrt{d-1}p^2}(\slashed{p}p_\mu \gamma_\nu-p_\mu
    p_\nu)\komma\\
    \left(P^{1/2}_{22}\right)_{\mu\nu}&=\frac{p_\mu p_\nu}{p^2}\komma\\
      \left(P^{1/2}_{11}\right)_{\mu\nu}&=\frac{1}{d-1}\gamma_\mu
    \gamma_\nu+\frac{1}{(d-1)p^2}(\slashed{p}\gamma_\mu p_\nu+p_\mu
    \gamma_\nu \slashed{p})-\frac{3}{d-1}\frac{p_\mu p_\nu}{p^2}\komma
  \end{aligned}
\end{equation}
which obey
\begin{equation}
  \label{eq0:117}
  \begin{aligned}
    \left(P^{3/2}_{33}\right)_{\mu\nu} +
    \left(P^{1/2}_{11}\right)_{\mu\nu} +
    \left(P^{1/2}_{22}\right)_{\mu\nu}
    = g_{\mu\nu}\komma\qquad
    \left(P^I_{ij}\right)_{\mu\nu}\left(P^J_{kl}\right)^{\nu\rho}=
    \delta^{IJ}\delta_{jk}\tensor{\left(P^I_{il}\right)}{_\mu^\rho}
  \end{aligned}
\end{equation}
and the isospin projection operators
\begin{equation}
  \label{eq0:122}
  \begin{aligned}
    \xi_{ij}^{3/2}&=\frac{2}{3}\delta_{ij}-\frac{\i}{3}\epsilon_{ijk}\tau^k\komma\qquad
    \xi_{ij}^{1/2}=\frac{1}{3}\delta_{ij}+\frac{\i}{3}\epsilon_{ijk}\tau^k\komma
  \end{aligned}
\end{equation}
which obey
\begin{equation}
  \label{eq0:121}
  \xi_{ij}^{3/2}+ \xi_{ij}^{1/2}= \delta_{ij}\komma\qquad \xi_{ij}^{I}\xi_{jk}^{J}=\delta^{IJ}\xi_{ik}^{J}\punkt
\end{equation}
The covariant pion-nucleon-$\Delta$ Lagrangian up to third order reads \cite{Hemmert:1997wz,ZoellerMaster}
  \begin{align*}
    \mathcal{L}_{\pi N \Delta}^{(1)}&= h\,\bar{\Psi}_\mu^i\Theta^{\mu\alpha}(z_0)w_\alpha^i\Psi
    +\mathrm{h.c.}\komma\\
    \mathcal{L}_{\pi N \Delta}^{(2)}&= \bar{\Psi}_\mu^i \Theta^{\mu
      \alpha}(z_1)\Big[ \i b_3 w_{\alpha\beta}^i\gamma^\beta +
      \frac{b_4}{2}w_\alpha^i w_\beta^j \gamma^\beta\gamma_5\tau^j +
      \frac{b_5}{2}w_\alpha^j w_\beta^i \gamma^\beta\gamma_5\tau^j-
      \frac{b_6}{m}  w_{\alpha\beta}^i  D^\beta \Big]\Psi
    +\mathrm{h.c.}\\
    \mathcal{L}_{\pi N \Delta}^{(3)}&=\bar{\Psi}_\mu^i\Theta^{\mu\nu}(z_2) \Big[
    \frac{h_4}{2}w_\nu^i\braket{\chi_+}+h_7 \i \braket{[D_\nu,\chi_-]\tau^i}
    +h_{58}\i\braket{[D_\alpha,u_\nu]\tau^i}w_\beta^j\tau^j\sigma^{\alpha\beta}\gamma_5
    \\&-\frac{h_{59}}{2m}\braket{[D_\alpha,u_\nu]\tau^i}w_\beta^j\tau^j\gamma^{\beta}\gamma_5D^\alpha
    +h_{60}\i\braket{[D_\alpha,u_\nu]\tau^j}w_\beta^i\tau^j\sigma^{\alpha\beta}\gamma_5    
    \numberthis\label{eq0:165}\\&-\frac{h_{61}}{2m}\braket{[D_\alpha,u_\nu]\tau^j}w_\beta^i\tau^j\gamma^{\alpha}\gamma_5D^\beta    
    -\frac{h_{62}}{2m}\braket{[D_\alpha,u_\beta]\tau^j}w_\nu^i\tau^j\gamma^{\alpha}\gamma_5D^\beta      
    \\&-\frac{h_{63}}{2m}\braket{[D_\alpha,u_\nu]\tau^j}w_\beta^i\tau^j\gamma^{\beta}\gamma_5D^\alpha      
    -\frac{h_{64}}{2m^2}\braket{[D_\nu,[D_\alpha,u_\beta]]\tau^i}D^{\alpha\beta}\\
    &-\frac{h_{65}}{2m^2}\braket{[D_\alpha,[D_\beta,u_\nu]]\tau^i}D^{\alpha\beta}
    \Big]\Psi+\mathrm{h.c.}\punkt
  \end{align*}
At this point we emphasize that the terms contributing to $\mathcal{L}_{\pi \Delta}^{(3)}$ and
$\mathcal{L}_{\pi \Delta}^{(4)}$ in Eq.~\eqref{eq0:L2} do not appear in Refs.~\cite{Hemmert:1997ye,Hemmert:1997wz},
but instead were constructed for our purpose based on the rules provided in these papers.
Also, the full Lagrangian $ \mathcal{L}_{\pi N \Delta}^{(3)}$ is constructed in the unpublished work \cite{ZoellerMaster} and only
the terms of interest are shown above. Furthermore, the Lagrangian $ \mathcal{L}_{\pi N \Delta}^{(4)}$ is omitted, because there
are no terms or only redundant ones which contribute to elastic pion-nucleon scattering.
For the sake of brevity, we refrain from showing the construction and verification of these terms in detail. 
However, the interested reader can check that these terms fulfill all the necessary symmetry and power counting requirements. 
Additionally, we would like to preempt, that these terms turn out to be redundant in the final pion-nucleon amplitude.
Next, we turn to the dependence on the off-shell parameters $z_i$, $y_i$ in Eqs.~\eqref{eq0:L2} and \eqref{eq0:165}.
In these equations, we constrained the off-shell parameters to be equal for all LECs at a given chiral order.
As shown in Ref.~\cite{Krebs:2009bf}, it is possible to redefine the LECs such that the dependence of the physical amplitude
on off-shell parameters is cancelled. Thus, we can ignore these parameters from the beginning.
However, it is advantageous to keep them in our Lagrangian such that we have a further check on our calculations
by demanding later on that the physical amplitudes should be free of off-shell parameters.
\\\\
In the HB approach, a light spin-$3/2$ and isospin-$3/2$ field is defined via
\begin{align}
  \label{eq0:120}
  T^\mu_i &= P^+_v \xi^{3/2}_{ij}\left(\hat P^{3/2}_{33}\right)^{\mu\nu}\Psi_\nu^j \,e^{\i m
    v\cdot x}
\end{align}
with the projection operator
\begin{equation}
  \label{eq0:119}
  \begin{aligned}
    \left(\hat P^{3/2}_{33}\right)_{\mu\nu} &= g_{\mu\nu} -
    v_\mu v_\nu - \frac{4}{1-d} S_\mu S_\nu\komma
  \end{aligned}
\end{equation}
whereas the other projections are treated as heavy contributions.
As mentioned before, we need to consider the HB Lagrangian only up to next-to-leading order explicitly.
The HB pion-$\Delta$ Lagrangian up to next-to-leading order is given by \cite{Hemmert:1997ye}
  \begin{align*}
    \mathcal{\hat{L}}_{\pi\Delta}^{(1)}&= -\bar{T}_i^\mu\Big[ \i
      v\cdot D^{ij} - \Delta_0 \delta^{ij} + {\bf g_1} S\cdot u^{ij} \Big]
    g_{\mu\nu} T_j^\nu\komma \\
    \mathcal{\hat{L}}_{\pi\Delta}^{(2)}&=-\bar{T}_i^\mu\Big[
c^\Delta_1\Braket{\chi_+} g_{\mu\nu}\delta^{ij}+
    \frac{{c}^\Delta_2}{4}\braket{(v\cdot
      u)^2} g_{\mu\nu}\delta^{ij}+ \frac{{c}^\Delta_3}{2}\Braket{u\cdot u}g_{\mu\nu}\delta^{ij}
+\frac{{c}^\Delta_4}{2} [S^\alpha, S^\beta][u_\alpha, u_\beta]g_{\mu\nu}\delta^{ij}
\\&+\frac{c_{11}^\Delta}{2}\braket{u^\mu u^\nu}\delta^{ij}
     +c_{12}^\Delta(w^i_\mu w^j_\nu+w^j_\mu w^i_\nu)
     -\frac{c_{13}^\Delta}{2}(w^i_\rho w^k_\lambda+w^i_\lambda w^k_\rho)v^{\rho}v^{\lambda}\delta^{kj}g_{\mu\nu}
\Big]T_j^\nu \numberthis\label{eq0:113}\\
&+\frac{1}{2m}\bar{T}^i_\mu \Big[
      [D_\alpha^{ik}D_\beta^{kj}g^{\alpha\beta}-v\cdot D^{ik}v\cdot D^{kj}]g^{\mu\nu}
      +{\bf g_1} \i (S\cdot D^{ik}v\cdot u^{kj}+v\cdot u^{ik}S\cdot D^{kj}) g^{\mu\nu}\\
      &- [S^\alpha,S^\beta](D_\alpha^{ik}D_\beta^{kj}-D_\beta^{ik}D_\alpha^{kj})g^{\mu\nu}
      + \frac{{\bf g_1}^2}{4}v\cdot u^{ik}v\cdot u^{kj}g^{\mu\nu}
    \Big] T^j_\nu\komma
  \end{align*}
where $u_\mu^{ij}=\xi_{3/2}^{il}\, u_\mu\, \xi_{3/2}^{lj}$. 
From the leading-order Lagrangian $\mathcal{\hat{L}}_{\pi\Delta}^{(1)}$ one can deduce the HB $\Delta$ propagator,
which reads
\begin{equation}
  \label{eq0:134}
   \mathcal{\hat{G}}_{ij ,\Delta}^{\mu\nu} (p)=\frac{- 1}{v\cdot
     k-\Delta_0}\left(\hat P^{3/2}_{33}\right)^{\mu\nu}\xi^{ij}_{3/2}\punkt
\end{equation}
The HB pion-nucleon-$\Delta$ Lagrangian up to next-to-leading order reads  \cite{Hemmert:1997ye}
\begin{equation}
  \label{eq0:126}
  \begin{aligned}
    \mathcal{\hat{L}}_{\pi N \Delta}^{(1)}&=h
      \bar{T}_i^\mu \,w_\mu^i\, N + \mathrm{h.c.}
    \komma\\
    \mathcal{\hat{L}}_{\pi N\Delta}^{(2)}&=\bar{T}_i^\mu \left[ (b_3+b_6)\, \i w_{\mu\nu}^i v^\nu + b_4\, w_\mu^i S\cdot u
  + b_5\, u_\mu S\cdot w^i
\right] N + \textrm{h.c.}\\
&-\frac{1}{2m}\bar{T^i_\mu}\Big[ 
 \frac{2}{d-1}h {\bf g_1} z_0 u_{ij}^\mu\xi^{jk}S\cdot w^k
+2 h  \i D^\mu_{ij} \xi^{jk} v\cdot w^k
\Big] N + \mathrm{h.c.}\punkt
  \end{aligned}
\end{equation}
Finally, one has additional contributions to the pion-nucleon Lagrangian
at next-to-leading order \cite{Hemmert:1997ye}
\begin{equation}
  \label{eq0:139}
  \begin{aligned}
\mathcal{\hat L}_{\pi N}^{(2)}&=
    -\frac{h^2}{2m}\bar N\Big[\frac{4}{d-1}(2z_0+(d-1)z_0^2)
    S\cdot w^i \xi^{ij} S\cdot w^j\\&+\frac{1}{d-1}(4(d-2)+2(d-3)z_0-z_0^2)
    v\cdot w^i \xi^{ij}v\cdot w^j \Big]N\punkt
  \end{aligned}
\end{equation}
Note that we restored the explicit $d$-dependence in the Lagrangians in Eqs.~\eqref{eq0:126} and \eqref{eq0:139}. 
In Ref.~\cite{Hemmert:1997ye} these terms are only given for the case $d=4$, which is sufficient at tree-level. 
However, for our loop-level analysis, where we employ dimensional regularization, the explicit $d$-dependence is necessary.
In particular, we restored it by matching covariant and HB amplitudes
for the reactions $\pi N\to\pi N$ and $\pi N\to\pi\pi N$ at tree-level and for the $\Delta$ mass up to loop-level.
We already mentioned that the Lagrangian in Ref.~\cite{Hemmert:1997ye} is effectively constructed with
the choice $A=0$ for the unphysical off-shell parameter, however, we employ $A=-1$ in the above Lagrangians.
The Lagrangian in Ref.~\cite{Hemmert:1997ye} can be matched tp the one presented here by applying the following relations
\begin{equation}
  \label{eq0:19}
  \hat z_0 = -\frac{1}{2}(1+z_0)\komma \quad {\bf \hat  g_2}=-{\bf g_1}\quad {\bf \hat  g_3}=-{\bf g_1}\komma
\end{equation}
where the parameters in Ref.~\cite{Hemmert:1997ye} are denoted by $\hat z_0$, ${\bf \hat g_2}$ and $ {\bf \hat g_3}$.

\section{Redundancy Shifts for the Pion-Nucleon Scattering Amplitude}
\label{cha:RedShiftsAmpl}
Treating the $\Delta$ resonance as an explicit degree of freedom introduces several new LECs and off-shell parameters
in the pion-nucleon scattering amplitude. Fortunately, the majority of these LECs and all of the off-shell parameters
turn out to be redundant. In the following we show the necessary redefinitions of LECs to get rid of the 
redundant dependence on the LECs $b_3$, $b_6$, $h_i$, $c_i^\Delta$ and the off-shell parameter $z_0$.
For the sake of brevity, we only show the required shifts in the HB framework and we also suppress the
off-shell parameters $y_i$ and $z_i$ for $i\in {1,2,3}$. 
Needless to say, we extracted the shifts for the covariant framework as well and 
including the off-shell parameters $y_i$ and $z_i$. The issue with the covariant approach is that
even after considering further powers of $1/m_N$ in the shifts of the LECs, the amplitude will still
depend on the off-shell parameters and the above mentioned redundant LECs at higher chiral orders. 
These terms can only be cancelled by higher-order LECs.
Thus, in the covariant framework we set $b_i=h_i=c_i^\Delta=y_i=z_i=0$ and demonstrate in the following
the independence on these redundant terms only in a strict chiral counting.
\\\\
At tree-level, the contributions proportional to $b_3$, $b_6$ and $h_i$ are cancelled by
\begin{align*}
  h&\to h -(b_{3}+b_{6}) \Delta +M_\pi^2 \left(-2 h_{4}+4
      h_{7}+\frac{b_{6}}{2 m_N}\right)\\&+\left(2
      (h_{64}+h_{65})-\frac{b_{6}}{2 m_N}\right) \Delta ^2+\frac{(4
      b_{6} c_1-2 (h_{64}+h_{65})) M_\pi^2 \Delta }{m_N}\\&+\frac{2
      (h_{64}+h_{65}) \Delta ^3}{m_N}\komma \\ 
  c_1&\to c_1\komma\\
    c_2&\to c_2+\frac{8}{9} (h_A+\delta h_A^{(3)}) (b_{3}+b_{6})-\frac{4}{9} \left((b_{3}+b_{6})^2+4 h_A (h_{64}+h_{65})\right) \Delta \\&-\frac{4 h_A (4 b_{3}+3 b_{6}) \Delta }{9 m_N}+\frac{16}{9} (b_{3}+b_{6}) (h_{64}+h_{65}) \Delta ^2+\frac{8 h_A (3 b_{3}+2 b_{6}) \Delta ^2}{9 m_N^2}\\&+\frac{4 ((b_{3}+b_{6}) (2 b_{3}+b_{6})+4 h_A (h_{64}+h_{65})) \Delta ^2}{9 m_N}\komma\\
    c_3&\to c_3-\frac{8}{9} (h_A+\delta h_A^{(3)}) (b_{3}+b_{6})+\frac{4}{9} \left((b_{3}+b_{6})^2+4 h_A (h_{64}+h_{65})\right) \Delta -\frac{4 h_A b_{6} \Delta }{9 m_N}\\&-\frac{16}{9} (b_{3}+b_{6}) (h_{64}+h_{65}) \Delta ^2+\frac{4 (b_{6} (b_{3}+b_{6})+4 h_A (h_{64}+h_{65})) \Delta ^2}{9 m_N}\komma\\
    c_4&\to c_4+\frac{4}{9} (h_A+\delta h_A^{(3)}) (b_{3}+b_{6})-\frac{2}{9} \left((b_{3}+b_{6})^2+4 h_A (h_{64}+h_{65})\right) \Delta +\frac{2 h_A b_{6} \Delta }{9 m_N}\\&+\frac{8}{9} (b_{3}+b_{6}) (h_{64}+h_{65}) \Delta ^2-\frac{2 (b_{6} (b_{3}+b_{6})+4 h_A (h_{64}+h_{65})) \Delta ^2}{9 m_N}\komma\\
    d_1+d_2&\to d_1+d_2+\frac{1}{9} \left((b_{3}+b_{6})^2-4 h_A (h_{64}+h_{65})\right)-\frac{2 h_A b_{3}}{9 m_N}+\frac{h_A (8 b_{3}+5 b_{6}) \Delta }{18 m_N^2}\\&+\frac{((b_{3}+b_{6}) (3 b_{3}+b_{6})-4 h_A (h_{64}+h_{65})) \Delta }{18 m_N}\komma\\
    d_3&\to d_3+\frac{1}{9} \left(-(b_{3}+b_{6})^2+4 h_A (h_{64}+h_{65})\right)-\frac{h_A b_{6}}{9 m_N}+\frac{2 h_A b_{6} \Delta }{9 m_N^2}\\&+\frac{2 \left((b_{3}+b_{6})^2-2 h_A (h_{64}+h_{65})\right) \Delta }{9 m_N}\komma\\
    d_5&\to d_5-\frac{h_A (2 b_{3}+3 b_{6})}{18 m_N}+\frac{h_A (8 b_{3}+7 b_{6}) \Delta }{36 m_N^2}\\&+\frac{((b_{3}+b_{6}) (b_{3}+3 b_{6})+12 h_A (h_{64}+h_{65})) \Delta }{36 m_N}\komma\\
    d_{14}-d_{15}&\to d_{14}-d_{15}+\frac{1}{9} \left(-2 (b_{3}+b_{6})^2+8 h_A (h_{64}+h_{65})\right)+\frac{2 h_A b_{6}}{9 m_N}\\&-\frac{2 h_A b_{6} \Delta }{9 m_N^2}+\frac{2 (b_{3}-b_{6}) (b_{3}+b_{6}) \Delta }{9 m_N}\komma\\
    e_{14}&\to e_{14}+\frac{h_A (4 b_{3}+3 b_{6})}{72 m_N^2}+\frac{b_{3}^2-4 b_{3} b_{6}-2 b_{6}^2+8 h_A (h_{64}+h_{65})}{72 m_N}\komma\\
    e_{15}&\to e_{15}+\frac{2}{9} (b_{3}+b_{6}) (h_{64}+h_{65})+\frac{h_A b_{6}}{18 m_N^2}\\&+\frac{2 b_{3}^2+6 b_{3} b_{6}+b_{6}^2-4 h_A (h_{64}+h_{65})}{36 m_N}\komma\\
    e_{16}&\to e_{16}-\frac{2}{9} (b_{3}+b_{6}) (h_{64}+h_{65})+\frac{b_{6}^2-4 h_A (h_{64}+h_{65})}{36 m_N}\komma\\
    e_{17}&\to e_{17}+\frac{h_A b_{6}}{36 m_N^2}+\frac{4 b_{3}^2+2 b_{3} b_{6}+b_{6}^2-4 h_A (h_{64}+h_{65})}{72 m_N}\komma\\
    e_{18}&\to e_{18}-\frac{1}{9} (b_{3}+b_{6}) (h_{64}+h_{65})-\frac{h_A b_{6}}{36 m_N^2}\\&+\frac{-2 b_{3}^2-b_{6}^2+4 h_A (h_{64}+h_{65})}{72 m_N}\komma\\    
2e_{19}-e_{22}-e_{36}&\to 2e_{19}-e_{22}-e_{36}-\frac{4 h_A (h_{64}+h_{65})}{9 m_N}\komma\\
e_{20}+e_{35}&\to e_{20}+e_{35}+\frac{h_A b_{6}}{18 m_N^2}+\frac{2 b_{3}^2+4 b_{3} b_{6}+3 b_{6}^2-4 h_A (h_{64}+h_{65})}{36 m_N}\komma\\
2e_{21}-e_{37}&\to 2e_{21}-e_{37}-\frac{h_A b_{6}}{18 m_N^2}+\frac{2 h_A (h_{64}+h_{65})}{9 m_N}\komma\\
e_{22}-4e_{38}&\to e_{22}-4e_{38}+\frac{h_A (4 b_{3}+5 b_{6})}{36 m_N^2}-\frac{b_{3}^2}{36 m_N}
  \end{align*}
and terms proportional to $z_0$ are cancelled by
\begin{align*}
     c_1&\to c_1\komma\\
    c_2&\to c_2+\frac{2(h_A^2+2h_A\delta h_A^{(3)})  z_0 (2+z_0)}{9 m_N}-\frac{4 h_A^2 z_0 (2+z_0) \Delta }{9 m_N^2}+\frac{2 h_A^2 z_0 (2+z_0) \Delta ^2}{3 m_N^3}\komma\\
    c_3&\to c_3-\frac{(h_A^2+2h_A\delta h_A^{(3)}) z_0 (2+3 z_0)}{9 m_N}+\frac{2 h_A^2 z_0 (1+2 z_0) \Delta }{9 m_N^2}-\frac{h_A^2 z_0 (2+5 z_0) \Delta ^2}{9 m_N^3}\komma\\
    c_4&\to c_4-\frac{(h_A^2+2h_A\delta h_A^{(3)}) z_0 (2+3 z_0)}{9 m_N}+\frac{2 h_A^2 z_0 (1+2 z_0) \Delta }{9 m_N^2}-\frac{h_A^2 z_0 (2+5 z_0) \Delta ^2}{9 m_N^3}\komma\\
    d_1+d_2&\to d_1+d_2-\frac{h_A^2 z_0 (2+z_0)}{36 m_N^2}+\frac{h_A^2 z_0 (2+z_0) \Delta }{18 m_N^3}\komma\\
    d_3&\to d_3\komma\\
    d_5&\to d_5-\frac{h_A^2 z_0}{36 m_N^2}+\frac{h_A^2 z_0 \Delta }{18 m_N^3}\komma\\
    d_{14}-d_{15}&\to  d_{14}-d_{15}+\frac{h_A^2 z_0 (2+z_0)}{9 m_N^2}-\frac{2 h_A^2 z_0 (2+z_0) \Delta }{9 m_N^3}\komma\\   
      e_{14}&\to e_{14}+\frac{h_A^2 z_0 (2+z_0)}{144 m_N^3}\komma\\
    e_{15}&\to e_{15}\komma\\
    e_{16}&\to e_{16}\komma\\
    e_{17}&\to e_{17}-\frac{h_A^2 z_0 (2+z_0)}{144 m_N^3}\komma\\
    e_{18}&\to e_{18}\komma\\    
2e_{19}-e_{22}-e_{36}&\to 2e_{19}-e_{22}-e_{36} -\frac{h_A^2 z_0 (2+z_0)}{36 m_N^3}\komma\\
e_{20}+e_{35}&\to e_{20}+e_{35}\komma\\
2e_{21}-e_{37}&\to 2e_{21}-e_{37}-\frac{h_A^2 z_0}{36 m_N^3}\komma\\
e_{22}-4e_{38}&\to e_{22}-4e_{38}+\frac{h_A^2 z_0 (2+z_0)}{36 m_N^3}\punkt
\end{align*}
Note that we set $d=4$ in the tree-level contributions above and that the terms
proportional to the redundant LECs from $\mathcal{L}_{\pi N\Delta}^{(4)}$ are suppressed.
\\\\
At loop-level, the contributions proportional to the LECs
$b_3$, $b_6$, $c_{11}^\Delta$, $c_{12}^\Delta$ and $c_{13}^\Delta$ are cancelled by
  \begin{align*}
    c_1&\to c_1\komma\\
    c_2&\to c_2+\frac{4(d-2)h_A(b_3+b_6)}{3(d-1)}\komma\\
    c_3&\to c_3-\frac{4(d-2)h_A(b_3+b_6)}{3(d-1)}\komma\\
    c_4&\to c_4+\frac{4h_A(b_3+b_6)}{3(d-1)}\komma\\
    b_4&\to b_4+\frac{((1+3d)g_1-3(d-1)g_A)(b_3+b_6)}{3(d-1)}\komma\\
    b_5&\to b_5-\frac{2(d+2)g_1(b_3+b_6)}{3(d-1)}\komma\\
    c^\Delta_1&\to c^\Delta_1\komma\\
    c^\Delta_2&\to c^\Delta_2 +\frac{-4 h_A (b_{3}+b_{6})+6 c^\Delta_{11}+4 c^\Delta_{12}+2 (d-1) c^\Delta_{13}}{3 (d-1)}\komma\\
    c^\Delta_3&\to c^\Delta_3 +\frac{2 h_A (b_{3}+b_{6})-3 c^\Delta_{11}-2 c^\Delta_{12}}{3 (d-1)}\komma\\
    c_4^\Delta&\to c_4^\Delta -\frac{2 (-7+3 d) h_A (b_{3}+b_{6})}{22+d (-3+(-4+d) d)}
  \end{align*}
and the off-shell parameter $z_0$ contributions are cancelled by
  \begin{align*}
    c_1&\to c_1\komma\\
    c_2&\to c_2+\frac{(d-2)h_A^2z_0(2+z_0)}{3(d-1)m_N}\komma\\
    c_3&\to c_3-\frac{h_A^2z_0(2+(d-1)z_0)}{3(d-1)m_N}\komma\\
    c_4&\to c_4-\frac{h_A^2z_0(2+(d-1)z_0)}{3(d-1)m_N}\komma\\
    b_4&\to b_4-\frac{2g_1 h_A z_0}{3(d-1)m_N}\komma\\
    b_5&\to b_5+\frac{g_1h_Az_0}{(d-1)m_N}\punkt
  \end{align*}
Finally, terms proportional to the LECs $c_i^\Delta$ are cancelled by performing
additional shifts for the LECs $c_i$ in the tree-level diagrams
  \begin{align*}
    c_1&\to c_1+\frac{2 (d-2) h_A^2 c^\Delta_{1} \left(A_0\left(M_\pi^2\right)+\Delta  J_0(-\Delta )\right)}{F_\pi^2}\komma\\
    c_2&\to c_2+\frac{(d-2) h_A^2 c^\Delta_{2} \left(A_0\left(M_\pi^2\right)+\Delta  J_0(-\Delta )\right)}{F_\pi^2}\komma\\
    c_3&\to c_3+\frac{2 (d-2) h_A^2 c^\Delta_{3} \left(A_0\left(M_\pi^2\right)+\Delta  J_0(-\Delta )\right)}{F_\pi^2}\komma\\
    c_4&\to c_4+\frac{10 \left(22-3 d-4 d^2+d^3\right) h_A^2 c^\Delta_{4} \left(A_0\left(M_\pi^2\right)+\Delta  J_0(-\Delta )\right)}{9 (d-1)^2 F_\pi^2}\punkt
  \end{align*}
Note that the shifts given above cancel the dependence on $z_0$ in the pion-nucleon scattering amplitude,
however, the renormalization of leading-order couplings is necessary to cancel the terms proportional to $b_3+b_6$ and
$c_i^\Delta$ at loop-level. In particular, the renormalization of the
coupling constant $h_A$ and the $\Delta$ mass $m_\Delta$ discussed in
Appendix~\ref{sec:renormalizationrules} are needed.
We have to emphasize that the LECs $b_3$ and $b_6$ 
should be redundant for every process involving a $\pi \bar N\Delta$-vertex.
However, the redundancy of the LECs $c_i^\Delta$ seems to be accidental
and it is not clear to us why it should hold for any reaction other than $\pi N\to\pi N$.
\\\\
In addition, we make use of the following linear combinations of LECs at tree-level
  \begin{align*}
    \bar c_1&\to \bar c_1 + 2M_\pi^2(\bar e_{22}-4 \bar e_{38}+ \bar c_1 \beta_{l_3}\bar l_3/(32\pi^2F_\pi^2))\komma\\
    \bar c_2&\to \bar c_2 - 8M_\pi^2(\bar e_{20}+\bar e_{35})\komma\\
    \bar c_3&\to \bar c_3 - 4M_\pi^2(2 \bar e_{19}-\bar e_{22}-\bar e_{36})\komma\\
    \bar c_4&\to \bar c_4 -4M_\pi^2 (2\bar e_{21}-\bar e_{37})\punkt
  \end{align*}
Note that these linear combinations refer to the renormalized LECs, which is denoted by the bar notation. 
The renormalization of the bare LECs is discussed in Appendix~\ref{sec:renormalizationrules}.

\section{Renormalization Rules}
\label{sec:renormalizationrules}

In this appendix, we list the renormalization rules for the leading-order coupling constants,
masses and higher-order LECs. Note that we employ the following integral notation
\begin{align*}
    A_0(m_0^2)&=\frac{1}{\i}\int \frac{\di^d l}{(2\pi)^d} \,
    \frac{1}{l^2-m_0^2}\komma\nonumber\\
    B_0(p^2,m_0^2,m_1^2)&=\frac{1}{\i}\int \frac{\di^d
      l}{(2\pi)^d} \, \frac{1}{(l^2-m_0^2)((l+p)^2-m_1^2)}\komma\\
    J_0(\omega)&=\frac{1}{\i}\int \frac{\di^d
      l}{(2\pi)^d} \, \frac{1}{(l^2-M_\pi^2)(\omega+v\cdot l)}\komma\nonumber\\
    C_0(p_1^2,(p_1-p_2)^2,p_2^2,m_0^2,m_1^2,m_2^2)&=\frac{1}{\i}\int \frac{\di^d
      l}{(2\pi)^d} \, \frac{1}{(l^2-m_0^2)((l+p_1)^2-m_1^2)((l+p_2)^2-m_2^2)}\nonumber
\end{align*}
where the $+\i\varepsilon$ prescription was suppressed. In particular, we only show the
HB expressions in the following sections, whereas the covariant expressions (as well as the HB ones) 
can be found in the supplementary material.

\subsection{Mesonic Sector}
In the meson sector, the renormalization rules for the pion mass, Z-factor and decay constant are given by
\begin{align*}
  M^2 &= M^2_\pi + \delta M^{(4)}\komma\\
  \delta M^{(4)} &=-\frac{2 l_3 M_\pi^4}{F_\pi^2}+\frac{M_\pi^2 A_0(M_\pi^2) }{2 F_\pi^2}\komma\\
    Z_\pi &= 1+ \delta Z_\pi^{(4)}\komma\\
    \delta Z_\pi^{(4)} &= -\frac{2 l_4 M_\pi^2}{F_\pi^2}-\frac{(-1+10 \alpha) A_0(M_\pi^2) }{F_\pi^2}\komma\\
 F &= F_\pi+ \delta F_\pi^{(4)}\komma\\
    \delta F_\pi^{(4)} &=-\frac{l_4 M_\pi^2}{F_\pi}-\frac{A_0(M_\pi^2)
    }{F_\pi}\punkt
  \end{align*}

\subsection{Heavy Baryon Chiral Perturbation Theory}
\label{sec:heavy-baryon}

\subsubsection{Renormalization of Masses and Couplings}
\label{sec:renorm-rulesHB}
The HB expression for the nucleon mass reads
\begin{align*}
  m &= m_N + \delta m^{(2)}+ \delta m^{(3)}+ \delta m^{(3,\Delta)}+ \delta m^{(4,\Delta)} \komma\\
 \delta m^{(2)} &=4 c_1 M_\pi^2 \komma\\
 \delta m^{(3)}  &= -\frac{3 g_A^2 M_\pi^2 J_0(0)}{4 F_\pi^2}\\
  \delta m^{(3,\Delta)} &=-\frac{h_A^2 M_\pi^2 \Delta }{12 F_\pi^2 \pi ^2}+\frac{h_A^2 \Delta ^3}{18 F_\pi^2 \pi ^2}+\frac{4 h_A^2 \Delta  A_0\left(M_\pi^2\right)}{3 F_\pi^2}+\left(-\frac{4 h_A^2 M_\pi^2}{3 F_\pi^2}+\frac{4 h_A^2 \Delta ^2}{3 F_\pi^2}\right) J_0(-\Delta ) \komma\\
 \delta m^{(4)}  &=M_\pi^4 \left(2 e_{115}+2 e_{116}+16 e_{38}-\frac{8 c_1 l_3}{F_\pi^2}-\frac{3 c_2}{128 F_\pi^2 \pi ^2}+\frac{3 g_A^2}{64 F_\pi^2 m_N \pi ^2}\right)\\&+M_\pi^2 \left(\frac{32 c_1-3 (c_2+4 c_3)}{4 F_\pi^2}-\frac{3 g_A^2}{4 F_\pi^2 m_N}\right) A_0\left(M_\pi^2\right) \komma\\
 \delta m^{(4,\Delta)}  &=\frac{49 h_A^2 M_\pi^4}{576 F_\pi^2 m_N \pi ^2}-\frac{7 h_A^2 M_\pi^2 \Delta ^2}{24 F_\pi^2 m_N \pi ^2}+\frac{7 h_A^2 \Delta ^4}{36 F_\pi^2 m_N \pi ^2}\\&+\left(\frac{h_A^2 M_\pi^2}{6 F_\pi^2 m_N}+\frac{2 h_A^2 \Delta ^2}{3 F_\pi^2 m_N}\right) A_0\left(M_\pi^2\right)+\left(-\frac{2 h_A^2 M_\pi^2 \Delta }{3 F_\pi^2 m_N}+\frac{2 h_A^2 \Delta ^3}{3 F_\pi^2 m_N}\right) J_0(-\Delta ) \komma
\end{align*}
the Z-Factor is given by
  \begin{align*}
    Z_N &= 1+ \delta Z_N^{(3)}+ \delta Z_N^{(3,\Delta)}+\delta Z_N^{(4)}+\delta Z_N^{(4,\Delta)} \komma\\
    \delta Z_N^{(3)} &= -\frac{3 g_A^2 M_\pi^2}{32 F_\pi^2 \pi
      ^2}+\frac{9 g_A^2 A_0(M_\pi^2) }{4 F_\pi^2}\komma\\
\delta Z_N^{(3,\Delta)} &=-\frac{h_A^2 M_\pi^2}{4 F_\pi^2 \pi ^2}+\frac{h_A^2 \Delta ^2}{2 F_\pi^2 \pi ^2}+\frac{4 h_A^2 A_0\left(M_\pi^2\right)}{F_\pi^2}+\frac{4 h_A^2 \Delta  J_0(-\Delta )}{F_\pi^2}\komma\\
    \delta Z_N^{(4)} &= -\frac{9 g_A^2 M_\pi^2 J_0(0)}{8 F_\pi^2 m_N}\komma\\
\delta Z_N^{(4,\Delta)} &= M_\pi^2 \left(-\frac{h_A (b_{3}+b_{6})}{6 F_\pi^2 \pi ^2}-\frac{2 h_A^2}{3 F_\pi^2 m_N \pi ^2}\right) \Delta +\left(\frac{h_A (b_{3}+b_{6})}{9 F_\pi^2 \pi ^2}+\frac{17 h_A^2}{18 F_\pi^2 m_N \pi ^2}\right) \Delta ^3 \\&+\left(\frac{8 h_A (b_{3}+b_{6})}{3 F_\pi^2}+\frac{8 h_A^2}{3 F_\pi^2 m_N}\right) \Delta  A_0\left(M_\pi^2\right) \\&+\left(M_\pi^2 \left(-\frac{8 h_A (b_{3}+b_{6})}{3 F_\pi^2}-\frac{2 h_A^2}{3 F_\pi^2 m_N}\right)+\left(\frac{8 h_A (b_{3}+b_{6})}{3 F_\pi^2}+\frac{8 h_A^2}{3 F_\pi^2 m_N}\right) \Delta ^2\right) J_0(-\Delta )\punkt
 \end{align*}
and the effective nucleon axial coupling constant reads
  \begin{align*}
    g &= g_A + \delta g^{(3)} + \delta g^{(3,\Delta)}+ \delta g^{(4)}+ \delta g^{(4,\Delta)} \komma\\
    \delta g^{(3)} &= M_\pi^2 \left(-4 d_{16}+2 d_{18}+\frac{g_A^3}{16 F_\pi^2 \pi ^2}\right)-\frac{\left(g_A+2 g_A^3\right) A_0(M_\pi^2) }{F_\pi^2}\komma\\
 \delta g_A^{(3,\Delta)} &= \frac{5 (-31 g_1+63 g_A) h_A^2 M_\pi^2}{972 F_\pi^2 \pi ^2}+\frac{(155 g_1-267 g_A) h_A^2 \Delta ^2}{486 F_\pi^2 \pi ^2}+\frac{4 (25 g_1-57 g_A) h_A^2 A_0\left(M_\pi^2\right)}{81 F_\pi^2}\\&+\frac{32 g_A h_A^2 M_\pi^2 J_0(0)}{27 F_\pi^2 \Delta }+\left(-\frac{32 g_A h_A^2 M_\pi^2}{27 F_\pi^2 \Delta }+\frac{4 (25 g_1-57 g_A) h_A^2 \Delta }{81 F_\pi^2}\right) J_0(-\Delta )\komma\\
   \delta g^{(4)} &= M_\pi^2 \left(-\frac{4 g_A (c_3-2 c_4)}{3
        F_\pi^2}+\frac{g_A+g_A^3}{F_\pi^2 m_N}\right)
    J_0(0)\komma\\
 \delta g^{(4,\Delta)} &=\frac{5 h_A b_{4} \Delta  \left(3 M_\pi^2-2 \Delta ^2\right)}{162 F_\pi^2 \pi ^2}+\frac{h_A^2 \Delta  \left(-140 g_1 M_\pi^2+252 g_A M_\pi^2+195 g_1 \Delta ^2-411 g_A \Delta ^2\right)}{486 F_\pi^2 m_N \pi ^2}\\&+\left(-\frac{8 h_A (13 b_{4}+12 b_{5}) \Delta }{27 F_\pi^2}+\frac{40 (5 g_1-9 g_A) h_A^2 \Delta }{243 F_\pi^2 m_N}\right) A_0\left(M_\pi^2\right) \\&+\frac{8 h_A (13 b_{4}+12 b_{5}) \left(M_\pi^2-\Delta ^2\right) J_0(-\Delta )}{27 F_\pi^2}\\&-\frac{2 h_A^2 \left(25 g_1 \left(M_\pi^2-4 \Delta ^2\right)+9 g_A \left(7 M_\pi^2+20 \Delta ^2\right)\right) J_0(-\Delta )}{243 F_\pi^2 m_N}\punkt
  \end{align*}
The $\Delta$ mass reads
\begin{align*}
  \mathbf{m} &= m_\Delta + \delta \mathbf{m}^{(2)}+ \delta \mathbf{m}^{(3)}+ \delta \mathbf{m}^{(4)} \komma\\
 \delta \mathbf{m}^{(2)} &=4 c^\Delta_{1} M_\pi^2 \komma\\
 \delta \mathbf{m}^{(3)}  &=-\frac{h_A^2 M_\pi^2 \Delta }{24 F_\pi^2 \pi ^2}+\frac{h_A^2 \Delta ^3}{36 F_\pi^2 \pi ^2}-\frac{h_A^2 \Delta  A_0\left(M_\pi^2\right)}{3 F_\pi^2}-\frac{25 g_1^2 M_\pi^2 J_0(0)}{108 F_\pi^2}\\&+\left(-\frac{h_A^2 M_\pi^2}{3 F_\pi^2}+\frac{h_A^2 \Delta ^2}{3 F_\pi^2}\right) J_0(\Delta )\komma\\
 \delta \mathbf{m}^{(4)}  &=M_\pi^4 \left(-2 e^\Delta_{115}-2 e^\Delta_{116}-16 e^\Delta_{38}-\frac{8 c^\Delta_{1} l_3}{F_\pi^2}-\frac{3 c^\Delta_{2}}{256 F_\pi^2 \pi ^2}+\frac{25 g_1^2-3 h_A^2}{768 F_\pi^2 m_N \pi ^2}\right) \\&+\frac{h_A^2 M_\pi^2 \Delta ^2}{24 F_\pi^2 m_N \pi ^2}-\frac{h_A^2 \Delta ^4}{36 F_\pi^2 m_N \pi ^2}+\left(\frac{5 h_A^2 M_\pi^2 \Delta }{6 F_\pi^2 m_N}-\frac{5 h_A^2 \Delta ^3}{6 F_\pi^2 m_N}\right) J_0(\Delta )\\&+\left(M_\pi^2 \left(\frac{64 c^\Delta_{1}-3 (c^\Delta_{2}+8 c^\Delta_{3})}{8 F_\pi^2}-\frac{5 \left(g_1^2+3 h_A^2\right)}{24 F_\pi^2 m_N}\right)+\frac{5 h_A^2 \Delta ^2}{6 F_\pi^2 m_N}\right) A_0\left(M_\pi^2\right) \\
\end{align*}
with the $\Delta$ width given by
\begin{align*}
  \frac{\Gamma}{2}=\frac{h_A^2 \left(-M_\pi^2+\Delta ^2\right) ^{3/2} }{12 F_\pi^2 \pi }-\frac{5 h_A^2 \Delta  \left(-M_\pi^2+\Delta ^2\right)^{3/2}}{24 F_\pi^2 m_N \pi }\komma
\end{align*}
the Z-Factor is given by
  \begin{align*}
    Z_\Delta &= 1+ \delta Z_\Delta^{(3)}+\delta Z_\Delta^{(4)}\komma\\
    \delta Z_\Delta^{(3)} &= -\frac{65 g_1^2 M_\pi^2}{864 F_\pi^2 \pi ^2}+\frac{\left(25 g_1^2+36 h_A^2\right) A_0\left(M_\pi^2\right)}{36 F_\pi^2}-\frac{h_A^2 \Delta  J_0(\Delta )}{F_\pi^2}\komma\\
    \delta Z_\Delta^{(4)} &= \frac{h_A^2 \Delta  \left(3 M_\pi^2-14 \Delta ^2\right)}{144 F_\pi^2 m_N \pi ^2}+\frac{h_A (b_{3}+b_{6}) \Delta  \left(3 M_\pi^2-2 \Delta ^2\right)}{36 F_\pi^2 \pi ^2}\\&+\left(\frac{2 h_A (b_{3}+b_{6}) \Delta }{3 F_\pi^2}-\frac{10 h_A^2 \Delta }{3 F_\pi^2 m_N}\right) A_0\left(M_\pi^2\right)-\frac{25 g_1^2 M_\pi^2 J_0(0)}{72 F_\pi^2 m_N}\\&+\left(-\frac{5 h_A^2 \left(M_\pi^2-4 \Delta ^2\right)}{6 F_\pi^2 m_N}+\frac{2 h_A (b_{3}+b_{6}) \left(M_\pi^2-\Delta ^2\right)}{3 F_\pi^2}\right) J_0(\Delta )\punkt\\
 \end{align*}
and the effective nucleon-$\Delta$ axial coupling constant is given by
  \begin{align*}
    h &= h_A + \delta h^{(3)} + \delta h^{(4)} \komma\\
 \delta h^{(3)} &=M_\pi^2 \left(\frac{h_A \left(155 g_1^2-390 g_1 g_A+459 g_A^2+624 h_A^2\right)}{5184 F_\pi^2 \pi ^2}\right) \\&+\left(-\frac{h_A \left(-5 g_1^2+27 g_A^2+240 h_A^2\right)}{972 F_\pi^2 \pi ^2}\right) \Delta ^2 \\&-\frac{h_A \left(475 g_1^2-1350 g_1 g_A+9 \left(216+171 g_A^2+532 h_A^2\right)\right) A_0\left(M_\pi^2\right)}{1944 F_\pi^2}\\&+\frac{\left(25 g_1^2-81 g_A^2\right) h_A M_\pi^2 J_0(0)}{243 F_\pi^2 \Delta }\\&+\left(-\frac{\left(50 g_1^2 h_A+9 h_A^3\right) M_\pi^2}{486 F_\pi^2 \Delta }+\frac{\left(50 g_1^2 h_A-963 h_A^3\right) \Delta }{486 F_\pi^2}\right) J_0(-\Delta ) \\&+\left(\frac{\left(18 g_A^2 h_A+h_A^3\right) M_\pi^2}{54 F_\pi^2 \Delta }+\frac{\left(-9 g_A^2 h_A+13 h_A^3\right) \Delta }{27 F_\pi^2}\right) J_0(\Delta )\komma\\
    \delta h^{(4)} &= \frac{h_A (189 h_A (b_{3}+b_{6})-2 (24 c^\Delta_{11}+36 c^\Delta_{12}+18 c_3-18 c^\Delta_{3}-9 c_4+5 c^\Delta_{4})) \Delta  \left(3 M_\pi^2-2 \Delta ^2\right)}{1296 F_\pi^2 \pi ^2}\\&-\frac{5 h_A \Delta  \left(3 \left(36+16 g_1^2-165 h_A^2\right) M_\pi^2-2 \left(36+16 g_1^2-327 h_A^2\right) \Delta ^2\right)}{7776 F_\pi^2 m_N \pi ^2}\\&+\frac{h_A (15 c^\Delta_{11}+45 c^\Delta_{12}-18 c_3+18 c^\Delta_{3}+9 c_4-25 c^\Delta_{4}) \Delta  A_0\left(M_\pi^2\right)}{27 F_\pi^2}\\&+\frac{h_A \left(25 g_1^2-81 g_A^2+60 \left(-3+h_A^2\right)\right) \Delta  A_0\left(M_\pi^2\right)}{162 F_\pi^2 m_N}\\&+\frac{\left(25 g_1 b_{4}+25 g_1^2 (b_{3}+b_{6})-9 g_A (-13 b_{4}-12 b_{5}+9 g_A (b_{3}+b_{6}))\right) M_\pi^2 J_0(0)}{108 F_\pi^2}\\&+\frac{\left(875 g_1^2-1350 g_1 g_A+1539 g_A^2\right) h_A M_\pi^2 J_0(0)}{3888 F_\pi^2 m_N}\\&+\frac{h_A (9 h_A (b_{3}+b_{6})-15 c^\Delta_{11}-45 c^\Delta_{12}-18 c^\Delta_{3}+25 c^\Delta_{4}) \left(M_\pi^2-\Delta ^2\right) J_0(-\Delta )}{27 F_\pi^2}\\&+\frac{h_A \left(\left(360-50 g_1^2+93 h_A^2\right) M_\pi^2+\left(-360+50 g_1^2-417 h_A^2\right) \Delta ^2\right) J_0(-\Delta )}{324 F_\pi^2 m_N}\\&+\frac{h_A (h_A (b_{3}+b_{6})-2 c_3+c_4) \left(M_\pi^2-\Delta ^2\right) J_0(\Delta )}{3 F_\pi^2}\\&+\frac{\left(h_A^3 \left(44 M_\pi^2-179 \Delta ^2\right)-54 g_A^2 h_A \left(M_\pi^2-\Delta ^2\right)\right) J_0(\Delta )}{108 F_\pi^2 m_N}
  \end{align*}
with the imaginary parts
  \begin{align*}
    \mathrm{Im}\; \delta h_A^{(3)}&=\frac{h_A \sqrt{-M_\pi^2+\Delta ^2} }{216 F_\pi^2 \pi  \Delta }\Big(18 g_A^2 \left(M_\pi^2-\Delta ^2\right)+h_A^2 \left(M_\pi^2+26 \Delta ^2\right)\Big)\komma\\
\mathrm{Im}\; \delta h_A^{(4)}&=\frac{h_A (2 c_3-c_4) \left(-M_\pi^2+\Delta ^2\right)^{3/2}}{12 F_\pi^2 \pi }\\&-\frac{h_A \sqrt{-M_\pi^2+\Delta ^2} }{432 F_\pi^2 m_N \pi }\Big(54 g_A^2 \left(M_\pi^2-\Delta ^2\right)+h_A^2 \left(-44 M_\pi^2+179 \Delta ^2\right)\Big)  \punkt
\end{align*}
We emphasize that in the expressions above we have already performed all the redundancy shifts discussed in
Appendix~\ref{cha:RedShiftsAmpl}. Note that the quantities $Z_N$, $Z_\Delta$ and $h_A$ however still depend on 
the linear combination $b_3+b_6$. This is crucial for the explicit redundancy of these LECs in the renormalized pion-nucleon 
scattering amplitude.

\subsubsection{Renormalization of LECs}
Next, we discuss the renormalization of the LECs $c_i$, $d_i$ and $e_i$ in the HB framework, see Eqs.~\eqref{eq2:23} and \eqref{eq2:25}. 
In particular, we list in the following the $\beta$-functions and the additional finite shifts.
The employed $\beta$-functions at order $Q^3$ read
  \begin{align*}
    \beta_{d_1}&= -\frac{g_A^4}{6}\komma \quad
    &\beta_{d_2}&= \frac{1}{12} \left(-1-5 g_A^2\right)\komma \\
    \beta_{d_3}&= \frac{1}{6} \left(3+g_A^4\right)\komma \quad
    &\beta_{d_4}&= 0\komma \quad\\
    \beta_{d_5}&= \frac{1}{24} \left(1+5 g_A^2\right)\komma \quad
    &\beta_{d_{10}}&= \frac{1}{2} \left(g_A+5 g_A^3+4 g_A^5\right)\komma \\
    \beta_{d_{11}}&= \frac{1}{6} \left(3 g_A-9 g_A^3-4 g_A^5\right)\komma 
    &\beta_{d_{12}}&= -g_A \left(2+g_A^2+2 g_A^4\right)\komma \\
    \beta_{d_{13}}&= g_A^3+\frac{2 g_A^5}{3}\komma \quad
    &\beta_{d_{14}}&= \frac{g_A^4}{3}\komma \\
    \beta_{d_{15}}&= 0\komma \quad
    &\beta_{d_{16}}&= \frac{g_A}{2}+g_A^3\komma \\
    \beta_{d_{18}}&= 0  
  \end{align*}
and at order $Q^4$ we get
  \begin{align*}
    \beta_{e_{10}}&=-\frac{1}{6} g_A \left(3+8 g_A^2\right) c_4-\frac{g_A \left(3+19 g_A^2+13 g_A^4\right)}{24 m_N}\komma\\
    \beta_{e_{11}}&=-\frac{g_A c_4}{3}+\frac{g_A \left(-7+35 g_A^2+12 g_A^4\right)}{48 m_N}\komma\\
    \beta_{e_{12}}&=\frac{4}{3} g_A \left(1+g_A^2\right) c_4+\frac{g_A \left(61+57 g_A^2+26 g_A^4\right)}{48 m_N}\komma\\
    \beta_{e_{13}}&=-\frac{2}{3} \left(g_A+2 g_A^3\right) c_4-\frac{g_A \left(73+54 g_A^2+21 g_A^4\right)}{24 m_N}\komma\\
    \beta_{e_{14}}&= \frac{1}{12} (-c_2-6 c_3)-\frac{g_A^2 \left(3+g_A^2\right)}{12 m_N}\komma\\
    \beta_{e_{15}}&= \frac{9+2 g_A^2+11 g_A^4}{24 m_N}\komma\\
    \beta_{e_{16}}&= \frac{-3-2 g_A^2-2 g_A^4}{4 m_N}\komma\\
    \beta_{e_{17}}&= -\frac{c_4}{12}+\frac{-1+7 g_A^2+4 g_A^4}{48 m_N}\komma\\
    \beta_{e_{18}}&= -\frac{2 g_A^2 c_4}{3}-\frac{g_A^2 \left(3+4 g_A^2\right)}{12 m_N}\komma\\
    2\beta_{e_{19}}-\beta_{e_{22}}-\beta_{e_{36}}&= 2 c_1-\frac{5
      c_2}{24}+\frac{3 c_3}{4}+\frac{-1+g_A^2-6 g_A^4}{8 m_N}\komma\\
    \beta_{e_{20}}+\beta_{e_{35}}&= \frac{c_2}{2}+\frac{6+16 g_A^2+15 g_A^4}{24 m_N}\komma\\
    2\beta_{e_{21}}-\beta_{e_{37}}&= \frac{1}{3} \left(2+9 g_A^2\right) c_4+\frac{2+16 g_A^2+9 g_A^4}{12 m_N}\komma\\
    \beta_{e_{22}}-4\beta_{e_{38}}&= \frac{1}{4} (-12 c_1+c_2+3
    c_3)\komma\\
    \beta_{e_{34}}&= \frac{2 g_A c_4}{3}+\frac{g_A-7 g_A^3-6 g_A^5}{24 m_N}\punkt
  \end{align*}
  Treating the $\Delta$ resonance as an explicit degree of freedom gives additional $\beta$ functions for the LECs $c_i$ at $\varepsilon^3$
  \begin{align*}
    \beta^\Delta_{c_i}&=0\\
    \beta^{(3,\Delta)}_{c_1}&=2 h_A^2 \Delta \komma\\
    \beta^{(3,\Delta)}_{c_2}&=-\frac{80 (5 g_1-9 g_A)^2 h_A^2 \Delta }{2187}\komma\\
    \beta^{(3,\Delta)}_{c_3}&=\frac{16 \left(125 g_1^2-450 g_1 g_A+81 \left(9+5 g_A^2\right)\right) h_A^2 \Delta }{2187}\komma\\
    \beta^{(3,\Delta)}_{c_4}&=-\frac{2 h_A^2 \left(972+125 g_1^2-2250 g_1 g_A+2349 g_A^2+1152 h_A^2\right) \Delta }{2187}
  \end{align*}
and the contributions at order $\varepsilon^4$ read
  \begin{align*}
    \beta^{(4,\Delta)}_{c_1}&=-16 h_A^2 c_1 \Delta ^2+\frac{h_A^2 \Delta ^2}{m_N}\komma\\
    \beta^{(4,\Delta)}_{c_2}&=-\frac{16}{243} h_A (-25 g_1 b_{4}+9 (13 g_A b_{4}+12 g_A b_{5}+27 h_A c_2)) \Delta ^2\\&-\frac{8 h_A^2 \left(-72+40 g_1^2-150 g_1 g_A-63 g_A^2+260 h_A^2\right) \Delta ^2}{243 m_N}\komma\\
    \beta^{(4,\Delta)}_{c_3}&= \frac{16}{243} h_A (-25 g_1 b_{4}+9 (13 g_A b_{4}+12 g_A b_{5}-27 h_A c_3)) \Delta ^2\\&+\frac{2 h_A^2 \left(1025 g_1^2-4050 g_1 g_A+9 \left(252-63 g_A^2+640 h_A^2\right)\right) \Delta ^2}{2187 m_N}\komma\\
    \beta^{(4,\Delta)}_{c_4}&=-\frac{4}{243} h_A (-25 g_1 (23 b_{4}+24 b_{5})+45 g_A (27 b_{4}+28 b_{5})+684 h_A c_4) \Delta ^2\\&+\frac{h_A^2 \left(-275 g_1^2+1350 g_1 g_A+9 \left(36+405 g_A^2-832 h_A^2\right)\right) \Delta ^2}{2187 m_N}\punkt
  \end{align*}
Additionally, one has corresponding finite shifts at $\varepsilon^3$
  \begin{align*}
    \delta\bar c_{1,f}^{(3,\Delta)} &=0\komma\\
    \delta\bar c_{2,f}^{(3,\Delta)} &=-\frac{\left(3575 g_1^2-8910 g_1 g_A+6399 g_A^2\right) h_A^2 \Delta }{52488 \pi ^2}\komma\\
    \delta\bar c_{3,f}^{(3,\Delta)} &=\frac{\left(3575 g_1^2-8910 g_1 g_A+6399 g_A^2\right) h_A^2 \Delta }{52488 \pi ^2}\komma\\
    \delta\bar c_{4,f}^{(3,\Delta)} &= \frac{h_A^2 \left(-4775 g_1^2+28350 g_1 g_A+27 \left(180-1317 g_A^2+64 h_A^2\right)\right) \Delta }{104976 \pi ^2}
  \end{align*}
and at $\varepsilon^4$
  \begin{align*}
    \delta\bar c_{1,f}^{(4,\Delta)} &=\frac{h_A^2 \Delta ^2}{16 m_N \pi ^2}\komma\\
    \delta\bar c_{2,f}^{(4,\Delta)} &=\frac{5 (7 g_1-3 g_A) h_A b_{4} \Delta ^2}{486 \pi ^2}\\&+\frac{h_A^2 \left(-4015 g_1^2+12510 g_1 g_A-9 \left(240+1839 g_A^2-304 h_A^2\right)\right) \Delta ^2}{34992 m_N \pi ^2}\komma\\
    \delta\bar c_{3,f}^{(4,\Delta)} &=\frac{5 (-7 g_1+3 g_A) h_A b_{4} \Delta ^2}{486 \pi ^2}\\&+\frac{h_A^2 \left(11405 g_1^2-33210 g_1 g_A+27 \left(744+1263 g_A^2-16 h_A^2\right)\right) \Delta ^2}{104976 m_N \pi ^2}\komma\\
    \delta\bar c_{4,f}^{(4,\Delta)} &=\frac{5 h_A (71 g_1 b_{4}+9 g_A b_{4}+68 g_1 b_{5}+48 g_A b_{5}-48 h_A c_4) \Delta ^2}{1944 \pi ^2}\\&+\frac{h_A^2 \left(-10355 g_1^2+81270 g_1 g_A-81 \left(4+1611 g_A^2-144 h_A^2\right)\right) \Delta ^2}{209952 m_N \pi ^2}\punkt
  \end{align*}
Analogously, for the LECs $d_i$ we obtain at $\varepsilon^3$
  \begin{align*}
    \beta^\Delta_{d_i}&=0\\
    \beta^{(3,\Delta)}_{d_1}+  \beta^{(3,\Delta)}_{d_2}&=-\frac{h_A^2 \left(125 g_1^2-450 g_1 g_A-9 \left(90+27 g_A^2-32 h_A^2\right)\right)}{2187}\komma\\
    \beta^{(3,\Delta)}_{d_3}&=\frac{h_A^2 \left(125 g_1^2-450 g_1 g_A-243 g_A^2+288 h_A^2\right)}{2187}\komma\\
    \beta^{(3,\Delta)}_{d_5}&=-\frac{5 h_A^2}{27}\komma\\
    \beta^{(3,\Delta)}_{d_{14}}-\beta^{(3,\Delta)}_{d_{15}}&=\frac{2 h_A^2 \left(125 g_1^2-450 g_1 g_A-243 g_A^2+288 h_A^2\right)}{2187}
  \end{align*}
and at $\varepsilon^4$
  \begin{align*}
      \beta^{(4,\Delta)}_{d_1}+\beta^{(4,\Delta)}_{d_2}&=\frac{4}{243} h_A (100 g_1 b_{4}-252 g_A b_{4}+75 g_1 b_{5}-198 g_A b_{5}+36 h_A c_4) \Delta \\&+\frac{h_A^2 \left(-175 g_1^2+750 g_1 g_A+3 \left(792+531 g_A^2-512 h_A^2\right)\right) \Delta }{1458 m_N}\komma\\
    \beta^{(4,\Delta)}_{d_3}&=-\frac{4}{243} h_A (100 g_1 b_{4}-252 g_A b_{4}+75 g_1 b_{5}-198 g_A b_{5}+36 h_A c_4) \Delta \\&+\frac{h_A^2 \left(275 g_1^2-750 g_1 g_A-9 \left(180+129 g_A^2-128 h_A^2\right)\right) \Delta }{729 m_N}\komma\\
    \beta^{(4,\Delta)}_{d_5}&=-\frac{h_A^2 \left(360+125 g_1^2-250 g_1 g_A-243 g_A^2+256 h_A^2\right) \Delta }{972 m_N}\komma\\
    \beta^{(4,\Delta)}_{d_{14}}-\beta^{(4,\Delta)}_{d_{15}}&=-\frac{8}{243} h_A (100 g_1 b_{4}-252 g_A b_{4}+75 g_1 b_{5}-198 g_A b_{5}+36 h_A c_4) \Delta\\& +\frac{8 h_A^2 \left(50 g_1^2-27 \left(6+33 g_A^2-16 h_A^2\right)\right) \Delta }{2187 m_N}\punkt
  \end{align*}
Additionally, the finite shifts at $\varepsilon^3$ have the form
  \begin{align*}
    \delta\bar d^{(3,\Delta)}_{1,f}+ \delta\bar d^{(3,\Delta)}_{2,f}&=\frac{h_A^2 \left(-925 g_1^2+3870 g_1 g_A-27 \left(-78+189 g_A^2+8 h_A^2\right)\right)}{104976 \pi ^2}\komma\\
    \delta\bar d^{(3,\Delta)}_{3,f}&=\frac{h_A^2 \left(925 g_1^2-3870 g_1 g_A+5103 g_A^2+216 h_A^2\right)}{104976 \pi ^2}\komma\\
    \delta\bar d^{(3,\Delta)}_{5,f}&=-\frac{13 h_A^2}{1296 \pi ^2}\komma\\
    \delta\bar d^{(3,\Delta)}_{14,f}-\delta\bar d^{(3,\Delta)}_{15,f}&=\frac{h_A^2 \left(3425 g_1^2-9090 g_1 g_A+5589 g_A^2+432 h_A^2\right)}{104976 \pi ^2}
  \end{align*}
and at $\varepsilon^4$
  \begin{align*}
    \delta\bar d^{(4,\Delta)}_{1,f}+ \delta\bar d^{(4,\Delta)}_{2,f}&=\frac{h_A (5 g_1 (38 b_{4}+31 b_{5})-6 (49 g_A b_{4}+31 g_A b_{5}+8 h_A c_4)) \Delta }{1944 \pi ^2}\\&-\frac{h_A^2 \left(4535 g_1^2-19950 g_1 g_A+27 \left(-232+965 g_A^2+64 h_A^2\right)\right) \Delta }{139968 m_N \pi ^2}\komma\\
    \delta\bar d^{(4,\Delta)}_{3,f}&=\frac{h_A (-5 g_1 (38 b_{4}+31 b_{5})+6 (49 g_A b_{4}+31 g_A b_{5}+8 h_A c_4)) \Delta }{1944 \pi ^2}\\&+\frac{h_A^2 \left(2605 g_1^2-10650 g_1 g_A+9 \left(504+1641 g_A^2+64 h_A^2\right)\right) \Delta }{69984 m_N \pi ^2}\komma\\
    \delta\bar d^{(4,\Delta)}_{5,f}&=\frac{h_A^2 \left(-1272-75 g_1^2+150 g_1 g_A-387 g_A^2+64 h_A^2\right) \Delta }{31104 m_N \pi ^2}\komma\\
    \delta\bar d^{(4,\Delta)}_{14,f}-\delta\bar d^{(4,\Delta)}_{15,f}&=\frac{h_A (393 g_A b_{4}+312 g_A b_{5}-5 g_1 (33 b_{4}+16 b_{5})+12 h_A c_4) \Delta }{972 \pi ^2}\\&+\frac{h_A^2 \left(7715 g_1^2-21150 g_1 g_A+81 \left(32+191 g_A^2-32 h_A^2\right)\right) \Delta }{104976 m_N \pi ^2}\punkt
  \end{align*}
Finally, the additional $\beta$-functions of the LECs $e_i$ at $\varepsilon^3$ read
  \begin{align*}
           \beta^{\Delta}_{e_{14}}&=\frac{2 h_A^2}{27 m_N}+\frac{5 h_A^2}{27 \Delta }\komma\\
          \beta^{\Delta}_{e_{15}}&=0\komma\\
          \beta^{\Delta}_{e_{16}}&=0\komma\\
          \beta^{\Delta}_{e_{17}}&=-\frac{h_A^2}{54 \Delta }\komma\\
          \beta^{\Delta}_{e_{18}}&=-\frac{4 g_A^2 h_A^2}{27 \Delta }\komma\\
          2\beta^{\Delta}_{e_{19}}-\beta^{\Delta}_{e_{22}}-\beta^{\Delta}_{e_{36}}&=-\frac{h_A^2}{27 m_N}+\frac{\left(-33+25 g_1^2-50 g_1 g_A+81 g_A^2\right) h_A^2}{162 \Delta }\komma\\
          \beta^{\Delta}_{e_{20}}+\beta^{\Delta}_{e_{35}}&=\frac{\left(25 g_1^2-50 g_1 g_A+81 g_A^2\right) h_A^2}{162 m_N}-\frac{\left(25 g_1^2-50 g_1 g_A+81 g_A^2\right) h_A^2}{324 \Delta }\komma\\
          2\beta^{\Delta}_{e_{21}}-\beta^{\Delta}_{e_{37}}&= \frac{\left(-24-25 g_1^2+50 g_1 g_A+135 g_A^2\right) h_A^2}{324 \Delta }\komma\\ 
          \beta^{\Delta}_{e_{22}}-4\beta^{\Delta}_{e_{38}}&=-\frac{2 h_A^2}{9 m_N}-\frac{2 h_A^2}{9 \Delta }
  \end{align*}
and at $\varepsilon^4$
  \begin{align*}
          \beta^{(4,\Delta)}_{e_{14}}&=-\frac{h_A^2 \left(125 g_1^2-450 g_1 g_A+9 \left(135-27 g_A^2+32 h_A^2\right)\right)}{4374 m_N}-\frac{5 h_A^2}{27 \Delta }\komma\\
          \beta^{(4,\Delta)}_{e_{15}}&=\frac{h_A^2 \left(1375 g_1^2-4950 g_1 g_A-9 \left(72+297 g_A^2-352 h_A^2\right)\right)}{8748 m_N}\komma\\
          \beta^{(4,\Delta)}_{e_{16}}&=\frac{h_A^2 \left(-125 g_1^2+450 g_1 g_A+9 \left(36+27 g_A^2-32 h_A^2\right)\right)}{729 m_N}\komma\\
          \beta^{(4,\Delta)}_{e_{17}}&=\frac{h_A^2 \left(125 g_1^2-450 g_1 g_A+9 \left(90-243 g_A^2+128 h_A^2\right)\right)}{17496 m_N}+\frac{h_A^2}{54 \Delta }\komma\\
          \beta^{(4,\Delta)}_{e_{18}}&=\frac{1}{486} h_A (5 (5 g_1-9 g_A) (7 b_{4}+8 b_{5})+288 h_A c_4)\\&-\frac{h_A^2 \left(-648+125 g_1^2+150 g_1 g_A-1971 g_A^2+384 h_A^2\right)}{2916 m_N}+\frac{4 g_A^2 h_A^2}{27 \Delta }\komma\\
          2\beta^{(4,\Delta)}_{e_{19}}-\beta^{(4,\Delta)}_{e_{22}}-\beta^{(4,\Delta)}_{e_{36}}&=2 h_A^2 (-2 c_1+c_3)\\&+\frac{h_A^2 \left(2025+20 \left(-35 g_1^2+90 g_1 g_A+81 g_A^2\right)-1872 h_A^2\right)}{2916 m_N}\\&+\frac{\left(33-25 g_1^2+50 g_1 g_A-81 g_A^2\right) h_A^2}{162 \Delta }\komma\\
          \beta^{(4,\Delta)}_{e_{20}}+\beta^{(4,\Delta)}_{e_{35}}&=h_A^2 c_2+\frac{h_A^2 \left(5 \left(37 g_1^2-210 g_1 g_A-351 g_A^2\right)+1152 \left(-1+h_A^2\right)\right)}{1944 m_N}\\&+\frac{\left(25 g_1^2-50 g_1 g_A+81 g_A^2\right) h_A^2}{324 \Delta }\komma\\
          2\beta^{(4,\Delta)}_{e_{21}}-\beta^{(4,\Delta)}_{e_{37}}&= \frac{1}{162} h_A (-25 g_1 (7 b_{4}+8 b_{5})+9 (31 g_A b_{4}+44 g_A b_{5}+4 h_A c_4))\\&+\frac{h_A^2 \left(575 g_1^2+450 g_1 g_A-9 \left(360+873 g_A^2-272 h_A^2\right)\right)}{5832 m_N}\\&+\frac{\left(24+25 g_1^2-50 g_1 g_A-135 g_A^2\right) h_A^2}{324 \Delta }\komma\\ 
          \beta^{(4,\Delta)}_{e_{22}}-4\beta^{(4,\Delta)}_{e_{38}}&=-4 h_A^2 c_1-\frac{h_A^2}{6 m_N}+\frac{2 h_A^2}{9 \Delta }
  \end{align*}
The finite pieces at $\varepsilon^3$ are given by
  \begin{align*}
    \delta\bar e^{(4,\Delta)}_{14,f}&=-\frac{h_A^2 \left(2725 g_1^2-8730 g_1 g_A+81 \left(216+141 g_A^2+16 h_A^2\right)\right)}{839808 m_N \pi ^2}\\&-\frac{13 h_A^2}{1296 \pi ^2 \Delta }\komma\\
    \delta\bar e^{(4,\Delta)}_{15,f}&=\frac{h_A (-25 g_1 b_{4}+9 g_A (13 b_{4}+12 b_{5}))}{11664 \pi ^2}\\&+\frac{h_A^2 \left(35725 g_1^2-116730 g_1 g_A+81 \left(64+1421 g_A^2+256 h_A^2\right)\right)}{1679616 m_N \pi ^2}\\&+\frac{\left(1225 g_1^2-4050 g_1 g_A+3969 g_A^2\right) h_A^2}{839808 \pi ^2 \Delta }\komma\\
    \delta\bar e^{(4,\Delta)}_{16,f}&=\frac{h_A (25 g_1 b_{4}-9 g_A (13 b_{4}+12 b_{5}))}{11664 \pi ^2}\\&-\frac{h_A^2 \left(34325 g_1^2-116010 g_1 g_A+27 \left(864+4695 g_A^2+832 h_A^2\right)\right)}{1679616 m_N \pi ^2}\\&-\frac{\left(1225 g_1^2-4050 g_1 g_A+3969 g_A^2\right) h_A^2}{839808 \pi ^2 \Delta }\komma\\
    \delta\bar e^{(4,\Delta)}_{17,f}&=\frac{h_A^2 \left(2125 g_1^2-10890 g_1 g_A+27 \left(66+999 g_A^2+16 h_A^2\right)\right)}{1679616 m_N \pi ^2}\\&+\frac{h_A^2}{5184 \pi ^2 \Delta }\komma\\
    \delta\bar e^{(4,\Delta)}_{18,f}&=-\frac{5 h_A (-98 g_1 b_{4}+162 g_A b_{4}-107 g_1 b_{5}+171 g_A b_{5}+36 h_A c_4)}{11664 \pi ^2}\\&-\frac{h_A^2 \left(8075 g_1^2-5670 g_1 g_A+81 \left(315 g_A^2+64 \left(3+h_A^2\right)\right)\right)}{1679616 m_N \pi ^2}\\&-\frac{h_A^2 \left(425 g_1^2-450 g_1 g_A+4617 g_A^2+3456 h_A^2\right)}{839808 \pi ^2 \Delta }\komma\\
    2 \delta\bar e^{(4,\Delta)}_{19,f}-\delta\bar e^{(4,\Delta)}_{22,f}-\delta\bar
    e^{(4,\Delta)}_{36,f}&=\frac{h_A (25 g_1 b_{4}-9 g_A (13 b_{4}+12 b_{5})+162 h_A (-2 c_1+c_3))}{1296 \pi ^2}\\&-\frac{h_A^2 \left(-2484+1255 g_1^2-5070 g_1 g_A+1647 g_A^2+3696 h_A^2\right)}{93312 m_N \pi ^2}\\&+\frac{\left(450-65 g_1^2+930 g_1 g_A+567 g_A^2\right) h_A^2}{46656 \pi ^2 \Delta }\komma\\
  \delta\bar e^{(4,\Delta)}_{20,f}+\delta\bar e^{(4,\Delta)}_{35,f}&=\frac{h_A (-25 g_1 b_{4}+9 (13 g_A b_{4}+12 g_A b_{5}+18 h_A c_2))}{2592 \pi ^2}\\&+\frac{h_A^2 \left(1584+2195 g_1^2-8310 g_1 g_A+9531 g_A^2+5712 h_A^2\right)}{186624 m_N \pi ^2}\\&+\frac{\left(65 g_1^2-930 g_1 g_A-567 g_A^2\right) h_A^2}{93312 \pi ^2 \Delta }\komma\\
  2 \delta\bar e^{(4,\Delta)}_{21,f}-\delta\bar e^{(4,\Delta)}_{37,f}&=\frac{h_A (-5 g_1 (307 b_{4}+308 b_{5})+9 (319 g_A b_{4}+296 g_A b_{5}+232 h_A c_4))}{15552 \pi ^2}\\&+\frac{h_A^2 \left(2880+355 g_1^2-870 g_1 g_A+3699 g_A^2+3648 h_A^2\right)}{186624 m_N \pi ^2}\\&+\frac{h_A^2 \left(24-95 g_1^2+590 g_1 g_A-567 g_A^2+384 h_A^2\right)}{31104 \pi ^2 \Delta }\komma\\
  \delta\bar e^{(4,\Delta)}_{22,f}-4 \delta\bar e^{(4,\Delta)}_{38,f}&=-\frac{h_A^2 c_1}{4 \pi ^2}+\frac{7 h_A^2}{432 \pi ^2 \Delta }\punkt
  \end{align*}

\clearpage
\section{Tables}
\vspace{8cm} 
\begin{table}[ht!]
  \centering
  \begin{tabular*}{0.9\textwidth}{@{\extracolsep{\fill}} c| c|c| r| r}
    $O(1/m_N)$&$\Gamma$[MeV]&$\mathrm{Re}\; h_A$
    &\multicolumn{2}{c}{$\mathrm{Im}\; h_A$} \\\hline
    \hline
0 & $0.079 h_A^2$&1.12 &$-0.242 h_A +0.288h_A^3$&$0.000 h_A c_2+0.079 h_A c_3-0.040 h_A c_4$\\
1 & $0.022 h_A^2$&2.14 &$-0.137 h_A +0.022h_A^3$ &$-0.023 h_A c_2+0.022 h_A c_3-0.011 h_A c_4$\\
2 & $0.054 h_A^2$&1.36 &$-0.174 h_A +0.208h_A^3$ &$-0.009 h_A c_2+0.054 h_A c_3-0.027 h_A c_4$\\
3 & $0.038 h_A^2$&1.62 &$-0.162 h_A +0.098h_A^3$ &$-0.016 h_A c_2+0.038 h_A c_3-0.019 h_A c_4$\\
4 & $0.045 h_A^2$&1.49 &$-0.166 h_A +0.156h_A^3$ &$-0.013 h_A c_2+0.045 h_A c_3-0.023 h_A c_4$\\
5 & $0.042 h_A^2$&1.54 &$-0.164 h_A +0.128h_A^3$ &$-0.014 h_A c_2+0.042 h_A c_3-0.021 h_A c_4$\\
6 & $0.043 h_A^2$&1.52 &$-0.165 h_A +0.141h_A^3$ &$-0.014 h_A c_2+0.043 h_A c_3-0.022 h_A c_4$\\
7 & $0.043 h_A^2$&1.52 &$-0.165 h_A +0.135h_A^3$ &$-0.014 h_A c_2+0.043 h_A c_3-0.022 h_A c_4$\\\hline \hline
Cov & $0.043 h_A^2$&1.52 &$-0.165 h_A +0.137h_A^3$ &$-0.014 h_A c_2+0.043 h_A c_3-0.022 h_A c_4$\\
  \end{tabular*}
  \caption{Convergence in $1/m_N$ of the quantities $\Gamma$, $\mathrm{Re}\;h_A$, and
    $\mathrm{Im}\;h_A$, where $\mathrm{Re}\;h_A$ is matched to the
    decay width $\Gamma_{\Delta\to\pi N}=100$ MeV \cite{Agashe:2014kda}, which is extracted
    from the complex-valued pole position of the $\Delta$ resonances. The column for
    $\mathrm{Im}\; h_A$ is divided in the contribution at order
    $\varepsilon^3$ (left) and $\varepsilon^4$ (right).
  }
\label{tab:Conv}
\end{table}

\newpage
\begin{table}[ht!]
\vspace{-0.6cm}
  \centering
  \begin{tabular*}{0.8\textwidth}{@{\extracolsep{\fill}} c| rr| rr| rr}
    \multicolumn{1}{c}{}&\multicolumn{2}{c}{HB-NN} &\multicolumn{2}{c}{HB-$\pi$N}&\multicolumn{2}{c}{Cov}\\\hline
    $\varepsilon^2$ &\multicolumn{1}{c}{$\pi$N}&\multicolumn{1}{c|}{$\pi$N+RS} &\multicolumn{1}{c}{$\pi$N}&\multicolumn{1}{c|}{$\pi$N+RS}&\multicolumn{1}{c}{$\pi$N}&\multicolumn{1}{c}{$\pi$N+RS}\\\hline 
$h_A$&1.34(0)&1.34(0)&1.37(0)&1.36(0)&1.40(0)&1.40(0)\\ $c_1$&-1.02(3)&-1.01(3)&-0.79(2)&-0.81(2)&-0.86(2)&-0.87(2)\\ $c_2$&0.27(5)&0.28(5)&0.80(5)&0.80(5)&0.45(3)&0.45(3)\\ $c_3$&-0.99(4)&-0.99(4)&-0.99(3)&-1.01(3)&-0.71(3)&-0.72(3)\\ $c_4$&0.51(3)&0.52(3)&1.09(3)&1.08(3)&0.87(2)&0.87(2)\\ 
\hline
$\chi^2_{\pi N}$/dof &0.65 &0.64 &0.61 &0.61 &0.55 &0.56\\\hline
$\bar\chi^2_{\pi N}$/dof &14.1 &14.0 &3.7 &3.7 &2.9 &3.0\\\hline
$\varepsilon^3$ &\multicolumn{1}{c}{$\pi$N}&\multicolumn{1}{c|}{$\pi$N+RS} &\multicolumn{1}{c}{$\pi$N}&\multicolumn{1}{c|}{$\pi$N+RS}&\multicolumn{1}{c}{$\pi$N}&\multicolumn{1}{c}{$\pi$N+RS}\\\hline 
$h_A$&1.46(1)&1.46(1)&1.45(1)&1.46(1)&1.46(1)&1.46(1)\\ $c_1$&-1.81(1)&-1.81(1)&-1.98(1)&-1.97(1)&-1.51(1)&-1.50(1)\\ $c_2$&1.69(9)&1.67(9)&1.20(6)&1.17(6)&0.67(3)&0.69(3)\\ $c_3$&-3.91(9)&-3.89(9)&-3.74(7)&-3.70(7)&-2.67(7)&-2.63(6)\\ $c_4$&1.71(5)&1.70(5)&1.50(4)&1.48(4)&1.26(5)&1.22(5)\\ $d_{1+2}$&0.23(6)&0.21(5)&0.59(6)&0.57(6)&0.57(5)&0.55(4)\\ $d_3$&-1.55(6)&-1.53(6)&-1.40(5)&-1.39(6)&-1.78(3)&-1.78(3)\\ $d_5$&0.82(3)&0.82(3)&0.53(3)&0.54(3)&0.70(3)&0.72(3)\\ $d_{14-15}$&-0.87(13)&-0.83(13)&-1.14(12)&-1.11(12)&-0.75(8)&-0.72(7)\\ $g_1$&-2.74(15)&-2.74(14)&-2.82(13)&-2.74(14)&-0.67(29)&-0.38(29)\\ 
\hline
$\chi^2_{\pi N}$/dof &1.16 &1.15 &1.26 &1.26 &1.20 &1.19\\\hline
$\bar\chi^2_{\pi N}$/dof &2.1 &2.1 &2.0 &2.0 &1.95 &1.95\\\hline
$\varepsilon^4$ &\multicolumn{1}{c}{$\pi$N}&\multicolumn{1}{c|}{$\pi$N+RS} &\multicolumn{1}{c}{$\pi$N}&\multicolumn{1}{c|}{$\pi$N+RS}&\multicolumn{1}{c}{$\pi$N}&\multicolumn{1}{c}{$\pi$N+RS}\\\hline 
$h_A$&1.38(1)&1.37(1)&1.39(1)&1.38(1)&1.42(1)&1.40(1)\\ $c_1$&-1.45(5)&-1.39(3)&-1.29(5)&-1.30(4)&-1.50(4)&-1.32(3)\\ $c_2$&0.39(13)&0.51(10)&1.66(13)&1.61(10)&0.52(7)&0.86(5)\\ $c_3$&-2.14(7)&-2.12(6)&-2.37(5)&-2.34(5)&-1.98(7)&-1.98(6)\\ $c_4$&2.47(10)&2.29(5)&2.56(10)&2.43(6)&2.31(7)&2.28(4)\\ $d_{1+2}$&2.12(7)&2.07(6)&1.98(7)&1.94(6)&1.67(5)&1.74(5)\\ $d_3$&-2.61(6)&-2.62(5)&-1.97(4)&-1.96(4)&-3.13(4)&-3.07(4)\\ $d_5$&0.36(3)&0.39(3)&0.13(3)&0.15(3)&0.90(3)&0.81(3)\\ $d_{14-15}$&-3.38(13)&-3.53(12)&-2.75(11)&-2.76(10)&-2.94(10)&-3.16(9)\\ $e_{14}$&2.10(15)&2.30(13)&1.76(14)&1.92(12)&1.76(12)&1.61(10)\\ $e_{15}$&-3.41(45)&-4.13(26)&-1.92(50)&-2.61(31)&-2.27(19)&-2.50(17)\\ $e_{16}$&2.55(48)&2.70(28)&-1.23(56)&-0.65(37)&1.40(18)&0.88(9)\\ $e_{17}$&-0.63(23)&-0.53(20)&-0.59(21)&-0.71(19)&-0.96(15)&-0.87(14)\\ $e_{18}$&-0.82(43)&-0.11(15)&-0.36(42)&0.30(21)&0.82(18)&1.03(10)\\ $g_1$&-2.41(20)&-2.52(19)&-2.55(19)&-2.60(17)&-2.35(21)&-2.32(20)\\ $b_4$&-1.33(34)&-1.45(29)&-1.44(31)&-1.56(28)&1.07(43)&1.55(28)\\ $b_5$&-1.24(37)&-1.39(32)&-1.31(35)&-1.39(32)&0.81(65)&1.35(32)\\ 
\hline
$\chi^2_{\pi N}$/dof &1.73 &1.73 &1.80 &1.80 &1.78 &1.80\\\hline
$\bar\chi^2_{\pi N}$/dof &1.91 &1.92 &1.92 &1.92 &1.91 &1.93\\\hline
  \end{tabular*}
  \caption{$K$-matrix approach: LECs extracted from fits at order $\varepsilon^2$,
    $\varepsilon^3$, and $\varepsilon^4$ with
    $T_\pi<125$~MeV. The labels $\pi$N and $\pi$N+RS denote fits
    without and with additional constraints $\chi^2_{\mathrm{RS}}$,
    see Eqs.~\eqref{eq2:31} and \eqref{eq2:48}, respectively. The labels
 HB-NN, HB-$\pi$N, and Cov (covariant) denote the different counting schemes
 of $1/m_N$ contributions, see section \ref{sec:power-count-renorm-Delta}.}
\label{tab:FitK}
\end{table}

\begin{table}[ht]
\vspace{-0.6cm}
  \centering
  \begin{tabular*}{0.8\textwidth}{@{\extracolsep{\fill}} c| rr| rr| rr}
    \multicolumn{1}{c}{}&\multicolumn{2}{c}{HB-NN} &\multicolumn{2}{c}{HB-$\pi$N}&\multicolumn{2}{c}{Cov}\\\hline
    $\varepsilon^2$ &\multicolumn{1}{c}{$\pi$N}&\multicolumn{1}{c|}{$\pi$N+RS} &\multicolumn{1}{c}{$\pi$N}&\multicolumn{1}{c|}{$\pi$N+RS}&\multicolumn{1}{c}{$\pi$N}&\multicolumn{1}{c}{$\pi$N+RS}\\\hline 
$h_A$&1.26(0)&1.26(0)&1.29(0)&1.29(0)&1.30(0)&1.30(0)\\ $c_1$&-0.88(2)&-0.89(2)&-0.68(2)&-0.69(2)&-0.80(1)&-0.80(1)\\ $c_2$&0.52(3)&0.52(3)&0.99(4)&0.98(4)&0.59(2)&0.59(2)\\ $c_3$&-1.16(3)&-1.16(3)&-1.13(2)&-1.13(2)&-0.90(2)&-0.90(2)\\ $c_4$&0.74(3)&0.73(3)&1.21(3)&1.20(3)&1.02(2)&1.01(2)\\ 
\hline
$\chi^2_{\pi N}$/dof &0.66 &0.66 &0.58 &0.58 &0.50 &0.51\\\hline
$\bar\chi^2_{\pi N}$/dof &35.6 &35.8 &7.1 &7.1 &4.9&4.9\\\hline
$\varepsilon^3$ &\multicolumn{1}{c}{$\pi$N}&\multicolumn{1}{c|}{$\pi$N+RS} &\multicolumn{1}{c}{$\pi$N}&\multicolumn{1}{c|}{$\pi$N+RS}&\multicolumn{1}{c}{$\pi$N}&\multicolumn{1}{c}{$\pi$N+RS}\\\hline 
$h_A$&1.37(0)&1.37(0)&1.37(0)&1.37(0)&1.42(0)&1.43(0)\\ $c_1$&-1.82(1)&-1.82(1)&-2.01(1)&-2.01(1)&-1.51(1)&-1.50(1)\\ $c_2$&1.66(5)&1.66(5)&1.04(3)&1.03(3)&0.51(2)&0.51(2)\\ $c_3$&-4.00(5)&-3.99(5)&-3.71(4)&-3.70(4)&-2.44(2)&-2.43(2)\\ $c_4$&1.83(2)&1.83(2)&1.48(2)&1.48(2)&1.14(2)&1.13(2)\\ $d_{1+2}$&0.10(3)&0.09(3)&0.47(3)&0.46(3)&0.39(2)&0.37(2)\\ $d_3$&-1.22(4)&-1.21(4)&-1.02(4)&-1.01(4)&-1.57(2)&-1.56(2)\\ $d_5$&0.68(2)&0.68(2)&0.35(2)&0.35(2)&0.66(1)&0.66(1)\\ $d_{14-15}$&-0.69(7)&-0.67(7)&-0.70(7)&-0.67(7)&-0.46(4)&-0.44(4)\\ $g_1$&-3.05(10)&-3.04(10)&-2.90(9)&-2.85(9)&-0.42(7)&-0.41(7)\\ 
\hline
$\chi^2_{\pi N}$/dof &0.97 &0.97 &1.04 &1.02 &0.98 &0.98\\\hline
$\bar\chi^2_{\pi N}$/dof &2.5 &2.5 &2.4 &2.4&1.87&1.87\\\hline
$\varepsilon^4$ &\multicolumn{1}{c}{$\pi$N}&\multicolumn{1}{c|}{$\pi$N+RS} &\multicolumn{1}{c}{$\pi$N}&\multicolumn{1}{c|}{$\pi$N+RS}&\multicolumn{1}{c}{$\pi$N}&\multicolumn{1}{c}{$\pi$N+RS}\\\hline 
$h_A$&1.47(0)&1.47(0)&1.63(0)&1.63(0)&1.45(0)&1.45(0)\\ $c_1$&-1.55(2)&-1.48(2)&-1.31(2)&-1.14(2)&-1.63(2)&-1.52(2)\\ $c_2$&-0.81(6)&-0.66(5)&-0.66(8)&-0.09(6)&-0.13(3)&0.02(2)\\ $c_3$&-1.00(4)&-0.99(4)&0.85(3)&0.95(2)&-1.30(3)&-1.21(3)\\ $c_4$&1.76(5)&1.84(4)&0.90(6)&1.29(4)&1.78(3)&1.82(3)\\ $d_{1+2}$&1.32(4)&1.39(3)&-1.15(6)&-0.82(5)&1.43(3)&1.38(2)\\ $d_3$&-1.89(3)&-1.93(3)&0.09(3)&-0.06(3)&-3.10(3)&-3.00(2)\\ $d_5$&0.34(2)&0.32(1)&0.62(2)&0.52(1)&1.01(1)&0.98(1)\\ $d_{14-15}$&-2.19(6)&-2.32(5)&2.13(9)&1.73(7)&-2.90(5)&-2.86(5)\\ $e_{14}$&2.25(9)&2.13(8)&1.01(10)&0.39(8)&1.56(6)&1.44(5)\\ $e_{15}$&-4.90(18)&-4.63(16)&-0.47(26)&1.16(21)&-3.32(9)&-3.18(8)\\ $e_{16}$&4.64(15)&4.21(13)&-0.24(27)&-2.24(20)&2.85(8)&2.43(6)\\ $e_{17}$&-1.00(8)&-0.97(8)&-0.82(7)&-0.66(7)&-0.07(5)&-0.05(5)\\ $e_{18}$&0.99(15)&0.79(10)&0.99(15)&0.40(12)&1.33(6)&1.28(5)\\ $g_1$&-2.63(18)&-2.73(17)&-2.01(21)&-3.08(12)&-2.13(13)&-2.27(12)\\ $b_4$&1.32(28)&1.36(23)&2.88(13)&3.09(12)&3.01(12)&3.26(11)\\ $b_5$&-0.85(51)&-1.14(36)&2.37(17)&2.48(16)&2.37(16)&2.45(16)\\ 
\hline
$\chi^2_{\pi N}$/dof &1.48 &1.48&1.59 &1.59 &1.62&1.62\\\hline
$\bar\chi^2_{\pi N}$/dof  &1.80 &1.80 &1.84 &1.86 &1.87&1.88\\\hline
  \end{tabular*}
  \caption{Complex mass approach: LECs extracted from fits at order $\varepsilon^2$,
    $\varepsilon^3$, and $\varepsilon^4$ with
    $T_\pi<200$~MeV. The labels $\pi$N and $\pi$N+RS denote fits
    without and with additional constraints $\chi^2_{\mathrm{RS}}$,
    see Eqs.~\eqref{eq2:31} and \eqref{eq2:48}, respectively. The labels
 HB-NN, HB-$\pi$N, and Cov (covariant) denote the different counting schemes
 of $1/m_N$ contributions, see section \ref{sec:power-count-renorm-Delta}.}
\label{tab:FitC}
\end{table}

\newpage

\newcolumntype{C}{>{\centering\arraybackslash}p{1.95em}}
\newcolumntype{D}{>{\centering\arraybackslash}p{2.5em}}
\begin{table}[ht]
  \centering
  \begin{tabular*}{1.0\textwidth}{@{\extracolsep{\fill}} D|| C C C C
   C C C C C C C C C C C C C}\hline\hline
$K$&$h_A$&$c_1$&$c_2$&$c_3$&$c_4$&$d_{1+2}$&$d_3$&$d_5$&$d_{14-15}$&$e_{14}$&$e_{15}$&$e_{16}$&$e_{17}$&$e_{18}$&$g_{1}$&$b_{4}$&$b_{5}$\\ \hline\hline$h_A$&1&-14&-29&45&-44&-89&71&39&93&33&15&-17&2&-12&-21&-29&-18\\ \cline{3-3}$c_1$&0&12&77&20&21&17&-5&-20&-10&-5&-1&-65&-10&0&0&14&9\\ \cline{4-4}$c_2$&-2&26&94&-44&9&19&-8&-20&-20&9&13&-76&-5&2&43&-8&-5\\ \cline{5-5}$c_3$&2&5&-28&42&6&-25&23&8&36&-5&-21&17&-7&-8&-67&23&15\\ \cline{6-6}$c_4$&-2&4&4&2&24&68&-30&-59&-43&-27&-9&7&-60&1&-20&-1&26\\ \cline{7-7}$d_{1+2}$&-5&4&12&-10&21&40&-79&-44&-86&-35&-14&17&-20&7&-2&34&22\\ \cline{8-8}$d_3$&3&-1&-4&8&-8&-27&30&-19&74&23&15&-23&-14&8&-2&-40&-20\\ \cline{9-9}$d_5$&1&-2&-5&1&-8&-7&-3&7&33&22&-1&7&48&-22&5&1&-5\\ \cline{10-10}$d_{14-15}$&10&-4&-24&29&-26&-68&50&11&154&30&12&-17&-7&-2&-2&-35&-18\\ \cline{11-11}$e_{14}$&4&-2&11&-4&-17&-28&16&8&48&163&-67&10&13&-13&-1&-20&-12\\ \cline{12-12}$e_{15}$&3&-1&34&-37&-11&-23&22&0&41&-227&700&-58&-2&3&11&-12&-8\\ \cline{13-13}$e_{16}$&-4&-64&-211&30&9&31&-36&5&-59&38&-439&805&2&5&-14&13&8\\ \cline{14-14}$e_{17}$&0&-7&-9&-9&-57&-25&-15&25&-17&33&-9&9&392&-66&4&8&7\\ \cline{15-15}$e_{18}$&-1&0&3&-7&1&6&6&-9&-4&-23&13&20&-191&216&2&-2&-14\\ \cline{16-16}$g_{1}$&-3&0&77&-81&-18&-2&-2&3&-5&-2&54&-73&14&6&344&-3&0\\ \cline{17-17}$b_{4}$&-7&14&-23&44&-1&62&-63&1&-128&-74&-90&111&48&-8&-18&845&-7\\ \cline{18-18}$b_{5}$&-5&10&-15&30&40&44&-34&-4&-70&-49&-65&74&43&-65&-2&-62&997\\ 
\hline\hline
$C$&$h_A$&$c_1$&$c_2$&$c_3$&$c_4$&$d_{1+2}$&$d_3$&$d_5$&$d_{14-15}$&$e_{14}$&$e_{15}$&$e_{16}$&$e_{17}$&$e_{18}$&$g_{1}$&$b_{4}$&$b_{5}$\\ \hline\hline$h_A$&0&-39&16&-58&0&-27&31&0&34&40&-8&6&6&-25&45&-23&-4\\ \cline{3-3}$c_1$&0&5&65&33&30&40&-31&-22&-26&-35&22&-63&-8&-10&-15&8&10\\ \cline{4-4}$c_2$&0&8&30&-50&-14&-1&1&-6&-2&-9&28&-70&-6&9&48&10&0\\ \cline{5-5}$c_3$&0&3&-12&18&58&50&-35&-25&-28&-27&-11&15&-1&-29&-76&-6&12\\ \cline{6-6}$c_4$&0&4&-4&13&27&76&-41&-62&-52&-24&3&-6&0&-75&-51&-34&13\\ \cline{7-7}$d_{1+2}$&0&4&0&8&15&15&-82&-46&-77&-38&22&-27&-4&-48&-66&-7&-17\\ \cline{8-8}$d_3$&0&-2&0&-5&-7&-10&10&-10&63&30&-21&23&5&18&57&-7&24\\ \cline{9-9}$d_5$&0&-1&0&-2&-5&-3&-1&2&42&21&-8&16&-3&51&24&19&-1\\ \cline{10-10}$d_{14-15}$&0&-4&-1&-8&-17&-19&13&4&41&38&-27&26&-4&30&57&-1&31\\ \cline{11-11}$e_{14}$&1&-7&-4&-10&-11&-13&8&3&21&75&-87&64&0&8&21&-9&-10\\ \cline{12-12}$e_{15}$&0&9&28&-9&2&16&-12&-2&-32&-138&339&-82&2&5&-5&7&1\\ \cline{13-13}$e_{16}$&0&-22&-60&9&-5&-16&11&4&25&85&-232&239&2&-3&-1&-8&-1\\ \cline{14-14}$e_{17}$&0&-1&-3&0&0&-1&1&0&-2&0&3&3&64&-53&0&-1&-3\\ \cline{15-15}$e_{18}$&-1&-3&8&-19&-61&-29&9&12&30&10&13&-8&-66&240&20&11&9\\ \cline{16-16}$g_{1}$&1&-6&47&-58&-48&-46&33&7&65&32&-18&-2&0&57&323&15&20\\ \cline{17-17}$b_{4}$&-1&5&16&-7&-49&-8&-6&8&-2&-20&37&-36&-2&49&74&760&-60\\ \cline{18-18}$b_{5}$&0&12&0&25&34&-33&39&-1&99&-45&13&-5&-13&72&182&-834&2573\\ 
\hline\hline
  \end{tabular*}
  \caption{The upper and lower table correspond to the $K$-matrix ($K$) and
    complex mass approach ($C$), respectively. 
    Correlation (upper triangle) and covariance (lower triangle)
    matrices for the fits denoted by $\pi$N+RS ($K$) and $\pi$N ($C$) at order
    $\varepsilon^4$ in the HB-NN counting. The units of the 
    correlation and covariance values are $10^{-2}$ and $10^{-4}$, respectively.}
\label{tab:Q4piNCorrCov1}
\end{table}
\newpage

\begin{table}[ht]
  \centering
  \begin{tabular*}{1.0\textwidth}{@{\extracolsep{\fill}} D|| C C C C
   C C C C C C C C C C C C C}\hline\hline
$K$&$h_A$&$c_1$&$c_2$&$c_3$&$c_4$&$d_{1+2}$&$d_3$&$d_5$&$d_{14-15}$&$e_{14}$&$e_{15}$&$e_{16}$&$e_{17}$&$e_{18}$&$g_{1}$&$b_{4}$&$b_{5}$\\ \hline\hline$h_A$&1&-8&0&42&-31&-89&74&59&95&24&37&-32&4&-14&-19&-24&-16\\ \cline{3-3}$c_1$&0&15&86&19&26&18&-4&-22&-4&-10&15&-70&-10&-10&-2&11&8\\ \cline{4-4}$c_2$&0&32&92&-10&29&13&7&-23&-1&-2&36&-86&-7&-12&-32&-8&-6\\ \cline{5-5}$c_3$&2&4&-5&25&6&-31&32&16&46&-8&-10&6&-9&-10&-27&34&25\\ \cline{6-6}$c_4$&-2&6&16&2&35&64&-24&-63&-31&-28&4&-12&-36&-42&-29&-4&19\\ \cline{7-7}$d_{1+2}$&-5&4&7&-9&23&36&-74&-72&-85&-31&-25&17&-18&-8&6&24&18\\ \cline{8-8}$d_3$&2&-1&2&6&-5&-16&13&8&75&14&33&-29&-15&12&-27&-30&-16\\ \cline{9-9}$d_5$&1&-2&-6&2&-10&-12&1&7&53&31&5&3&38&-3&11&-6&-9\\ \cline{10-10}$d_{14-15}$&8&-2&-1&24&-19&-52&28&15&107&22&31&-26&-8&-1&-18&-27&-14\\ \cline{11-11}$e_{14}$&2&-5&-2&-5&-20&-22&6&10&28&150&-62&20&20&-6&10&-16&-11\\ \cline{12-12}$e_{15}$&10&18&107&-15&7&-47&38&4&100&-241&990&-72&-5&-7&-29&-13&-9\\ \cline{13-13}$e_{16}$&-10&-100&-307&11&-27&38&-40&3&-101&93&-843&1371&2&16&27&11&7\\ \cline{14-14}$e_{17}$&1&-8&-12&-8&-41&-20&-11&20&-16&46&-29&16&367&-57&5&7&6\\ \cline{15-15}$e_{18}$&-3&-9&-25&-10&-53&-11&9&-2&-2&-17&-44&128&-234&453&6&0&-8\\ \cline{16-16}$g_{1}$&-3&-1&-50&-23&-28&6&-16&5&-30&20&-151&164&17&19&273&-3&-1\\ \cline{17-17}$b_{4}$&-6&12&-20&46&-6&40&-30&-5&-76&-55&-114&109&39&1&-16&761&-3\\ \cline{18-18}$b_{5}$&-4&10&-18&39&36&34&-18&-8&-45&-43&-91&87&37&-53&-7&-30&1009\\ 
\hline\hline
$C$&$h_A$&$c_1$&$c_2$&$c_3$&$c_4$&$d_{1+2}$&$d_3$&$d_5$&$d_{14-15}$&$e_{14}$&$e_{15}$&$e_{16}$&$e_{17}$&$e_{18}$&$g_{1}$&$b_{4}$&$b_{5}$\\ \hline\hline$h_A$&0&-9&-8&15&4&-48&61&21&65&13&22&-20&6&-43&8&-37&-21\\ \cline{3-3}$c_1$&0&6&83&39&52&59&-53&-47&-48&-44&42&-71&10&-17&-50&23&15\\ \cline{4-4}$c_2$&0&16&60&45&70&80&-70&-66&-70&-65&68&-90&13&-13&-86&25&12\\ \cline{5-5}$c_3$&0&3&9&7&81&49&-28&-54&-26&-41&35&-44&23&-60&-65&22&24\\ \cline{6-6}$c_4$&0&8&33&13&37&77&-54&-79&-59&-61&54&-65&23&-62&-80&20&21\\ \cline{7-7}$d_{1+2}$&-1&8&36&7&27&33&-92&-80&-93&-67&48&-61&17&-15&-83&40&23\\ \cline{8-8}$d_3$&0&-4&-18&-2&-11&-18&11&52&89&54&-34&47&-8&-7&68&-43&-20\\ \cline{9-9}$d_5$&0&-2&-9&-2&-8&-8&3&3&72&62&-49&57&-23&37&71&-24&-17\\ \cline{10-10}$d_{14-15}$&1&-10&-47&-6&-31&-46&26&11&75&60&-37&47&-9&-4&70&-39&-18\\ \cline{11-11}$e_{14}$&0&-11&-52&-11&-38&-40&19&11&54&108&-90&78&-16&19&68&-27&-17\\ \cline{12-12}$e_{15}$&1&28&139&25&87&74&-30&-23&-85&-248&701&-91&16&-21&-69&15&9\\ \cline{13-13}$e_{16}$&-1&-48&-189&-32&-107&-96&43&27&111&220&-653&740&-17&24&79&-19&-10\\ \cline{14-14}$e_{17}$&0&2&7&4&10&7&-2&-3&-6&-12&30&-33&52&-64&-17&5&3\\ \cline{15-15}$e_{18}$&-1&-6&-16&-24&-58&-14&-4&10&-5&30&-87&99&-71&237&21&2&0\\ \cline{16-16}$g_{1}$&0&-26&-138&-36&-100&-99&47&26&126&147&-380&446&-25&67&430&-27&-16\\ \cline{17-17}$b_{4}$&-1&7&26&8&16&31&-19&-6&-45&-37&54&-69&5&4&-76&178&-12\\ \cline{18-18}$b_{5}$&-1&6&16&11&22&22&-12&-5&-27&-31&40&-48&4&0&-57&-26&293\\ 
\hline\hline
  \end{tabular*}
  \caption{The upper and lower table correspond to the $K$-matrix ($K$) and
    complex mass approach ($C$), respectively. 
    Correlation (upper triangle) and covariance (lower triangle)
    matrices for the fits denoted by $\pi$N+RS ($K$) and $\pi$N ($C$) at order
    $\varepsilon^4$ in the HB-$\pi$N counting. The units of the 
    correlation and covariance values are $10^{-2}$ and $10^{-4}$, respectively.}
\label{tab:Q4piNCorrCov2}
\end{table}
\newpage

\begin{table}[ht]
  \centering
  \begin{tabular*}{1.0\textwidth}{@{\extracolsep{\fill}} D|| C C C C
   C C C C C C C C C C C C C}\hline\hline
$K$&$h_A$&$c_1$&$c_2$&$c_3$&$c_4$&$d_{1+2}$&$d_3$&$d_5$&$d_{14-15}$&$e_{14}$&$e_{15}$&$e_{16}$&$e_{17}$&$e_{18}$&$g_{1}$&$b_{4}$&$b_{5}$\\ \hline\hline$h_A$&1&-31&-54&35&-27&-69&54&34&86&47&20&-6&-5&-7&-32&-40&-17\\ \cline{3-3}$c_1$&-1&7&78&26&12&3&34&-41&-10&-22&-38&-50&2&-7&15&28&9\\ \cline{4-4}$c_2$&-2&10&25&-35&12&35&-8&-42&-42&2&-41&-39&-4&4&38&4&-1\\ \cline{5-5}$c_3$&2&5&-12&45&14&-34&65&-8&35&-26&-2&-27&12&-22&-65&45&11\\ \cline{6-6}$c_4$&-1&1&3&4&19&49&-14&-35&-47&-23&-8&-9&-50&0&-39&-7&25\\ \cline{7-7}$d_{1+2}$&-3&0&8&-11&10&22&-54&-58&-81&-20&-15&10&-17&15&-12&8&-19\\ \cline{8-8}$d_3$&2&3&-2&17&-2&-10&15&-33&58&6&-3&-40&-6&4&-34&6&-11\\ \cline{9-9}$d_5$&1&-3&-5&-1&-4&-7&-3&7&38&19&18&25&27&-24&22&-11&29\\ \cline{10-10}$d_{14-15}$&6&-2&-18&21&-18&-33&19&8&76&31&16&-5&0&-4&-2&-20&-16\\ \cline{11-11}$e_{14}$&5&-7&1&-20&-11&-11&3&6&31&130&-55&35&2&-3&-4&-48&-17\\ \cline{12-12}$e_{15}$&3&-17&-34&-3&-6&-12&-2&8&24&-105&287&-40&-4&6&-1&-5&-3\\ \cline{13-13}$e_{16}$&0&-12&-18&-17&-4&4&-14&6&-4&37&-63&87&-7&14&18&-19&-2\\ \cline{14-14}$e_{17}$&-1&1&-3&11&-32&-12&-4&10&-1&3&-10&-9&215&-71&3&36&17\\ \cline{15-15}$e_{18}$&-1&-2&2&-15&0&7&2&-6&-4&-4&11&14&-107&105&5&-26&-14\\ \cline{16-16}$g_{1}$&-5&7&37&-85&-33&-11&-26&11&-3&-8&-5&32&9&10&377&-12&0\\ \cline{17-17}$b_{4}$&-17&38&10&155&-15&20&12&-14&-89&-279&-43&-90&268&-136&-117&2593&-11\\ \cline{18-18}$b_{5}$&-5&8&-1&23&35&-29&-13&25&-46&-63&-16&-7&80&-46&0&-184&1061\\ 
\hline\hline
$C$&$h_A$&$c_1$&$c_2$&$c_3$&$c_4$&$d_{1+2}$&$d_3$&$d_5$&$d_{14-15}$&$e_{14}$&$e_{15}$&$e_{16}$&$e_{17}$&$e_{18}$&$g_{1}$&$b_{4}$&$b_{5}$\\ \hline\hline$h_A$&0&7&-7&12&-18&-43&24&17&57&37&-22&10&-12&0&14&-9&-2\\ \cline{3-3}$c_1$&0&4&85&35&11&-19&56&-60&14&-34&14&-78&3&-14&4&20&11\\ \cline{4-4}$c_2$&0&5&8&-11&14&21&15&-57&-14&-12&5&-64&-12&-7&19&7&1\\ \cline{5-5}$c_3$&0&2&-1&12&30&-56&76&-19&12&-48&19&-49&39&-26&-68&30&7\\ \cline{6-6}$c_4$&0&1&1&3&10&30&-14&-11&-66&-21&0&-24&-1&-65&-46&-5&0\\ \cline{7-7}$d_{1+2}$&0&-1&2&-5&2&7&-76&-23&-68&8&-5&17&-14&-12&9&-21&-41\\ \cline{8-8}$d_3$&0&3&1&7&-1&-5&7&-41&45&-35&18&-55&25&5&-30&29&11\\ \cline{9-9}$d_5$&0&-2&-2&-1&0&-1&-1&2&16&34&-16&52&-8&9&5&-12&25\\ \cline{10-10}$d_{14-15}$&0&1&-2&2&-11&-9&6&1&27&13&-2&6&3&22&42&22&16\\ \cline{11-11}$e_{14}$&0&-4&-2&-9&-4&1&-5&3&4&33&-87&72&-28&18&26&-26&-14\\ \cline{12-12}$e_{15}$&0&2&1&6&0&-1&4&-2&-1&-42&73&-64&23&-2&-14&18&10\\ \cline{13-13}$e_{16}$&0&-12&-14&-13&-6&3&-11&5&2&31&-41&57&-25&18&31&-30&-11\\ \cline{14-14}$e_{17}$&0&0&-2&6&0&-2&3&-1&1&-8&9&-9&23&-48&-37&14&17\\ \cline{15-15}$e_{18}$&0&-2&-1&-5&-12&-2&1&1&6&6&-1&8&-13&33&13&-12&-6\\ \cline{16-16}$g_{1}$&0&1&7&-29&-19&3&-10&1&27&19&-15&29&-22&9&159&-5&18\\ \cline{17-17}$b_{4}$&0&5&3&13&-2&-7&10&-2&14&-18&19&-28&8&-8&-7&152&-22\\ \cline{18-18}$b_{5}$&0&4&1&4&0&-17&5&5&14&-13&14&-14&13&-6&38&-44&266\\ 
\hline\hline
  \end{tabular*}
  \caption{The upper and lower table correspond to the $K$-matrix ($K$) and
    complex mass approach ($C$), respectively. 
    Correlation (upper triangle) and covariance (lower triangle)
    matrices for the fits denoted by $\pi$N+RS ($K$) and $\pi$N ($C$) at order
    $\varepsilon^4$ in the covariant counting. The units of the 
    correlation and covariance values are $10^{-2}$ and $10^{-4}$, respectively.}
\label{tab:Q4piNCorrCov3}
\end{table}
\newpage

\newcolumntype{F}{>{\centering\arraybackslash}p{6.5em}}
\begin{table}[ht]
\vspace{-0.6cm}
  \centering
  \begin{tabular*}{0.9\textwidth}{@{\extracolsep{\fill}} c| F |  F|| F |  F||
     c}
  \multicolumn{1}{c}{}  &\multicolumn{2}{c}{$K$-matrix approach}
    &\multicolumn{2}{c}{Complex mass approach} \\\hline
   $\varepsilon^4$  &\multicolumn{1}{c}{$\pi$N} &\multicolumn{1}{c}{$\pi$N+RS} &\multicolumn{1}{c}{$\pi$N} &\multicolumn{1}{c}{$\pi$N+RS} &\multicolumn{1}{c}{RS} \\\hline
$d_{00}^+[M_\pi^{-1}]$&-1.15(7)(9)&-1.15(4)(9)&-0.71(3)(5)&-1.07(3)(9)&-1.36(3)\\ $d_{10}^+[M_\pi^{-3}]$&1.14(11)(13)&1.10(6)(13)&0.50(4)(5)&0.84(4)(10)&1.16(2)\\ $d_{01}^+[M_\pi^{-3}]$&1.24(3)(1)&1.19(2)(0)&1.07(1)(0)&1.22(1)(1)&1.16(2)\\ $d_{20}^+[M_\pi^{-5}]$&0.04(5)(6)&0.07(3)(5)&0.30(2)(1)&0.29(1)(2)&0.196(3)\\ $d_{11}^+[M_\pi^{-5}]$&0.10(2)(2)&0.14(1)(1)&0.21(1)(0)&0.22(1)(1)&0.185(3)\\ $d_{02}^+[M_\pi^{-5}]$&0.046(4)(3)&0.052(3)(3)&0.052(2)(3)&0.045(2)(3)&0.0336(6)\\ $b_{00}^+[M_\pi^{-3}]$&-1.88(10)(24)&-2.24(1)(16)&-1.77(8)(16)&-2.95(6)(9)&-3.45(7)\\ $d_{00}^-[M_\pi^{-2}]$&1.03(1)(9)&1.06(1)(9)&0.95(1)(9)&1.03(1)(9)&1.41(1)\\ $d_{10}^-[M_\pi^{-4}]$&0.21(2)(10)&0.17(2)(9)&0.27(1)(10)&0.17(1)(9)&-0.159(4)\\ $d_{01}^-[M_\pi^{-4}]$&-0.155(3)(3)&-0.158(3)(3)&-0.139(3)(4)&-0.189(2)(6)&-0.141(5)\\ $b_{00}^-[M_\pi^{-2}]$&11.31(43)(13)&10.51(11)(5)&9.08(20)(16)&10.12(11)(9)&10.49(11)\\ $b_{10}^-[M_\pi^{-4}]$&-0.50(30)(22)&-0.01(11)(12)&0.82(11)(2)&0.75(7)(3)&1.00(3)\\ $b_{01}^-[M_\pi^{-4}]$&0.24(8)(3)&0.22(7)(2)&0.36(3)(5)&0.32(3)(5)&0.21(2)\\ 
\hline
  \end{tabular*}
  \begin{tabular*}{0.9\textwidth}{@{\extracolsep{\fill}} c| F |  F|| F |  F||
     c}\hline
$d_{00}^+[M_\pi^{-1}]$&-1.35(8)(20)&-1.28(5)(18)&-0.58(5)(6)&-0.96(4)(14)&-1.36(3)\\ $d_{10}^+[M_\pi^{-3}]$&1.47(13)(27)&1.33(8)(24)&0.42(7)(8)&0.97(5)(19)&1.16(2)\\ $d_{01}^+[M_\pi^{-3}]$&1.23(3)(3)&1.19(2)(2)&0.88(2)(4)&1.00(1)(2)&1.16(2)\\ $d_{20}^+[M_\pi^{-5}]$&-0.10(6)(9)&-0.03(4)(7)&0.14(3)(5)&-0.044(20)(86)&0.196(3)\\ $d_{11}^+[M_\pi^{-5}]$&0.11(3)(2)&0.15(2)(1)&0.18(1)(1)&0.10(1)(3)&0.185(3)\\ $d_{02}^+[M_\pi^{-5}]$&0.051(4)(3)&0.056(3)(3)&0.037(2)(3)&0.024(2)(6)&0.0336(6)\\ $b_{00}^+[M_\pi^{-3}]$&-0.70(10)(53)&-0.91(9)(49)&0.37(6)(77)&0.36(6)(76)&-3.45(7)\\ $d_{00}^-[M_\pi^{-2}]$&0.93(1)(11)&0.94(1)(10)&0.85(1)(11)&0.89(1)(10)&1.41(1)\\ $d_{10}^-[M_\pi^{-4}]$&0.51(2)(15)&0.49(2)(15)&0.60(1)(17)&0.50(1)(15)&-0.159(4)\\ $d_{01}^-[M_\pi^{-4}]$&-0.059(3)(24)&-0.063(3)(23)&0.016(2)(42)&0.01(2)(40)&-0.141(5)\\ $b_{00}^-[M_\pi^{-2}]$&11.28(42)(34)&10.59(20)(20)&7.51(17)(33)&8.23(14)(19)&10.49(11)\\ $b_{10}^-[M_\pi^{-4}]$&0.17(30)(9)&0.63(15)(1)&1.66(11)(16)&1.33(9)(9)&1.00(3)\\ $b_{01}^-[M_\pi^{-4}]$&0.28(7)(4)&0.32(6)(5)&0.28(2)(4)&0.23(2)(3)&0.21(2)\\ 
\hline
  \end{tabular*}
  \begin{tabular*}{0.9\textwidth}{@{\extracolsep{\fill}} c| F | F|| F |  F||
    c}\hline
$d_{00}^+[M_\pi^{-1}]$&-1.06(5)(1)&-1.23(2)(2)&-0.84(2)(5)&-0.98(2)(2)&-1.36(3)\\ $d_{10}^+[M_\pi^{-3}]$&0.99(6)(2)&1.16(3)(4)&0.61(2)(5)&0.76(2)(2)&1.16(2)\\ $d_{01}^+[M_\pi^{-3}]$&1.22(1)(0)&1.20(1)(0)&1.21(1)(0)&1.21(1)(0)&1.16(2)\\ $d_{20}^+[M_\pi^{-5}]$&0.12(2)(2)&0.10(1)(2)&0.30(1)(1)&0.27(0)(0)&0.196(3)\\ $d_{11}^+[M_\pi^{-5}]$&0.13(1)(0)&0.15(1)(0)&0.24(0)(1)&0.23(0)(1)&0.185(3)\\ $d_{02}^+[M_\pi^{-5}]$&0.048(3)(3)&0.044(2)(2)&0.042(1)(2)&0.039(1)(2)&0.0336(6)\\ $b_{00}^+[M_\pi^{-3}]$&-2.68(7)(13)&-2.84(6)(9)&-4.41(4)(13)&-4.36(4)(13)&-3.45(7)\\ $d_{00}^-[M_\pi^{-2}]$&1.39(1)(3)&1.38(1)(3)&1.47(0)(2)&1.46(0)(2)&1.41(1)\\ $d_{10}^-[M_\pi^{-4}]$&-0.05(0)(3)&-0.07(0)(3)&-0.12(0)(2)&-0.12(0)(2)&-0.159(4)\\ $d_{01}^-[M_\pi^{-4}]$&-0.117(2)(5)&-0.120(2)(4)&-0.178(2)(4)&-0.173(2)(3)&-0.141(5)\\ $b_{00}^-[M_\pi^{-2}]$&11.17(27)(11)&10.79(11)(5)&10.36(12)(5)&10.45(8)(5)&10.49(11)\\ $b_{10}^-[M_\pi^{-4}]$&0.64(13)(8)&0.87(7)(3)&1.29(4)(2)&1.27(3)(1)&1.00(3)\\ $b_{01}^-[M_\pi^{-4}]$&0.35(5)(2)&0.32(5)(2)&0.01(2)(4)&0.00(2)(4)&0.21(2)\\ 
\hline
  \end{tabular*}
  \caption{Comparison of the subthreshold parameters determined at
    order $\varepsilon^4$ with the RS values \cite{Hoferichter:2015hva} in both unitarization
    approaches. The LECs from Tables~\ref{tab:FitK} and \ref{tab:FitC} are taken as input. The results in the HB-NN, HB-$\pi$N, and covariant counting
    are given in the upper, middle, and lower table, respectively.
    The first and second bracket denote the statistical and
    theoretical uncertainty, respectively.}
\label{tab:SubThrParaKvsC}
\end{table}
\newpage

\newcolumntype{G}{>{\centering\arraybackslash}p{7em}}
\begin{table}[ht]
  \centering
  \begin{tabular*}{0.95\textwidth}{@{\extracolsep{\fill}} c| G |  G|| G | G||
     c}
  \multicolumn{1}{c}{}  &\multicolumn{2}{c}{$K$-matrix approach}
    &\multicolumn{2}{c}{Complex mass approach} \\\hline\hline
   $\varepsilon^4$  &\multicolumn{1}{c}{$\pi$N} &\multicolumn{1}{c}{$\pi$N+RS}&\multicolumn{1}{c}{$\pi$N} &\multicolumn{1}{c}{$\pi$N+RS} &\multicolumn{1}{c}{RS} \\\hline\hline
$a_{0+}^+[M_\pi^{-1}10^{-3}]$&3.8(6)(6)&2.8(5)(8)&7.4(4)(4)&4.5(4)(8)&-0.9(1.4)\\ $a_{0+}^-[M_\pi^{-1}10^{-3}]$&86.1(2)(3)&86.1(2)(3)&85.1(1)(3)&85.0(1)(3)&85.4(9)\\ $a_{1+}^+[M_\pi^{-3}10^{-3}]$&128.7(3)(3)&129.1(3)(3)&123.1(3)(2)&135.1(3)(5)&131.2(1.7)\\ $a_{1+}^-[M_\pi^{-3}10^{-3}]$&-79.4(4)(3)&-78.9(2)(2)&-77.1(2)(2)&-82.0(2)(3)&-80.3(1.1)\\ $a_{1-}^+[M_\pi^{-3}10^{-3}]$&-45.9(4)(3)&-47.0(3)(3)&-50.9(3)(4)&-47.5(2)(3)&-50.9(1.9)\\ $a_{1-}^+[M_\pi^{-3}10^{-3}]$&-7.0(8)(9)&-8.7(3)(5)&-10.9(4)(4)&-9.3(3)(4)&-9.9(1.2)\\ $b_{0+}^+[M_\pi^{-3}10^{-3}]$&-59.7(2.2)(2.0)&-57.2(1.6)(2.5)&-64.7(1.2)(1.6)&-61.9(1.1)(2.0)&-45.0(1.0)\\ $b_{0+}^-[M_\pi^{-3}10^{-3}]$&15.3(3)(1.2)&15.1(3)(1.2)&16.6(1)(1.1)&17.0(1)(1.1)&4.9(8)\\ 
\hline\hline
  \end{tabular*}
\vskip 5pt
  \begin{tabular*}{0.95\textwidth}{@{\extracolsep{\fill}} c| G |  G|| G | G||
     c}\hline\hline
$a_{0+}^+[M_\pi^{-1}10^{-3}]$&3.4(6)(1.0)&3.4(6)(1.0)&6.1(4)(9)&4.3(4)(1.2)&-0.9(1.4)\\ $a_{0+}^-[M_\pi^{-1}10^{-3}]$&85.8(2)(1.0)&85.8(2)(1.0)&84.6(1)(9)&84.5(1)(9)&85.4(9)\\ $a_{1+}^+[M_\pi^{-3}10^{-3}]$&128.5(3)(4)&128.4(3)(5)&115.4(3)(3.5)&117.0(3)(3.2)&131.2(1.7)\\ $a_{1+}^-[M_\pi^{-3}10^{-3}]$&-79.3(3)(2)&-79.0(2)(2)&-73.5(2)(1.3)&-74.7(2)(1.1)&-80.3(1.1)\\ $a_{1-}^+[M_\pi^{-3}10^{-3}]$&-45.9(4)(6)&-46.7(3)(4)&-55.9(3)(1.3)&-55.0(2)(1.2)&-50.9(1.9)\\ $a_{1-}^+[M_\pi^{-3}10^{-3}]$&-6.7(7)(1.0)&-8.0(4)(8)&-12.7(4)(2)&-11.4(3)(4)&-9.9(1.2)\\ $b_{0+}^+[M_\pi^{-3}10^{-3}]$&-59.6(2.2)(3.8)&-60.1(1.9)(3.6)&-62.5(1.1)(4.0)&-57.1(1.1)(5.0)&-45.0(1.0)\\ $b_{0+}^-[M_\pi^{-3}10^{-3}]$&15.9(3)(4)&15.9(3)(4)&15.1(1)(4)&15.2(1)(4)&4.9(8)\\ 
\hline\hline
  \end{tabular*}
\vskip 5pt
  \begin{tabular*}{0.95\textwidth}{@{\extracolsep{\fill}} c| G | G|| G | G||
     c}\hline\hline
$a_{0+}^+[M_\pi^{-1}10^{-3}]$&3.5(6)(4)&0.9(4)(5)&5.4(4)(4)&3.5(3)(4)&-0.9(1.4)\\ $a_{0+}^-[M_\pi^{-1}10^{-3}]$&88.8(2)(7)&88.2(2)(7)&89.4(1)(7)&89.0(1)(7)&85.4(9)\\ $a_{1+}^+[M_\pi^{-3}10^{-3}]$&131.1(3)(6)&131.4(3)(6)&140.2(3)(9)&140.2(2)(10)&131.2(1.7)\\ $a_{1+}^-[M_\pi^{-3}10^{-3}]$&-80.0(3)(5)&-79.9(2)(5)&-83.6(2)(5)&-83.6(2)(5)&-80.3(1.1)\\ $a_{1-}^+[M_\pi^{-3}10^{-3}]$&-48.6(4)(7)&-49.1(3)(6)&-51.9(2)(6)&-51.7(2)(6)&-50.9(1.9)\\ $a_{1-}^+[M_\pi^{-3}10^{-3}]$&-7.5(7)(3)&-8.4(3)(2)&-11.2(3)(3)&-10.7(3)(2)&-9.9(1.2)\\ $b_{0+}^+[M_\pi^{-3}10^{-3}]$&-58.5(2.0)(7)&-50.0(1.2)(2.0)&-66.1(1.0)(1.1)&-60.0(8)(8)&-45.0(1.0)\\ $b_{0+}^-[M_\pi^{-3}10^{-3}]$&4.6(4)(1.6)&6.1(3)(1.3)&3.7(2)(2.0)&4.5(2)(1.8)&4.9(8)\\ 
\hline\hline
  \end{tabular*}
  \caption{Comparison of the threshold parameters determined at
    order $\varepsilon^4$ with the RS values \cite{Hoferichter:2015hva} in both unitarization
    approaches. The LECs from Tables~\ref{tab:FitK} and \ref{tab:FitC} are taken as input. The results in the HB-NN, HB-$\pi$N, and covariant counting
    are given in the upper, middle, and lower table, respectively.
    The first and second bracket denote the statistical and
    theoretical uncertainty, respectively.}
\label{tab:ThrParaKvsC}
\end{table}

\clearpage
\section{Figures}

\vspace{6cm} 

\newpage

\newpage

\newpage

\newpage

\newpage

\newpage

\newpage
\begin{figure}[ht!]
  \centering
  \includegraphics[width=0.6\textwidth]{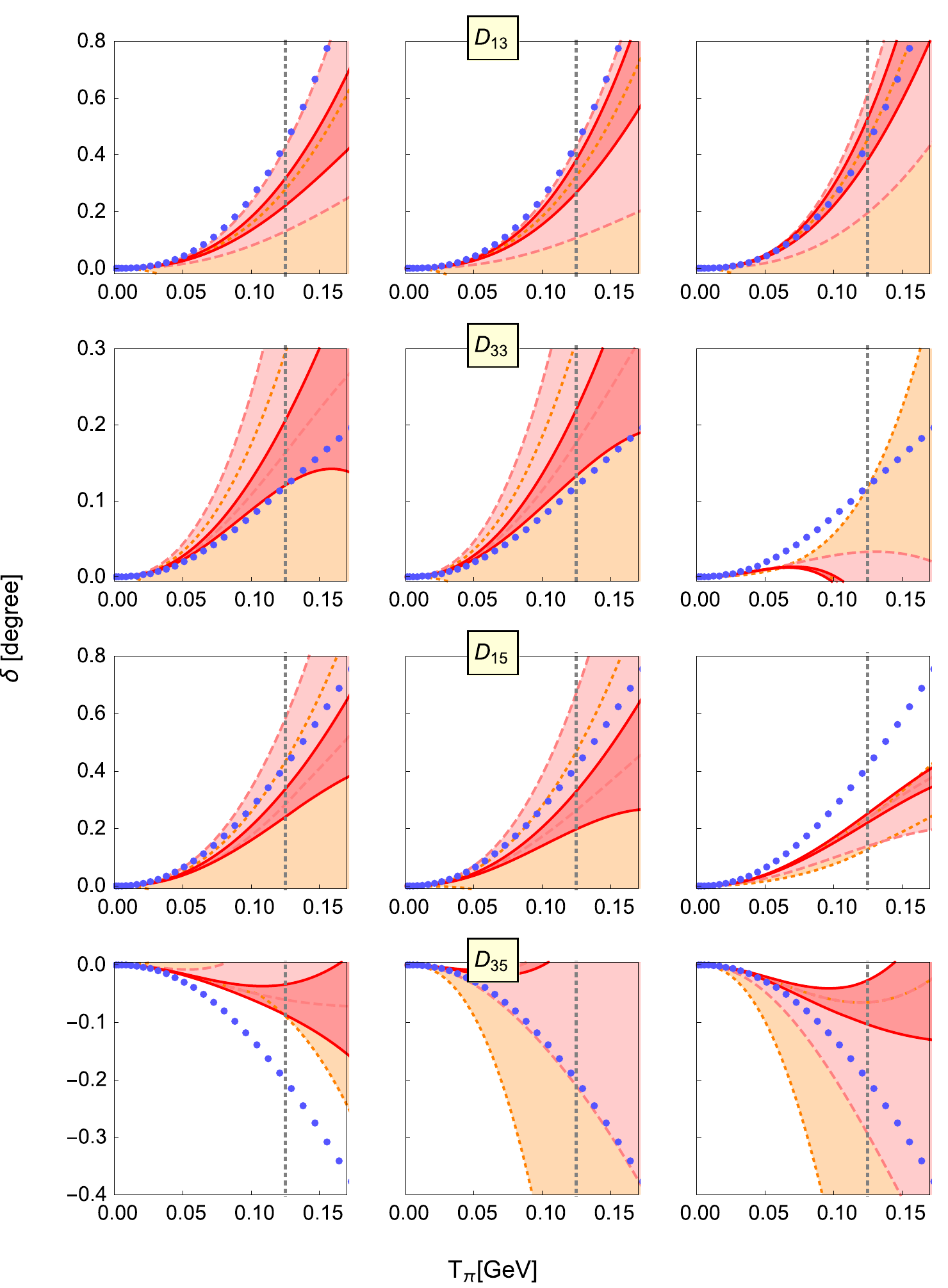}
\caption{$K$-matrix approach: Predicted $D$-wave phase shifts
  including theoretical uncertainties up to pion energies
  $T_\pi=170$~MeV. The data are taken from the GWU-SAID PWA \cite{Workman:2012hx,Igor}. For notations see Fig.~\ref{fig:SnPwavesK}.}
\label{fig:DwavesK}
\end{figure}

\newpage
\begin{figure}[ht!]
  \centering
  \includegraphics[width=0.6\textwidth]{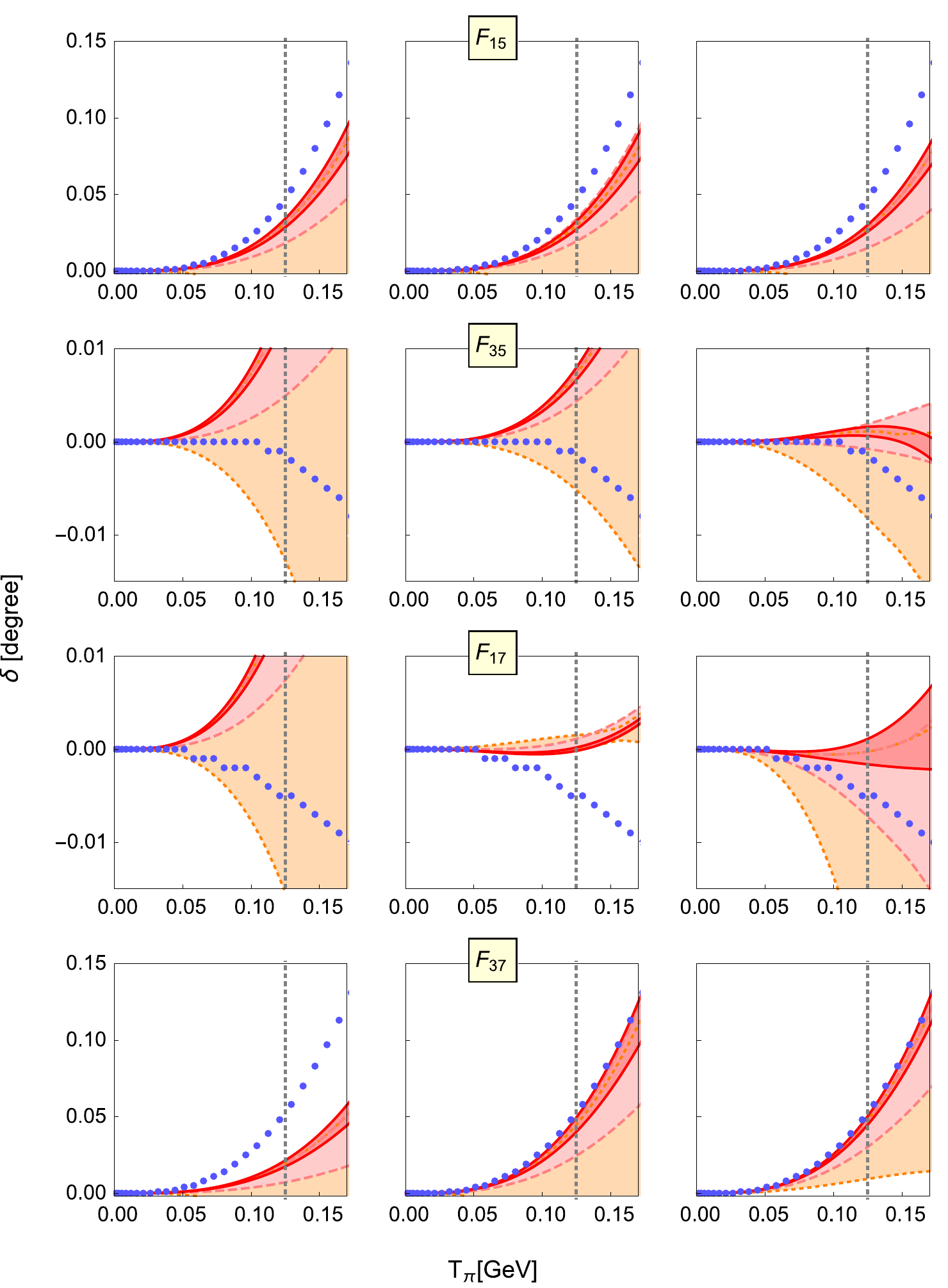}
\caption{$K$-matrix approach: Predicted $F$-wave phase shifts
  including theoretical uncertainties up to pion energies
  $T_\pi=170$~MeV. The data are taken from the GWU-SAID PWA \cite{Workman:2012hx,Igor}. For notations see Fig.~\ref{fig:SnPwavesK}.}
\label{fig:FwavesK}
\end{figure}

\newpage
\begin{figure}[ht!]
  \centering
\includegraphics[width=0.7\textwidth]{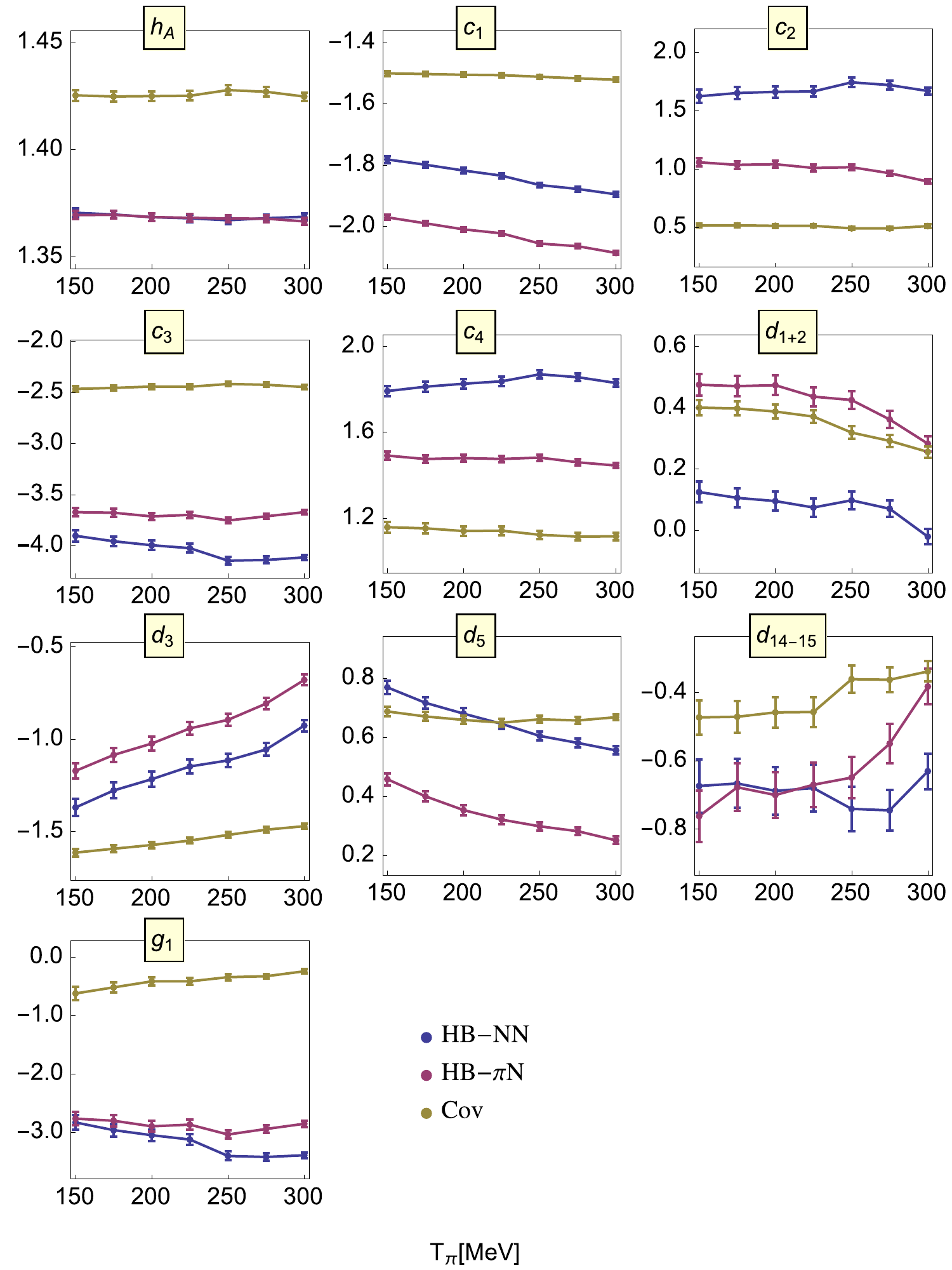}
  \caption{Complex mass approach: LECs extracted at order
    $\varepsilon^3$ as functions of maximum fit energy
    $T_\pi$, see Eq.~\eqref{eq2:31}. The labels
 HB-NN, HB-$\pi$N, and Cov (covariant) denote the different counting schemes
 of $1/m_N$ contributions, see section \ref{sec:power-count-renorm-Delta}.}
  \label{fig:LECsQ3C}
\end{figure}

\newpage
\begin{figure}[ht!]
  \centering
\includegraphics[width=0.65\textwidth]{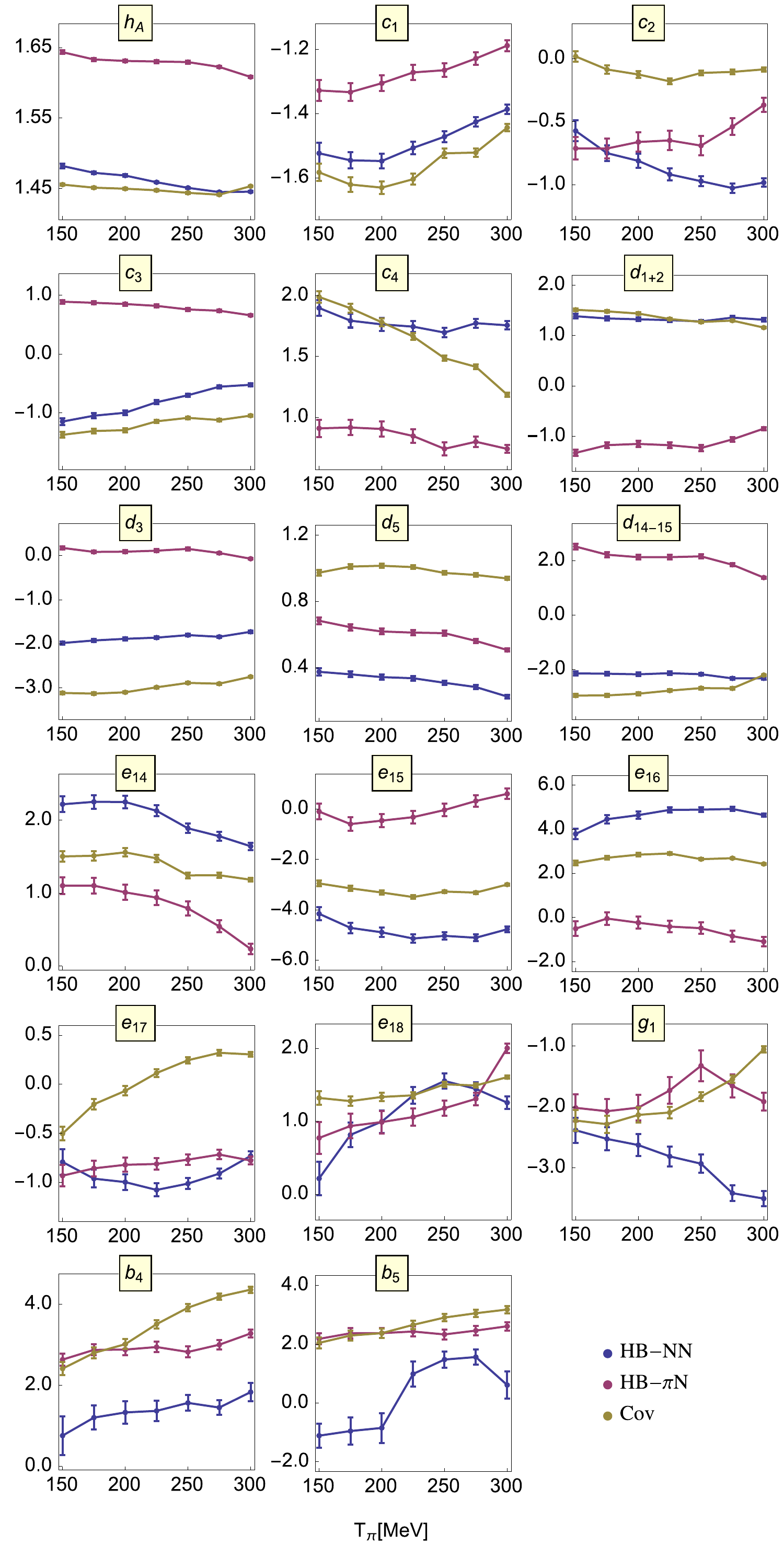}
  \caption{Complex mass approach: LECs extracted at order
    $\varepsilon^4$ as functions of maximum fit energy
    $T_\pi$, see Eq.~\eqref{eq2:31}. The labels
 HB-NN, HB-$\pi$N, and Cov (covariant) denote the different counting schemes
 of $1/m_N$ contributions, see section \ref{sec:power-count-renorm-Delta}.}
  \label{fig:LECsQ4C}
\end{figure}

\newpage

\newpage

\newpage
\clearpage
\newpage

\clearpage

\newpage

\newpage
\begin{figure}[ht]
  \centering
  \includegraphics[width=0.6\textwidth]{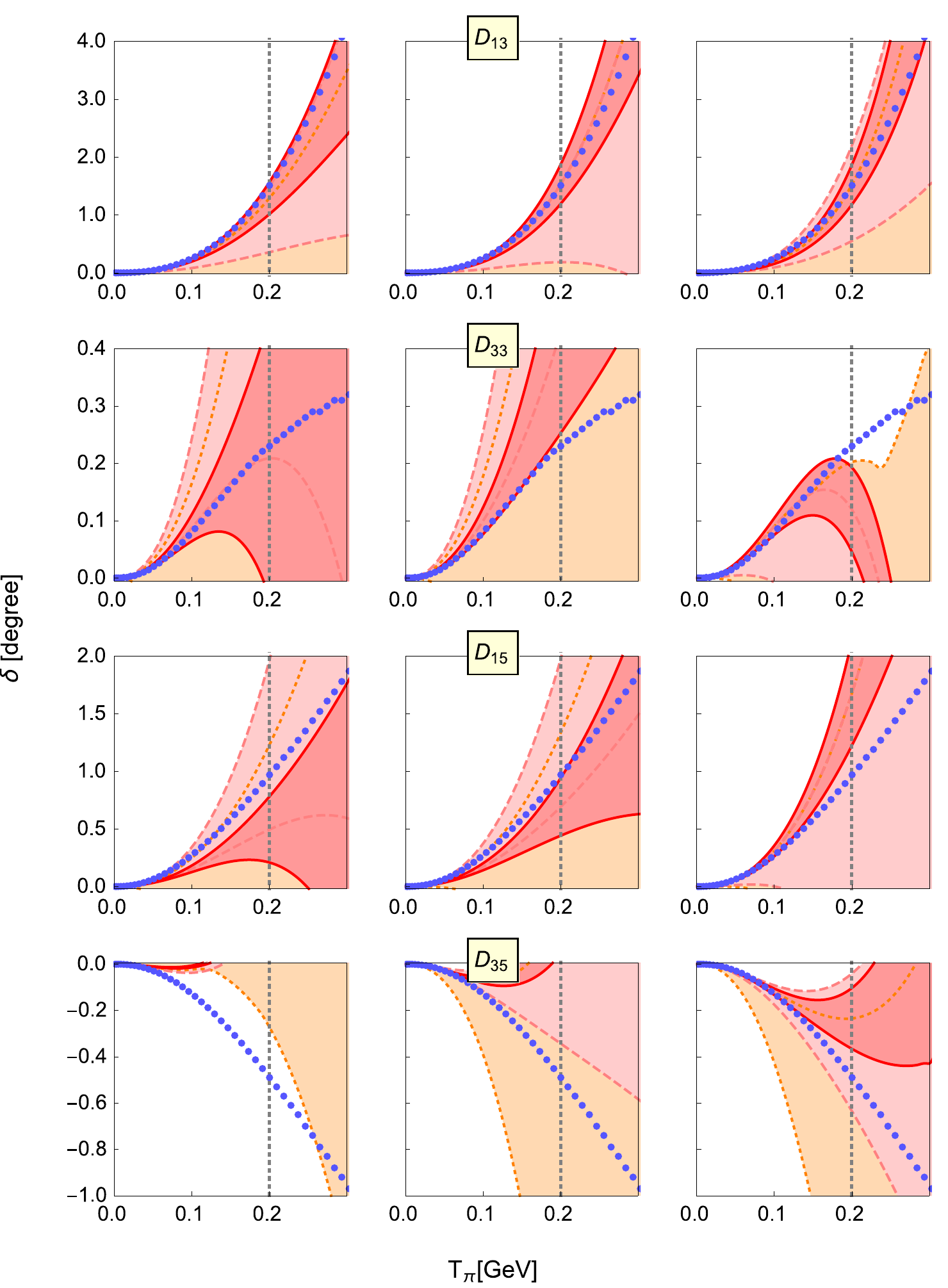}
\caption{Complex mass approach: Predicted $D$-wave phase shifts
  including theoretical uncertainties up to pion energies
  $T_\pi=300$~MeV. The data are taken from the GWU-SAID PWA \cite{Workman:2012hx,Igor}. For notations see Fig.~\ref{fig:SnPwavesC}.}
\label{fig:DwavesC}
\end{figure}

\newpage
\begin{figure}[ht]
  \centering
  \includegraphics[width=0.6\textwidth]{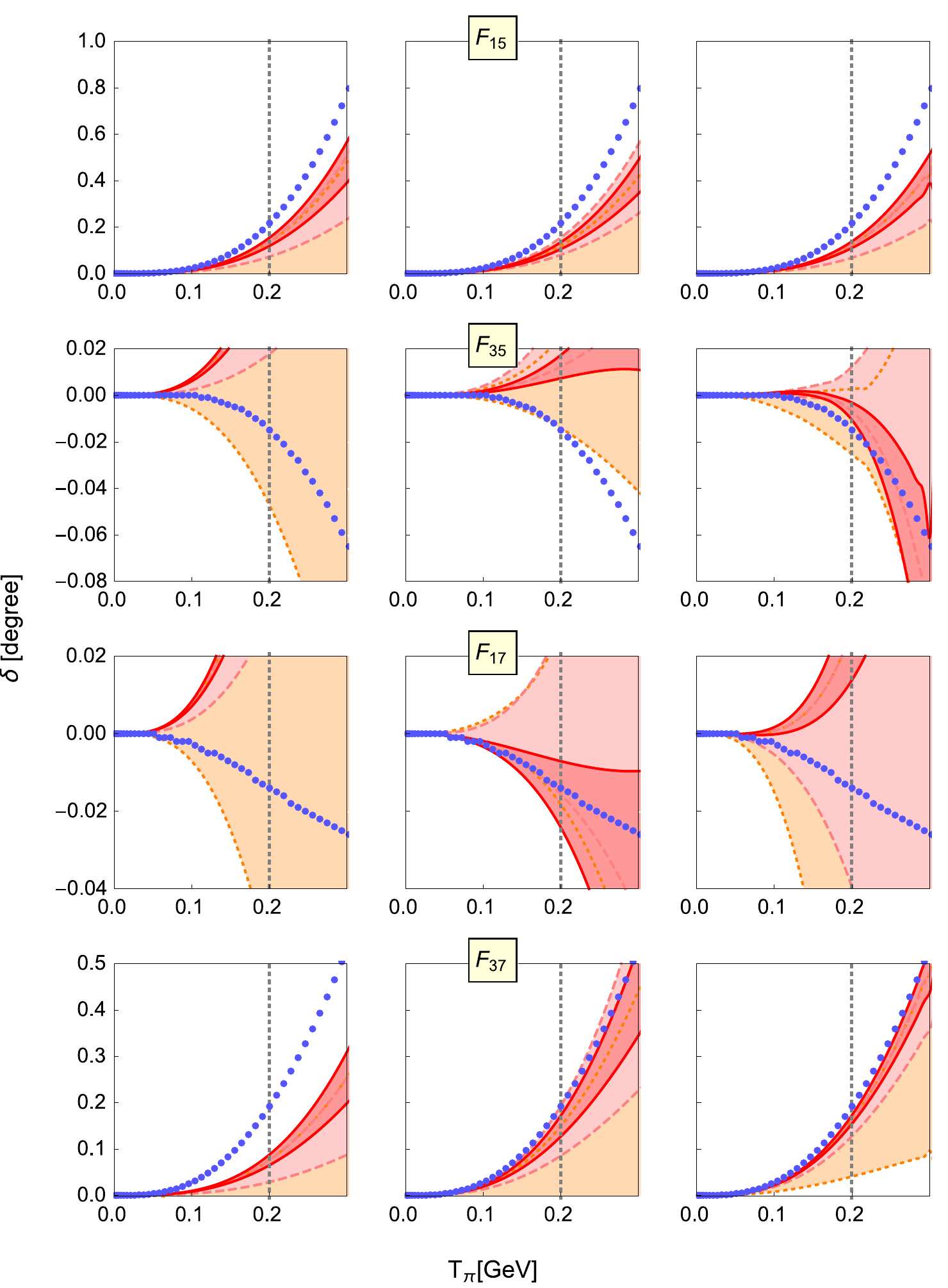}
\caption{Complex mass approach: Predicted $F$-wave phase shifts
  including theoretical uncertainties up to pion energies
  $T_\pi=300$~MeV. The data are taken from the GWU-SAID PWA \cite{Workman:2012hx,Igor}. For notations see Fig.~\ref{fig:SnPwavesC}.}
\label{fig:FwavesC}
\end{figure}

\end{document}